\documentclass[12pt]{elsarticle}
\usepackage{graphicx} 
\usepackage{amsfonts}
\usepackage{mystyle}

\usepackage{titlesec}
\titleformat{\paragraph}
{\normalfont\normalsize\bfseries}{\theparagraph}{1em}{}
\titlespacing*{\paragraph}
{0pt}{3.25ex plus 1ex minus .2ex}{1.5ex plus .2ex}

\usepackage{mathrsfs}
\usepackage{amssymb}
\usepackage{amsmath}
\usepackage{graphicx}
\usepackage{lipsum}
\usepackage{multirow}
\usepackage[colorlinks=True]{hyperref}
\usepackage{xcolor}
\usepackage{comment}
\usepackage[normalem]{ulem}
\usepackage[modulo]{lineno}
\newcommand{\p}{\partial}
\usepackage{subfigure}

\newcommand\figref[1]{Figure \ref{fig:#1}} 
\newcommand\tabref[1]{Table \ref{tab:#1}} 
\newcommand\eref[1]{Eq. (\ref{eq:#1})} 

\begin{document}

\begin{frontmatter}

\title{ECLEIRS: Exact conservation law embedded identification of reduced states for parameterized partial differential equations from sparse and noisy data}
\author[LANL]{Aviral Prakash\corref{cor1}}
\ead{aviralp@lanl.gov}\cortext[cor1]{Corresponding author}
\author[LANL]{Ben S. Southworth}
\author[LANL]{Marc L. Klasky}

\address[LANL]{Theoretical Division, Los Alamos National Laboratory, Los Alamos, NM 87545, USA}


\begin{abstract}
    Multi-query applications such as parameter estimation, uncertainty quantification and design optimization for parameterized PDE systems are expensive due to the high computational cost of high-fidelity simulations. Reduced/Latent state dynamics approaches for parameterized PDEs offer a viable method where high-fidelity data and machine learning techniques are used to reduce the system's dimensionality and estimate the dynamics of low-dimensional reduced states. These reduced state dynamics approaches rely on high-quality data and struggle with highly sparse spatiotemporal noisy measurements typically obtained from experiments. Furthermore, there is no guarantee that these models satisfy governing physical conservation laws, especially for parameters that are not a part of the model learning process. In this article, we propose a reduced state dynamics approach, which we refer to as ECLEIRS, that satisfies conservation laws exactly even for parameters unseen in the model training process. ECLEIRS is demonstrated for two applications: 1) obtaining clean solution signals from sparse and noisy measurements of parametric systems, and 2) predicting dynamics for unseen system parameters. We compare ECLEIRS with other reduced state dynamics approaches, those that do not enforce any physical constraints and those with physics-informed loss functions, for three shock-propagation problems: 1-D advection, 1-D Burgers and 2-D Euler equations. The numerical experiments conducted in this study demonstrate that ECLEIRS provides the most accurate prediction of dynamics for unseen parameters even in the presence of highly sparse and noisy data. We also demonstrate that ECLEIRS yields solutions and fluxes that satisfy the governing conservation law up to machine precision for unseen parameters, while the other methods yield much higher errors and do not satisfy conservation laws.
\end{abstract}

\begin{keyword}
 Exact conservation \sep Reduced/Latent state dynamics \sep Sparse and noisy data \sep Shock-propagation problems \sep Scientific machine learning 
 \end{keyword}

\end{frontmatter}

\section{Introduction}

Over the past few decades, advances in scientific computation have enabled the prediction of complex physical phenomena by solving relevant partial differential equations with accurate discretization techniques. As these simulation methods have advanced predictive capabilities with high accuracy, we referred to them as high-fidelity models. Despite the availability of numerous accurate high-fidelity models, these simulations have a high computational expense that make these models expensive for multi-query forward and inverse problems such as parameter estimation, design space exploration or optimization and uncertainty quantification. These multi-query scenarios benefit from lower-fidelity models that provide reasonably accurate results at a small fraction of the cost of high-fidelity methods. This article focuses on lower-fidelity models in the context of time-dependent dynamical systems obtained for parameterized partial differential equations (PDEs). 

Recently, there has been a growing interest in machine learning techniques that utilize large amounts of data to predict spatiotemporal solution dynamics. Although there are several methods to tackle this problem, this article focuses on approaches that reduce high-dimensional systems into a low-dimensional reduced states space and determine dynamics for this low-dimensional state space. This strategy is also followed for reduced order models (ROMs) \cite{Sirovich1987, Aubry1988, Benner2015, Prakash2024a}. ROMs rely on large amounts of high-fidelity simulation data to reduce the spatial dimensionality of a PDE problem while evolving the resulting low-dimensional reduced states using the underlying physical equations. As the physical equations are used for evolving the dynamics of reduced states, these methods can be robust and sufficiently accurate in some scenarios. However, ROMs are often tied to the underlying high-fidelity simulation code, for obtaining consistent gradients and projecting the governing equations, which makes these methods intrusive. Therefore, these methods are less practical for scenarios where the data is from sparse experimental measurements or situations where access to underlying simulation code is unavailable, for example, situations involving commercial and national security software. Recently, there has been growing interest in developing nonintrusive projection-based ROMs \cite{Peherstorfer2016, Gkimisis2024, Padovan2024, Puri2025, Prakash2024b, Prakash2025} that overcome this drawback, while still using relevant governing physical equations for dynamics prediction. As these projection-based ROMs rely on physical equations to determine the dynamical equations for reduced states, applying them to scenarios where governing equations are unknown or exhibit a complex nonlinear form is non-trivial. For such scenarios, there is recent interest in methods determining low-dimensional dynamical equations from data.

\subsection{Summary of reduced/latent state dynamics methods}

Reduced or latent state dynamical models use high-fidelity data to construct an approximate model of full spatiotemporal dynamics, instead of using governing equations. Therefore, these methods offer a nonintrusive approach for modeling spatiotemporal dynamics. Similar to ROMs, reduced state dynamics involves a two-step process: 1) reduction of a high-dimensional system arising from PDE solutions to a low-dimensional system using modern artificial neural network architectures such as autoencoders or implicit neural representations, and 2) learning the dynamics of the low-dimensional system using different dynamical system identification methods such as long-short term memory (LSTM), SINDy \cite{Brunton2016}, neural ODEs \cite{Chen2018, Lee2021a} and transformers \cite{Vaswani2017, Chen2022}. Early work in this area for parameterized problems \cite{Lee2021a} used autoencoder architecture to reduce dimensionality and parametrized neural ODEs (PNODEs) for learning dynamics of latent states. Another common approach, latent space dynamics identification (LaSDI) \cite{Fries2022} and its variants \cite{Bonneville2024}, has also been successfully used for parameterized PDE problems. 
This method is similar to \cite{Lee2021a} with the main difference being the approach for identification of latent space dynamics using more interpretable ODE identification methods such as SINDy and its variants \cite{Messenger2021}. As these methods rely on autoencoder-decoder architectures, these methods are limited to fixed spatial grid data. To overcome this issue Dynamics-aware Implicit Neural Representation (DINo) is introduced in \cite{Yin2023} where implicit neural representations (INRs) are used to reduce dimensionality while neural ODEs are used to learn the dynamics in the latent space. While these serve as a suitable alternative to grid-based reduced state dynamics models discussed earlier, the applicability of the proposed frameworks was not demonstrated for parametrized systems. INRs-based reduced state dynamics models that use a coefficient variant of PNODE, called HyperPNODE, were proposed in \cite{Wen2023} to enable modeling solutions of parameterized PDEs while being applicable to arbitrary spatiotemporal data locations. 

\subsection{Adding physical conservation constraints to data-driven lower-fidelity models}

Several approaches exist for adding physical conservation law constraints in data-driven lower-fidelity models. These methods can be divided into three categories: 1) adding physical constraints in the optimization problem to determine these lower-fidelity models, 2) adding a projection step that ensures that the lower-fidelity solution is projected onto a space that satisfies conservation laws and 3) embedding the physical constraints in the lower-fidelity model form.

The first approach typically involves formulating a PDE-constrained optimization problem to obtain the unknown lower-fidelity model. A common approach for solving this optimization problem is by using a penalty method approach where the data loss function is augmented with an additional penalized physical constraint term, as popularized in physics-informed neural networks (PINNs) \cite{Raissi2019}. For reduced states dynamics problems, this formulation is shown to be useful for fine-tuning the prediction during dynamics forecasting \cite{Wen2023}. However, this approach will not guarantee the satisfaction of conservation laws at parameters or time instances that are not used in the model training process. In the context of a reduced state dynamics approach, the ability of the model to perform well for parameters not included in the model training is essential, thereby highlighting the need for approaches that could satisfy conservation laws also during the inference stage for arbitrary system parameters. 

The second approach involves projecting the obtained solution onto a solution space that satisfies the conservation law exactly. Earlier work in this area \cite{Carlberg2018, Lee2021b}, in the context of ROMs, ensured that the evolution of dynamical equations of reduced states involves a projection step that solves an equality-constrained optimization problem to ensure conservation over local subdomains. Similar approaches are also proposed in \cite{Negiar2022, Duruisseaux2024}, where conservation is enforced by projecting the solution to a function space that satisfies conservation by solving equality-constrained optimization problems. The latter approach gives flexibility in selecting arbitrary grid points at which linear differential constraints are satisfied. 

The third approach involves the selection of a model form ensuring that the lower-fidelity model obeys the underlying conservation laws by construction for any parameters, even those that are outside the training dataset. Early work in this direction involves the work by \cite{Muller2022} to learn the Lagrangian of the problem that allows exact preservation of the conservation laws derived following Noether's theorem. The work by \cite{Mohan2023}, uses the vector calculus identities to identify the form of velocity representation that satisfies divergence-free conditions for incompressible fluid flows. This approach is similar to the work by \cite{Richter2022}, where the continuity equation was embedded in the model form for mass conservation. In \cite{Richter2022}, conservation law is viewed as a space-time divergence-free condition and suitable vectors are identified that ensure this condition. This approach follows the comprehensive work in \cite{Barbarosie2011, Kelliher2021} to identify divergence-free forms for higher-dimensional vector fields. 

\subsection{Proposed method and contributions}

This article focuses on reduced/latent state dynamics identification from data. To our knowledge, there is no prior work on exactly enforcing conservation laws for reduced state dynamics approaches. Furthermore, there are limited studies that rigorously evaluate the performance of reduced state dynamics approaches in the presence of spatiotemporal sparse data and with added noise. Sparse and noisy data scenarios commonly exist in experimental measurements of physical phenomena and it is integral to develop lower-fidelity models that can perform well in these scenarios.  In this article, we propose an INR-based reduced state dynamics identification approach, which we refer to as exact conservation law-embedded identification of reduced states (ECLEIRS), that applies to arbitrarily distributed space-time noisy data and ensures exact satisfaction of conservation laws at any spatial, temporal and system parameters, even those that they may be not be included in the training dataset. We achieve this goal by combining ideas from the space-time divergence-free formulation proposed in \cite{Kelliher2021, Richter2022} with INR-based reduced state dynamics approaches \cite{Yin2023, Wen2023}. ECLEIRS can be used for two important applications in approximating solutions: 1) identifying clean solution representation from sparse and noisy data and 2) as a reduced state dynamics model for predicting solutions of parameterized PDEs. The performance of ECLEIRS is rigorously evaluated for scenarios with spatiotemporal sparsity in data and added Gaussian noise of different standard deviations. This assessment of performance includes a comparison of ECELIRS with other INR-based reduced state dynamics methods, those without conservation law information and a physics-informed variant, both of which can by definition also handle sparse data with noise. By comparing the performance of these three reduced system approaches for three example problems: 1) 1-D advection equation, 2) 1-D Burgers equation and 3) 2-D Euler equations, we demonstrate that ECLEIRS yields generally superior performance compared to the other methods for identifying clean solution representation from sparse and noisy data, while also providing more accurate performance for dynamics predictions for parameters that are outside the training dataset, and exactly satisfying local conservation laws.

\subsection{Outline}

The detailed outline of the article is given below. In Section \ref{sec:MathBack}, we give the mathematical background for the problem setting and reduced state dynamics modeling while highlighting the state-of-the-art approaches used in a mesh-free setting. In Section \ref{sec:ECELIRSFormulation}, we provide a mathematical formulation of a solution representation that exactly satisfies conservation laws. We also provide details on including this formulation in ECLEIRS and proving that it ensures global/sub-domain conservation of solution while being robust to sparse and noisy data. In Section \ref{sec:Implementation}, we discuss some key aspects of the computational implementation of this model. In Section \ref{sec:Results}, we compare the performance of ECLEIRS against other common reduced state dynamics approaches for clean solution representation and dynamics predictions for unseen parameters in the presence of sparse and noisy data. Lastly, in Section \ref{sec:Conclusions}, we conclude this article by highlighting the key observations and suggesting directions for future research.



\section{Mathematical background}
\label{sec:MathBack}

\subsection{System description}

We consider a solution vector $\pmb{q} (\pmb{x}, t; \pmb{\mu}) \in \mathbb{R}^{d_e}$ that evolves in time $t \in \mathbb{R}^+$ within a domain $\pmb{x} \in \mathcal{D}_x \subset \mathbb{R}^d$, where $d$ is the dimensionality of the spatial domain, subject to the parameters $\pmb{\mu} \in \mathcal{D}_{\mu} \subset \mathbb{R}^{d_{\mu}}$, where $d_{\mu}$ is the dimensionality of the parameterized domain. For many physical phenomena of interest, the evolution of the solution vector is governed by conservation laws of the form
\begin{equation}
    \frac{\p \pmb{q} (\pmb{x}, t; \pmb{\mu})}{\p t} + \nabla_x \cdot \pmb{f}(\pmb{q} (\pmb{x}, t; \pmb{\mu})) = 0, 
    \label{eq:cons_law}
\end{equation}
where $\pmb{f}(\pmb{q} (\cdot))$ is defined as the flux vector. The divergence operator considered in this article is of the form $\nabla_x = \frac{\p}{\p x_1} \hat{\pmb{e}}_{x_1} + \frac{\p}{\p x_2} \hat{\pmb{e}}_{x_2} + \frac{\p}{\p x_3} \hat{\pmb{e}}_{x_3}$, where $\hat{\pmb{e}}_{x_1}$, $\hat{\pmb{e}}_{x_2}$ and $\hat{\pmb{e}}_{x_3}$ are unit vectors along the Cartesian coordinate directions for $d = 3$. 


\subsection{Reduced state dynamics modeling}
\label{sec:CommonMethod}

The two key components of a reduced state dynamics framework are dimensionality reduction and identification of dynamics of the reduced states.  In Section \ref{sec:DimReduction}, we describe two common implicit neural representation-based architectures for dimensionality reduction. In Section \ref{sec:ReducedDynamics}, we discuss a method for modeling parameterized ODEs for the reduced states obtained through the dimensionality reduction approach. The model testing stage combines these two components to provide an inexpensive prediction of full-dimensional solution dynamics for parameterized PDEs. 

\subsubsection{Reduction of system's dimensionality}
\label{sec:DimReduction}

Several strategies for reducing the dimensionality of a PDE system exist. The most common strategy follows proper orthogonal decomposition (POD) that identifies low-dimensional affine subspaces for optimally representing solutions based on solution data \cite{Sirovich1987, Aubry1988}. While this strategy has been successfully applied in model order reduction, it is shown to be non-optimal for problems exhibiting a slow decay of Kolmogorov $n$-width. For such problems, nonlinear alternate approaches, such as kernel-based methods \cite{Diez2021}, polynomial manifolds \cite{Geelen2023, Barnett2022} and artificial neural networks architectures \cite{Lee2020, Kim2022}, have been shown to yield efficient dimensionality reduction.

These common dimensionality reduction approaches rely on a discrete representation of solutions on a grid. For these methods, the high-fidelity solution is approximated as
\begin{equation}
    \bar{\pmb{q}}(t; \pmb{\mu}) \approx \bar{\pmb{q}}^m ( \tilde{\pmb{q}} (t, \pmb{\mu})),
\end{equation}
where $\bar{\pmb{q}} (t; \pmb{\mu}) \in \mathbb{R}^{n_p}$ is the discrete solution vector field on $n_p$ spatial coordinate locations that could be distributed in either structured or unstructured way. In the above expression, $\bar{\pmb{q}}^m (t; \pmb{\mu}) \in \mathbb{R}^{n_p}$ is the modeled approximation (denoted as superscript $m$) of the high-fidelity discrete solution field and $\tilde{\pmb{q}} (t; \pmb{\mu}) \in \mathbb{R}^{n_r}$ is low-dimensional latent representation field, where $n_r \ll n_p$. As observed from this expression, the dimensionality reduction approaches are constrained to the coordinate grid used in the training data. Furthermore, there are no direct approaches for combining data from different grids, which can arise in parameterized PDEs due to spatiotemporal adaptivity.

Another nonlinear dimensionality reduction approach growing in prominence involves implicit neural representations or neural fields (INRs). INRs \cite{Chen2019b, Park2019, Sitzmann2020} offer an ideal approach for nonlinear dimensionality reduction for systems with slow decay of Kolmogorov $n$-width as demonstrated in \cite{Chen2023, Yin2023, Puri2025}. In particular, \cite{Pan2023, Yin2023} highlighted that INRs can be used to reduce the dimensionality of systems that may not have a suitable underlying grid structure as needed for autoencoder-based methods and can even provide more accurate representations than these grid-based methods \cite{Puri2025}. Using INRs, the high-fidelity solution is approximated as 
\begin{equation}
    \pmb{q} (\pmb{x}, t; \pmb{\mu}) \approx \pmb{q}^m (\tilde{\pmb{q}}(\pmb{\mu}, t), \pmb{x}),
    \label{eq:INR_basic}
\end{equation}
where $\pmb{q}^m (\pmb{x}, t; \pmb{\mu}) \in \mathbb{R}^{d_e}$ is the modeled solution field that depends on reduced states $\tilde{\pmb{q}} \in \mathcal{M}$, where $\mathcal{M}$ is a lower dimensional manifold where these states evolve in time for different parameters. There are two approaches for formulating INRs to represent solutions for parameterized PDE problems: 

\paragraph{Standard INR-based dimensionality reduction}
\label{sec:INRStandard}

Using autodecoder-based INRs \cite{Park2019} for dimensionality reduction involves approximating the solution as 
\begin{equation}
    \pmb{q}^m (\pmb{x}, t; \pmb{\mu}) = \pmb{d}_{\pmb{\theta}_d} (\tilde{\pmb{q}} (\pmb{\mu},t), \pmb{x}),
    \label{eq:INR_uncons}
\end{equation}
where $\tilde{\pmb{q}} \in \mathcal{M}$ are reduced states and $\pmb{d}_{\pmb{\theta}_d}$ is an multilayer perceptron (MLP) parametrized with weights $\pmb{\theta}_d$. The reduced states are represented as a continuous function of system parameters and time using an MLP
\begin{equation}
    \tilde{\pmb{q}} (\pmb{\mu},t) = \pmb{h}_{\pmb{\theta}_h}(\pmb{\mu},t),
    \label{eq:latent_hypernetwork}
\end{equation}
where the weights $\pmb{\theta}_h$ parameterize this hypernetwork $\pmb{h}_{\pmb{\theta}_h}$. Given the data for high-fidelity solution $\pmb{q}(\pmb{x}, t; \pmb{\mu})$, the unknown parameters $\pmb{\theta}_d$ and $\pmb{\theta}_h$ are identified by solving a nonconvex optimization problem
\begin{equation}
    \pmb{\theta}_d, \pmb{\theta}_h = \underset{\hat{\pmb{\theta}}_d, \hat{\pmb{\theta}}_h}{\text{arg min}} \Big\vert \Big\vert \pmb{d}_{\hat{\pmb{\theta}}_d} (\pmb{h}_{\hat{\pmb{\theta}}_h}(\pmb{\mu},t), \pmb{x}) -  \pmb{q}(\pmb{x}, t; \pmb{\mu}) \Big\vert \Big\vert_2^2.
    \label{eq:IRS_opt}
\end{equation}
In this article, we refer to reduced/latent states identified using this method for dimensionality reduction as identified reduced states (IRS). This approach is similar to the methods in \cite{Yin2023, Wen2023, Regazzoni2024} with some differences in the neural network architecture.

\paragraph{Physics-informed INR-based dimensionality reduction}
\label{sec:PI-INRStandard}

The dimensionality reduction method used in IRS does not utilize any information on the governing conservation laws and is solely based on solution data. While the addition of physics conservation information during dimensionality reduction may not seem essential, the governing equations serve as a good regularization method for low-quality solution data. Therefore, this ``physics-informed" approach involves the same solution representation used in \eref{INR_uncons} and \eref{latent_hypernetwork} while modifying the optimization problem to be informed on the underlying conservation laws. This approach for making reduced states identification ``physics-informed" augments the loss function of the optimization problem in \eref{IRS_opt} with a conservation law-based constraint. A common approach for solving this optimization problem is by using penalty formation that converts the constrained optimization problem to an unconstrained optimization problem making it simpler to solve using common gradient-based optimization methods such as ADAM \cite{Kingma2017}. This strategy is akin to the formulation of loss function for PINNs. 

Following this strategy, given the data for high-fidelity solution $\pmb{q}(\pmb{x}, t; \pmb{\mu})$, the unknown parameters $\pmb{\theta}_d$ and $\pmb{\theta}_h$ are identified by solving a nonconvex optimization problem
\begin{equation}
    \pmb{\theta}_d, \pmb{\theta}_h = \underset{\hat{\pmb{\theta}}_d, \hat{\pmb{\theta}}_h}{\text{arg min}} \underbrace{\Big\vert \Big\vert \pmb{d}_{\hat{\pmb{\theta}}_d} (\pmb{h}_{\hat{\pmb{\theta}}_h}(\pmb{\mu},t), \pmb{x}) -  \pmb{q}(\pmb{x}, t; \pmb{\mu}) \Big\vert \Big\vert_2^2}_{\text{Data loss}} + \lambda \underbrace{\Bigg\vert \Bigg\vert \frac{\p \pmb{d}_{\hat{\pmb{\theta}}_d} (\pmb{h}_{\hat{\pmb{\theta}}_h}(\pmb{\mu},t), \pmb{x})}{\p t} + \nabla \cdot \pmb{f}(\pmb{d}_{\hat{\pmb{\theta}}_d} (\pmb{h}_{\hat{\pmb{\theta}}_h}(\pmb{\mu},t), \pmb{x})) \Bigg\vert \Bigg\vert_2^2}_{\text{Conservation law loss}},
    \label{eq:PI-IRS_opt}
\end{equation}
\noindent where $\lambda$ is a user-input penalty parameter that governs the strictness of enforcement of the conservation law constraint. Lower values of $\lambda$ imply that the conservation loss function is less strictly enforced, while $\lambda \to 0$ returns to the formulation in Section \ref{sec:INRStandard}. At higher values of $\lambda$, the conservation law loss function is more strictly enforced at the cost of reduced importance of the data loss function. In addition to the availability of solution data $\pmb{q}$, this strategy also requires data for the flux $\pmb{f} (\pmb{q})$. For example, if only density in the Euler equation is being modeled, this approach will also require velocity data to assemble the conservation loss. 

While data and conservation law losses can be evaluated at different spatial and temporal coordinates during the solution of the optimization problem in \eref{IRS_opt}, we elect to use the same coordinates in this article. It can be argued that a much larger selection of spatial and temporal coordinates can be used for conservation loss in conjunction with a small number of data points to yield accurate representations. However, there is no guarantee for satisfying conservation laws for unseen parameters, which is the main focus of this article. In this article, we refer to reduced states identified using this dimensionality reduction method as physics-informed identified reduced states (PI-IRS). This approach is similar to the method in \cite{Wen2023} with some differences in the neural network architecture.

\subsubsection{Identification of reduced state dynamics}
\label{sec:ReducedDynamics}

Once the hypernetwork for reduced states is identified by determining \eref{latent_hypernetwork}, any system parameter and time instance can be probed to obtain reduced states. However, as this hypernetwork provides a mapping of reduced states as a function of time, it may not yield accurate reduced states for long time periods, which is needed for dynamics forecasting problems. In such a scenario, it is valuable to identify a dynamical equation for the evolution of reduced states. This design decision compromises the inexpensive evaluation of hypernetwork to provide more robust dynamics forecasting.

Over the years, several strategies have been proposed to learn dynamical systems from data. Some of the most common strategies are DMD \cite{Schmid2010}, SINDy \cite{Brunton2016} and neural ODEs \cite{Chen2018}. In this article, we use parametrized neural ODEs-based representations to determine the dynamical equations for reduced states of the form
\begin{equation}
    \frac{d \tilde{\pmb{q}} (t, \pmb{\mu})}{d t} = \pmb{g}_{\pmb{\theta}_g} (\tilde{\pmb{q}} (t), \pmb{\mu}),
    \label{eq:NODE_ls}
\end{equation}
where $\pmb{g}_{\pmb{\theta}_g}$ is an MLP with $\pmb{\theta}_g$ weights. Neural ODEs are commonly used for scenarios where only the data for $\tilde{\pmb{q} (t, \pmb{\mu})}$ is available but the temporal derivatives of states $\frac{d \tilde{\pmb{q}} (t, \pmb{\mu})}{d t}$ is unavailable. Such situations require adjoints-based implementations. In this article, as we already approximate the representation for $\tilde{\pmb{q}} (t, \pmb{\mu})$ as \eref{latent_hypernetwork}, we can obtain $\frac{d \tilde{\pmb{q}} (t, \pmb{\mu}) }{d t}$ using automatic differentiation, thereby simplifying the model training process. In such a scenario, the weights of the hypernetwork $\pmb{\theta}_g$ are obtained by solving nonconvex optimization problem
\begin{equation}
    \pmb{\theta}_g = \underset{\hat{\pmb{\theta}}_g}{\text{arg min}} \Big\vert \Big\vert \frac{d \tilde{\pmb{q}} (t, \pmb{\mu}) }{d t} - \pmb{g}_{\hat{\pmb{\theta}}_g} (\tilde{\pmb{q}} (t), \pmb{\mu})  \Big\vert \Big\vert_2^2.
    \label{eq:Opt_reddyn}
\end{equation}
The above formulation assumes sufficiently accurate time resolution of data is available to ensure that $\frac{d \tilde{\pmb{q}} (t, \pmb{\mu}) }{d t}$ is accurate. If this scenario does not hold, then adjoint-based neural ODE implementations can be used. 

\section{ECLEIRS formulation}
\label{sec:ECELIRSFormulation}

In this section, we provide the mathematical formulation of key components in ECLEIRS. In Section \ref{sec:ConsvExact}, we briefly describe the mathematical theory of the approach used to identify the underlying structure of solution and fluxes that exactly satisfy conservation laws.  In Section \ref{sec:ECLEIRS_dimred}, this mathematical formulation is used in conjunction with a dimensionality reduction approach, autodecoder-based INRs, to present a solution representation that exactly satisfies conservation laws. In Section \ref{sec:ECLEIRS_dynamics}, we present the formulation for identifying equations for the dynamical evolution of reduced states. In Section \ref{sec:ConservationProof}, we show that ECLEIRS also guarantee global and subdomain conservation of solution. Lastly, in Section \ref{sec:SparseNoiseApplication}, we describe the applicability of ECLEIRS for identifying clean solution representation and corresponding dynamics from sparse and noisy data. 

\subsection{Enforcement of conservation laws}
\label{sec:ConsvExact}

Many numerical techniques have been developed to conserve \eref{cons_law} in an integral sense over the spatial domain, one prominent example being finite volume methods \cite{LeVeque2002}. Of particular interest here are structure-preserving techniques that exactly preserve conservation laws. For example, \textit{div-free} discretization methods \cite{Nedelec1980, Karakashian1998} have been developed to ensure that $\nabla_x \cdot \pmb{u} (\pmb{x}, t) = 0$, that is a zero-divergence condition for a solution vector field $\pmb{u} (\pmb{x}, t) \in \mathbb{R}^p$ where $p \in \{2, 3\}$, is exactly enforced at each point in the spatial domain. These constraints are often encountered while simulating incompressible fluid and structural dynamics. The main workhorse of these approaches lies in the identification of appropriate function space for $\pmb{u}$ that ensures the \textit{div-free} constraint is exactly satisfied. Furthermore, vector calculus identities can also be used to identify suitable functional representations that exactly enforce this constraint. For example, we know that if there exists a vector field $\pmb{v} (\pmb{x}, t) \in \mathbb{R}^p$ such that
\begin{equation}
\pmb{u} (\pmb{x}, t) = \nabla_x \times \pmb{v} (\pmb{x}, t), \quad \text{then} \quad \nabla_x \cdot \nabla_x \times \pmb{v} (\pmb{x}, t) = 0.    \label{eq:vecidenini}
\end{equation}
While this vector calculus identity exists for $p \in \{2, 3\}$, this idea can be extended to higher values of $p$ using concepts from exterior calculus.

For the rest of the formulation, we consider vector fields of dimensionality $d_e = 1$ to simplify the notation. Therefore the solution is represented as a scalar field $q(\pmb{x},t;\pmb{\mu})$. The proposed method is also applicable for $d_e > 1$ as each equation can be independently treated. Inspired by the vector calculus identity described above, it has been shown in \cite{Richter2022} that conservation laws of the form in \eref{cons_law} can be written as
\begin{equation}
    \nabla_w \cdot \pmb{z} (\pmb{w}) = 0,
    \label{eq:veciden}
\end{equation}
where $\pmb{z} (\pmb{w}) = [q (\pmb{w}) \; \pmb{f} (q (\pmb{w}))]^T \in \mathbb{R}^{d+1}$ is the lifted vector field and $\nabla_w = \frac{\p}{\p t} \hat{t} + \frac{\p}{\p x_1} \hat{e}_{x_1} + \frac{\p}{\p x_2} \hat{e}_{x_2} + \frac{\p}{\p x_3} \hat{e}_{x_3}$ is used to redefine the divergence operator with respect to space-time coordinate vector $\pmb{w} = [t \; \pmb{x}] \in \mathbb{R}^{d+1}$.  Note that the vector calculus identity in \eref{vecidenini} holds only for calculus over $d+1 = \{2, 3\}$-dimensional vector fields. For high-dimensional vector fields, similar identities can be determined by using concepts from exterior calculus. Precisely, \eref{veciden} corresponds to the vanishing of the square of exterior derivative in the De Rham chain complex when $d+1 \leq 3$. For high-dimensional vector fields, the form of $\pmb{z}$ for $d+1 > 3$ was studied in \cite{Barbarosie2011, Kelliher2021}. In particular, \cite{Kelliher2021} identified that if there exists a skew-symmetric matrix field $\pmb{A} (\pmb{w}) \in \mathbb{R}^{d+1 \times d+1}$ such that $\pmb{z}$ defined as row-wise divergence of $\pmb{A} (\pmb{w})$ exactly satisfies \eref{veciden}. The above statement implies that $\pmb{z} (\pmb{w})$ has the form
\begin{equation}
    \pmb{z} (\pmb{w}) = \begin{bmatrix}
    \nabla_w \cdot \bar{\pmb{A}}_1 (\pmb{w}) \\
     \cdot \\
     \cdot \\
    \nabla_w \cdot \bar{\pmb{A}}_{d+1} (\pmb{w})\\
    \end{bmatrix},
    \label{eq:Amat}
\end{equation}
where $\bar{\pmb{A}}_i (\pmb{w})$ denotes the $i$th row of $\pmb{A} (\pmb{w})$. For $d+1=3$, this definition reduces to \eref{vecidenini}, where elements of vector $\pmb{v}$ define a skew-symmetric matrix field $\pmb{A} (\pmb{w})$. The detailed derivation of the condition in \eref{Amat} can be found in \cite{Barbarosie2011, Kelliher2021, Richter2022}.

\subsection{Embedding conservation law constraint in the dimensionality reduction method}

\label{sec:ECLEIRS_dimred}

When high-dimensional solutions are represented using INR described in Sections \ref{sec:INRStandard} and \ref{sec:PI-INRStandard}, there is no guarantee that conservation laws will be satisfied. Therefore, this article focuses on choosing a suitable solution representation to ensure that conservation laws are exactly satisfied. This is enabled by designing solution $q^m (\pmb{w}; \pmb{\mu})$ and flux $\pmb{f}^m (\pmb{w}; \pmb{\mu})$ representations to have the form
\begin{equation}
    \pmb{z}^m (\pmb{w}; \pmb{\mu}) = \begin{bmatrix}
    q^m (\pmb{w}; \pmb{\mu})\\
    \pmb{f}^m (\pmb{w}; \pmb{\mu})\\
    \end{bmatrix} = \begin{bmatrix}
    \nabla_w \cdot \bar{\pmb{A}}_1 (\pmb{w}; \pmb{\mu})\\
     \cdot \\
     \cdot \\
    \nabla_w \cdot \bar{\pmb{A}}_{d+1} (\pmb{w}; \pmb{\mu})\\
    \end{bmatrix},
    \label{eq:Amat_CINR}
\end{equation}
where $\pmb{z}^m (\pmb{w}; \pmb{\mu})$ is the modeled approximation of $\pmb{z} (\pmb{w}; \pmb{\mu})$ and $\bar{\pmb{A}}_i (\pmb{w}; \pmb{\mu})$ is the $i$th row of the skew-symmetric matrix field $\pmb{A} (\pmb{w}; \pmb{\mu}) \in \mathbb{R}^{(d+1) \times (d+1)}$ . The $\frac{d (d+1)}{2}$ components of $\pmb{A} (\pmb{w}; \pmb{\mu})$, represented as $\pmb{a} (\pmb{w}; \pmb{\mu}) \in \mathbb{R}^{\frac{d (d+1)}{2}}$, are modeled using implicit neural representation of the form
\begin{equation}
    \pmb{a} (\pmb{w}; \pmb{\mu})= \pmb{d}_{\pmb{\theta}_d} (\tilde{\pmb{q}} (\pmb{\mu},t), \pmb{x}),
    \label{eq:INR_cons}
\end{equation}
where $\pmb{d}_{\pmb{\theta}_d}$ is the MLP defined in \eref{INR_uncons}. For a $1$-D spatial domain system
\begin{equation}
    \pmb{A} (\pmb{w}; \pmb{\mu})= \begin{bmatrix}
    0 & a (\pmb{w}; \pmb{\mu})\\
    -a (\pmb{w}; \pmb{\mu}) & 0 \\
    \end{bmatrix},
\end{equation}
implying that $\pmb{a} (\pmb{w}; \pmb{\mu})$ is a scalar represented as $a (\pmb{w}; \pmb{\mu}) \in \mathbb{R}$ and 
\begin{equation}
    \pmb{z}^m (\pmb{w}; \pmb{\mu})= \begin{bmatrix}
    q^m \\
    f^m \\
    \end{bmatrix} = \begin{bmatrix}
    \frac{\p a}{\p x_1} \\
    - \frac{\p a}{\p t} \\
    \end{bmatrix}.
    \label{eq:z_m_1D}
\end{equation}
Similarly, for a $2$-D spatial domain system, 
\begin{equation}
    \pmb{A} (\pmb{w}; \pmb{\mu}) = \begin{bmatrix}
    0 & a_1 (\pmb{w}; \pmb{\mu}) & a_2 (\pmb{w}; \pmb{\mu})\\
    -a_1 (\pmb{w}; \pmb{\mu})& 0 & a_3 (\pmb{w}; \pmb{\mu})\\
    -a_2 (\pmb{w}; \pmb{\mu})& -a_3 (\pmb{w}; \pmb{\mu})& 0 
    \end{bmatrix},
\end{equation}
such that $\pmb{a} (\pmb{w}; \pmb{\mu}) = [a_1 (\pmb{w}; \pmb{\mu})\; a_2 (\pmb{w}; \pmb{\mu})\; a_3 (\pmb{w}; \pmb{\mu})]^T \in \mathbb{R}^3$ and
\begin{equation}
    \pmb{z}^m (\pmb{w}; \pmb{\mu})= \begin{bmatrix}
    q^m \\
    f^m_1 \\
    f^m_2 \\
    \end{bmatrix} = \begin{bmatrix}
    \frac{\p a_1}{\p x_1} + \frac{\p a_2}{\p x_2} \\
    -\frac{\p a_1}{\p t} + \frac{\p a_3}{\p x_2} \\
    -\frac{\p a_2}{\p t} - \frac{\p a_3}{\p x_1} \\
    \end{bmatrix}.
    \label{eq:z_m_2D}
\end{equation}
Lastly, for a system with a $3$-D spatial domain,
\begin{equation}
    \pmb{A} (\pmb{w}; \pmb{\mu}) = \begin{bmatrix}
    0 & a_1 (\pmb{w}; \pmb{\mu})& a_2 (\pmb{w}; \pmb{\mu})& a_3 (\pmb{w}; \pmb{\mu})\\
    -a_1 (\pmb{w}; \pmb{\mu})& 0 & a_4 (\pmb{w}; \pmb{\mu})& a_5 (\pmb{w}; \pmb{\mu})\\
    -a_2 (\pmb{w}; \pmb{\mu})& -a_3 (\pmb{w}; \pmb{\mu})& 0 & a_6 (\pmb{w}; \pmb{\mu})\\
    -a_3 (\pmb{w}; \pmb{\mu})& -a_5 (\pmb{w}; \pmb{\mu})& -a_6 (\pmb{w}; \pmb{\mu})& 0
    \end{bmatrix},
\end{equation}
such that $\pmb{a} (\pmb{w}; \pmb{\mu}) = [a_1 (\pmb{w}; \pmb{\mu})\; a_2 (\pmb{w}; \pmb{\mu})\; \cdot \cdot \cdot \; a_6 (\pmb{w}; \pmb{\mu})]^T \in \mathbb{R}^6$ and
\begin{equation}
    \pmb{z}^m (\pmb{w}; \pmb{\mu})= \begin{bmatrix}
    q^m \\
    f^m_1 \\
    f^m_2 \\
    f^m_3 \\
    \end{bmatrix} = \begin{bmatrix}
    \frac{\p a_1}{\p x_1} + \frac{\p a_2}{\p x_2} + \frac{\p a_3}{\p x_3}  \\
    -\frac{\p a_1}{\p t} + \frac{\p a_4}{\p x_2} + \frac{\p a_5}{\p x_3} \\
    -\frac{\p a_2}{\p t} - \frac{\p a_3}{\p x_1} + \frac{\p a_6}{\p x_3}\\
    -\frac{\p a_3}{\p t} - \frac{\p a_5}{\p x_1} - \frac{\p a_6}{\p x_2}\\
    \end{bmatrix}.
    \label{eq:z_m_3D}
\end{equation}
Through algebraic manipulation, it is observed that these satisfy $\nabla_w \cdot \pmb{z} (\pmb{w}; \pmb{\mu}) = 0$. Following the same definition of $\tilde{\pmb{q}} (t, \pmb{\mu}) = \pmb{h}_{\pmb{\theta}_h}(\pmb{\mu},t)$ as in \eref{latent_hypernetwork}, the problem reduces to determining the unknown parameters $\pmb{\theta}_d$ and $\pmb{\theta}_h$. Therefore, given data for the high-fidelity solution, these unknown parameters are identified by solving a nonconvex optimization problem
\begin{equation}
    \pmb{\theta}_d, \pmb{\theta}_h = \underset{\hat{\pmb{\theta}}_d, \hat{\pmb{\theta}}_h}{\text{arg min}} \Big\vert \Big\vert \pmb{z}^m (\pmb{w}; \pmb{\mu}) -  \pmb{z}(\pmb{x}, t; \pmb{\mu})  \Big\vert \Big\vert_2^2,
    \label{eq:ECLEIRS_opt}
\end{equation}
where $\pmb{z}^m (\pmb{w}; \pmb{\mu})$ is chosen as \eref{z_m_1D}, \eref{z_m_2D} and \eref{z_m_3D} for systems that are spatially $1$-D, $2$-D and $3$-D respectively. Similar to PI-IRS, this strategy also requires relevant data for flux $\pmb{f} (\pmb{q})$ in addition to data for the solution $\pmb{q}$. An important thing to note is that the ECLEIRS approach models both solution and fluxes. As the flux is also modeled as $\pmb{f}^m$, therefore $\pmb{f}^m \ne \pmb{f} (\pmb{q^m})$. As a result, there is an inconsistency between the true flux and the modeled flux. This inconsistency introduced by additional degrees of freedom as $\pmb{f}^m$ allows more flexibility in the model to achieve local space-time satisfaction of the conservation laws. As mentioned earlier, we call the latent dynamics models identified using this novel approach Exact Conservation Law-Embedded Identified Reduced States (ECLEIRS).

\subsection{Identification of reduced state dynamics}
\label{sec:ECLEIRS_dynamics}

Once the low-dimensional reduced states are identified through the dimensionality reduction approach in Section \ref{sec:ECLEIRS_dimred}, we need to identify equations for the dynamical evolution of reduced states $\tilde{\pmb{q}}$. This procedure for obtaining the dynamical evolution of these reduced states is similar to those used for IRS and PI-IRS and described in \ref{sec:ReducedDynamics}. The only difference between the dynamics identification of ECLEIRS compared to IRS and PI-IRS is that the data used for identifying dynamics is generated using a different dimensionality reduction approach. Once the underlying ODE for the evolution of reduced states is identified by solving the optimization problem in \eref{Opt_reddyn}, this equation can be used in the inference stage to give predictions of reduced states at different times and parameter instances. These predicted reduced states can then be used in conjunction with \eref{Amat_CINR} and \eref{INR_cons} to predict solution dynamics. As the reduced states obtained using ECLEIRS always lie in the kernel of space-time divergence-free functions, the satisfaction of conservation law is guaranteed by construction even for forecasting problems and system parameters that are not included in the training dataset. 

\subsection{Conservation guarantees for the learned reduced dynamical system}
\label{sec:ConservationProof}

As the conservation laws are exactly satisfied, the conservation law for the modeled equation can be represented as
\begin{equation}
    \frac{\p \pmb{q}^m (\pmb{t,x; \mu})}{\p t} + \nabla \cdot \pmb{f}^m (\pmb{t,x; \mu}) = \mathcal{O}(\epsilon), 
    \label{eq:cons_law_model}
\end{equation}
where $\epsilon$ is the machine precision. Integrating \eref{cons_law_model} over a subdomain $\Omega_s \subset \Omega$ with subdomain boundaries $\Gamma_s$, we get
\begin{equation}
    \frac{d \bar{\pmb{q}}^m}{d t} + \Big\vert \pmb{f}^m (\pmb{t,x; \mu}) \Big\vert_{\Gamma_s} = \mathcal{O}(\epsilon), 
    \label{eq:cons_law_model_subD}
\end{equation}
where $\bar{\pmb{q}}^m = \int_{\Omega_s} \pmb{q}^m (\pmb{t,x; \mu}) d \pmb{x}$ is the subdomain integrated solution and the second term on the left-hand side is obtained by using Gauss-divergence theorem. This equation implies that the integrated solution over each spatial subdomain is also conserved. If $\Omega_s = \Omega$ and $\Gamma_s = \Gamma$, which is the domain boundary, we can show that the above equation reduces to the equation for $\bar{\pmb{q}}^m = \int_{\Omega} \pmb{q}^m (\pmb{t,x; \mu}) d \pmb{x}$ given as
\begin{equation}
    \frac{d \bar{\pmb{q}}^m}{d t} + \Big\vert \pmb{f}^m (\pmb{t,x; \mu}) \Big\vert_{\Gamma} = \mathcal{O}(\epsilon), 
    \label{eq:cons_law_model_BigD}
\end{equation}
thus also implying global conservation of the integrated solution. These results imply that exact satisfaction of the conservation laws also ensures local and global conservation of the integrated quantity.

\subsection{Applicability for sparse and noisy training data}
\label{sec:SparseNoiseApplication}

In many scenarios, especially for experimental campaigns, high-quality data may not be available. In such scenarios, the data is expected to be sparse spatially, due to limitations in placement of sensors, and temporally due to limitations in data collection frequency for the given sensor. Even when dealing with numerical simulation data, storage of large amounts of spatiotemporal data may not be possible, especially for large degrees of freedom problems. This scenario also demands storing sparse spatiotemporal data and will benefit from reduced state dynamics methods that can leverage sparse data. 

Reduced state dynamics approaches relying on neural network architectures, such as convolutional autoencoders and graph-based networks, require a fixed spatial resolution during the learning and evaluation phase. This limits the applicability of these architectures to data with fixed spatial grids for both offline and online stages of model training and testing. This drawback is overcome by our selection of autodecoder-based INRs for dimensionality reduction. As the reduced state dynamics approach using this architecture learns a continuous function between the spatiotemporal domain and solution, this approach is directly applicable to data presented without any fixed spatiotemporal data structure. Furthermore, the offline and online stages of the model could be performed on different spatiotemporal grid resolutions, which provides flexibility in evaluating the model in specific regions of interest. 

These reduced dynamics methods are also naturally suitable for obtaining a clean solution representation from sparse and noisy datasets. Both PI-IRS and ECLEIRS exhibit two mechanisms of obtaining a denoised solution representation: a low-dimensional solution representation and enforcement of physical conservation laws. The low-dimensional solution representation by itself also denoises to a certain extent by not permitting high-dimensional noisy solutions to be represented accurately. We demonstrate this applicability through numerical experiments in the later sections. However, solely using a low-dimensional solution representation may not be sufficient for high noise and sparsity levels. Therefore, enforcement of conservation laws acts as an additional regularization mechanism for obtaining clean solution representation.

\section{Computational implementation}
\label{sec:Implementation}

The training and testing of reduced state dynamics models involve two stages: offline and online stages. The offline stage includes all the computationally expensive workloads such as obtaining data from relevant high-fidelity data sources, cleaning and assembling relevant data into a training dataset, identifying relevant model hyperparameters and using the data to learn the reduced state dynamics models. Once the model is learned using relevant data and physics in the offline stage, the online stage refers to the testing stage where the model is evaluated for parameters and time instances outside the training dataset. 


\subsection{Offline stage - Model learning}

There are two main approaches for learning the reduced state dynamics methods described in Sections \ref{sec:CommonMethod} and \ref{sec:ECELIRSFormulation}: 1) segregated and 2) integrated approaches. In the segregated approach, the dimensionality reduction (identification of $\theta_d$ and $\theta_h$) and dynamics identification of reduced states (identification of $\theta_g$) are treated separately. Conversely, dimensionality reduction and dynamics identification are treated simultaneously for the integrated approach. The former approach is simpler to implement as two neural networks, one each for dimensionality reduction and dynamics prediction, can be learned separately, whereas the latter approach requires identifying parameters for both neural networks, that is $\pmb{\theta}_d$, $\pmb{\theta}_h$ and $\pmb{\theta}_g$ in the same training process. In this article, we follow the segregated approach to demonstrate the effectiveness of the ECLEIRS framework. The segregated approach is divided into three steps:
\begin{itemize}
    \item[1.] Given data for $\bm{q}$ (and $\bm{f}$ for PI-IRS and ECLEIRS), solve the optimization problems \eref{IRS_opt} for IRS, \eref{PI-IRS_opt} for PI-IRS and \eref{ECLEIRS_opt} for ECLEIRS to provide dimensionality reduction model in \eref{INR_uncons} for IRS and PI-IRS and \eref{INR_cons} for ECLEIRS. This data for $\bm{q}$ and $\bm{f}$ may be sparsely sampled in space-time and exhibit noise. 
    \item[2.] The learned hypernetwork $\tilde{\pmb{q}} = \pmb{h}_{\hat{\pmb{\theta}}_h}(\pmb{\mu},t)$ is used to generate samples of low-dimensional reduced states for the time and parameter instances in the training dataset. Using auto-differentiation, the learned hypernetwork is also evaluated to estimate the temporal derivative of these reduced states $\frac{d \tilde{\pmb{q}}}{d t}$. 
    \item[3.] Using the data for reduced states $\tilde{\pmb{q}}$ and its temporal derivative $\frac{d \tilde{\pmb{q}}}{d t}$, optimization problem in \eref{Opt_reddyn} is solved to obtain the dynamical equation for reduced states given in \eref{NODE_ls}.
\end{itemize}

\subsection{Online stage - Model testing}

Once the reduced state dynamics model is learned in the offline stage, using either segregated or integrated approaches, the model is tested for predicting solution dynamics for unseen parameters. The testing of the model for unseen parameters involves the following steps:
\begin{itemize}
    \item[1.] In most practical scenario, only the initial solution is available and we need to obtain the initial values of the reduced states $\tilde{\pmb{q}} (t = 0, \pmb{\mu})$. Therefore, given the initial condition of the solution at a certain system parameter $\pmb{q} (\pmb{x}, t = 0; \pmb{\mu})$ (and fluxes $\pmb{f} (\pmb{x}, t = 0; \pmb{\mu})$ for PI-IRS and ECLEIRS), a nonlinear least-squares problem
    \begin{equation}
        \tilde{\pmb{q}} (t = 0, \pmb{\mu}) = \underset{\hat{\pmb{q}}}{\text{arg min}} \Big\vert \Big\vert \pmb{d}_{\hat{\pmb{\theta}}_d} (\hat{\pmb{q}}, \pmb{x}) -  \pmb{q}(\pmb{x} , t = 0; \pmb{\mu}) \Big\vert \Big\vert_2^2
    \end{equation}
    is solved for IRS and PI-IRS to obtain $\tilde{\pmb{q}} (t = 0, \pmb{\mu})$. A similar optimization problem
    \begin{equation}
        \tilde{\pmb{q}} (t = 0, \pmb{\mu}) = \underset{\hat{\pmb{q}}}{\text{arg min}} \Big\vert \Big\vert \pmb{z}^m (\pmb{x}, \hat{\pmb{q}}) -  \pmb{z}(\pmb{x}, t = 0; \pmb{\mu})  \Big\vert \Big\vert_2^2
    \end{equation}    
    is solved for ECLEIRS to obtain $\tilde{\pmb{q}} (t = 0, \pmb{\mu})$. These optimization problems are solved using the Gauss-Newton method with the initial guess $\hat{\pmb{q}}_0$ obtained using the hypernetwork $\hat{\pmb{q}}_0 = \pmb{h}_{\hat{\pmb{\theta}}_h}(t = 0, \pmb{\mu})$. 
    \item[2.] The initial reduced states $\tilde{\pmb{q}} (t = 0, \pmb{\mu})$ are used in conjunction with reduced state dynamics models in \eref{NODE_ls} and an appropriate time integration scheme to obtain $\tilde{\pmb{q}} (t, \pmb{\mu})$. 
    \item[3.] The reduced states provide a compact representation of the solution at all time instances. This compact representation is decoded using \eref{INR_uncons} for IRS and PI-IRS, or \eref{INR_cons} for ECLEIRS to provide high-dimensional solutions at any time and parameter instances.     
\end{itemize}

The selection of the time-integration scheme in Step 2 can be independent of the schemes used to generate the data. ECLEIRS is designed to exactly satisfy conservation laws, irrespective of this selection. Furthermore, the solution can be obtained in Step 3 for only a select spatial region if desired by providing appropriate values of $\pmb{x}$. This ability is useful for large-scale simulations where evaluating solutions over the entire spatial domain can be expensive.

\section{Results}
\label{sec:Results}

\begin{table}[t]
    \centering
    \caption{Abbreviations for different latent space dynamics models considered in this study.}
    \begin{tabular}{|c|c|}
    \hline
         \textbf{Abbreviations} & \textbf{Full name} \\
    \hline
          IRS & Identified Reduced States \\
          PI-IRS & Physics-informed Identified Reduced States \\
          ECLEIRS & Exact Conservation Law-Embedded Identified Reduced States \\
        \hline
    \end{tabular}
    \label{tab:Abbrev_models}
\end{table}

In this section, we assess and compare the performance of different reduced state dynamics approaches for: 1) the 1-D advection problem, 2) the 1-D Burgers problem and 3) the 2-D Euler problem. The abbreviations for different reduced state dynamics approaches are summarized in \tabref{Abbrev_models}. The details about the neural network architecture for various sub-networks, that is hypernetwork, decoder and low-dimensional dynamics model, are given in \tabref{hyper_tab}. For all the numerical experiments, we ensure that the neural network architecture remains the same between three comparison models IRS, PI-IRS and ECLEIRS. Furthermore, minimal hyperparameter tuning is done between various numerical experiments, highlighting the robustness of these methods. SIREN activation functions are used for all sub-networks, following details mentioned in \cite{Sitzmann2020}. The optimization problems for obtaining the parameters for these neural networks are solved using ADAM \cite{Kingma2017} with a starting learning rate of $0.001$ and decreasing the learning rate by a factor of $0.3$ if convergence loss asymptotes. The parameters for the optimization problem are also kept constant for all the test cases and different reduced state dynamics approaches. Lastly, the dimensionality of the reduced space is selected to be $n_r = d_{\mu} + 2$, where the factor $2$ includes an account for temporal dependence and is higher than the intrinsic manifold dimension of the solution space by $1$.  


\begin{table}[]
    \centering
    \caption{Hyperparameters in the neural network architectures for different problems.}    
    \begin{tabular}{|c|c|c|c|}
        \hline
         \textbf{Test case} & \textbf{Hypernetwork} ($\pmb{h}_{\pmb{\theta}_h}$) & \textbf{Decoder} ($\pmb{d}_{\pmb{\theta}_d}$) & \textbf{Dynamics} ($\pmb{g}_{\pmb{\theta}_g}$)  \\
        \hline
         1-D advection & 4 layers, 64 neurons & 4 layers, 40 neurons & 4 layers, 32 neurons \\
         1-D Burgers & 4 layers, 64 neurons & 4 layers, 40 neurons & 4 layers, 32 neurons \\
         2-D Euler & 4 layers, 100 neurons & 4 layers, 100 neurons & 4 layers, 100 neurons \\
         \hline
    \end{tabular}
    \label{tab:hyper_tab}
\end{table}

\subsection{1-D advection problem}

For the first numerical experiment, we consider the 1-D parameterized scalar advection equation 
\begin{equation}
    \frac{\p q}{\p t} + \frac{\p c q}{\p x} = 0,
\end{equation}
with the initial conditions 
\begin{equation}
    q(x,0) = q_0 \quad \forall \quad x = [0, \;x_{in}],
\end{equation}
where $q_0$ and $x_0$ defines the parameterized initial condition of the wave, while $c$ is a parametrized advection velocity. Therefore, the numerical experiment comprises the parameter vector $\pmb{\mu} = [c, \; q_0, \; x_{in}] \in \mathcal{D}^{\mu} \subset \mathbb{R}^3$. The high-fidelity simulations that generate the data involve a finite-volume scheme with WENO-JS spatial discretization \cite{Jiang1996} with third-order TVD Runge-Kutta scheme for time integration \cite{Shu1988}. These simulations are carried out at a CFL number of 0.5, which is within the range of commonly employed CFL numbers for simulations of hyperbolic systems. This combination of spatial and temporal discretization yields unconditionally stable results, which is essential for evaluating parameterized systems. The details about the dataset size used for learning and validating reduced dynamics approaches assessed in this study are shown in \tabref{Advec_dataset}. The validation dataset considered in this experiment has no overlap with the learning dataset. Furthermore, the validation dataset enables assessment of the learned model performance for parameters interpolating within the parameter space $\mathcal{D}^{\mu}$ or extrapolating outside this parameter space.

\begin{table}[t]
    \centering
        \caption{Details of the learning and validation datasets for 1-D advection problem. The learning dataset is further randomly divided between training ($75 \%$) and testing set ($25 \%$) to ensure that the model is not overfitted.}
    \begin{tabular}{|c|c|c|}
    \hline
         \textbf{Dataset Name} & \textbf{Number of data points} & \textbf{Parameter values} \\
         & ($n_x \times n_t$) & ($c \times q_0 \times x_{in}$) \\
    \hline
         Learning & $101 \times 190$ & $c \in \{0.8, 0.9, 1.0, 1.1, 1.2 \}$ \\
        & & $q_0 \in \{0.9, 0.95, 1.0, 1.05, 1.1\}$ \\
        & & $x_{in} \in \{0.1, 0.15, 0.2, 0.25, 0.3\}$ \\
         \hline
         Validation & $101 \times 190$ & $c \in \{0.75, 0.85, 0.95, 1.05, 1.15, 1.25 \}$ \\
        & & $q_0 \in \{0.88, 0.93, 0.98, 1.03, 1.08, 1.13\}$ \\
        & & $x_{in} \in \{0.07, 0.12, 0.17, 0.22, 0.27, 0.32\}$ \\
        \hline
    \end{tabular}
    \label{tab:Advec_dataset}
\end{table}

\subsubsection{Denoising and sparse reconstruction capability on the learning dataset}

\begin{figure}[t]
    \centering
    \subfigure[\label{fig:STnRatio_0p2}]{\includegraphics[width=0.32\textwidth, trim={0.2cm 0cm 1.0cm 0.5cm},clip]{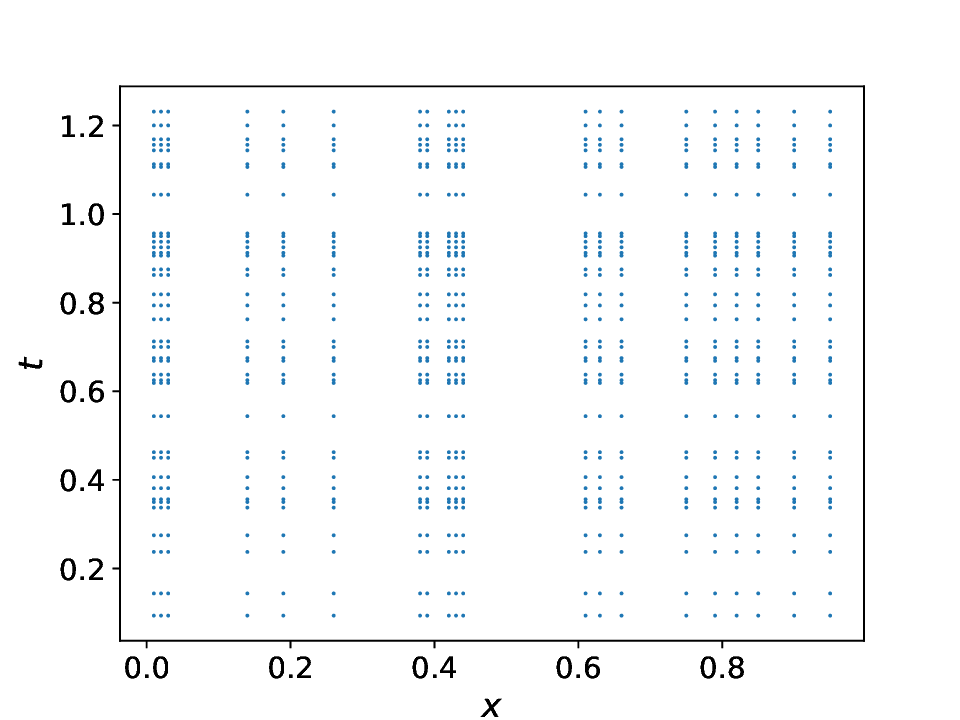}}
    \subfigure[\label{fig:STnRatio_0p6}]{\includegraphics[width=0.32\textwidth, trim={0.2cm 0cm 1.0cm 0.5cm},clip]{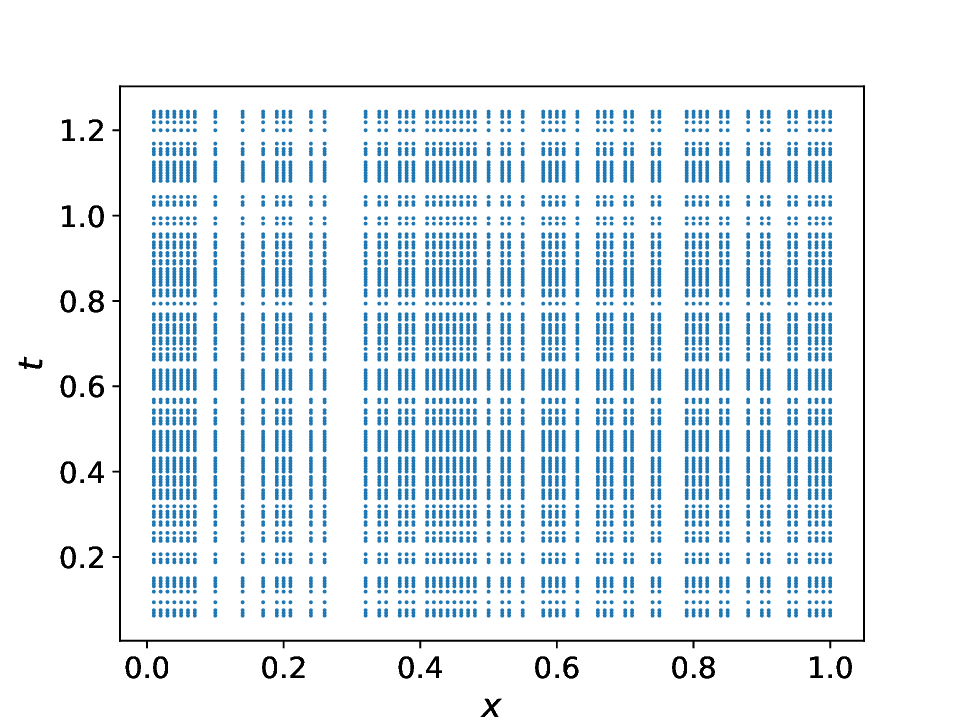}}
    \subfigure[\label{fig:STnRatio_1p0}]{\includegraphics[width=0.32\textwidth, trim={0.2cm 0cm 1.0cm 0.5cm},clip]{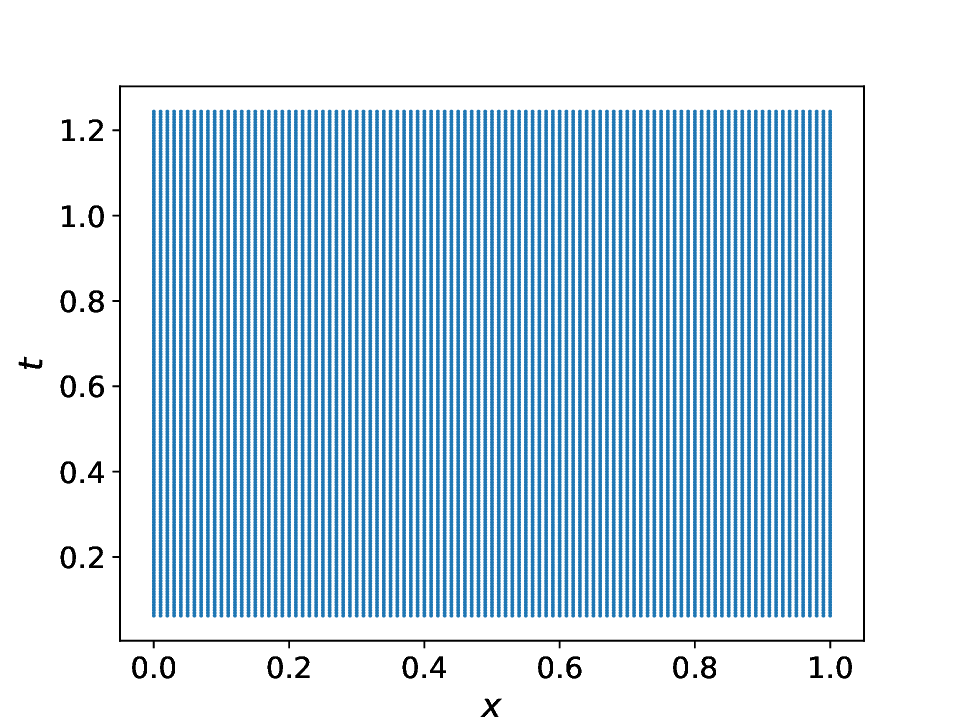}}
    \vspace{-3mm}
    \caption{1-D advection problem: Space-time points for (a) $20\%$, (b) $60\%$ and (e) $100\%$ sparsity in data used to learn different reduced state dynamics models. Note that $x\%$ sparsity in both both space and time implies $x^2 \%$ sparsity in the entire dataset.}
    \label{fig:STnRatio}
\end{figure}

We first assess the ability of ECLEIRS to learn the representation of the true solution $q(x,t; \pmb{\mu})$ for scenarios where sparse and noisy spatiotemporal data is available. In such situations, there is no information on the ground truth signal, labeled or unlabeled data, to denoise the solution signal. We generate sparsely observed data by randomly sampling the full simulation data in space and time. These different spatiotemporal sparsity levels are shown in \figref{STnRatio}. As observed from these figures, we choose tensor product spatiotemporal data because we expect data to be collected from high-fidelity simulations or experimental data to not be present in some completely arbitrary random spatiotemporal grid but at some regular frequency, especially in time. We refer to $x\%$ sparsity as $x\%$ data is sampled randomly in spatial direction and $x\%$ data is also randomly sampled in temporal direction following a tensor product selection procedure. Therefore, $20\%$ sparsity implies $4\%$ of total spatiotemporal data is used. In addition to sparsity, we add Gaussian noise with zero mean and standard deviation of $\sigma_N$, which is varied to assess the performance for different noise levels. 

\begin{figure}[t]
    \centering
    \subfigure[\label{fig:SolutionNoiseAdvec}]{\includegraphics[width=0.49\linewidth, trim={0.5cm 0cm 1.5cm 1.0cm},clip]{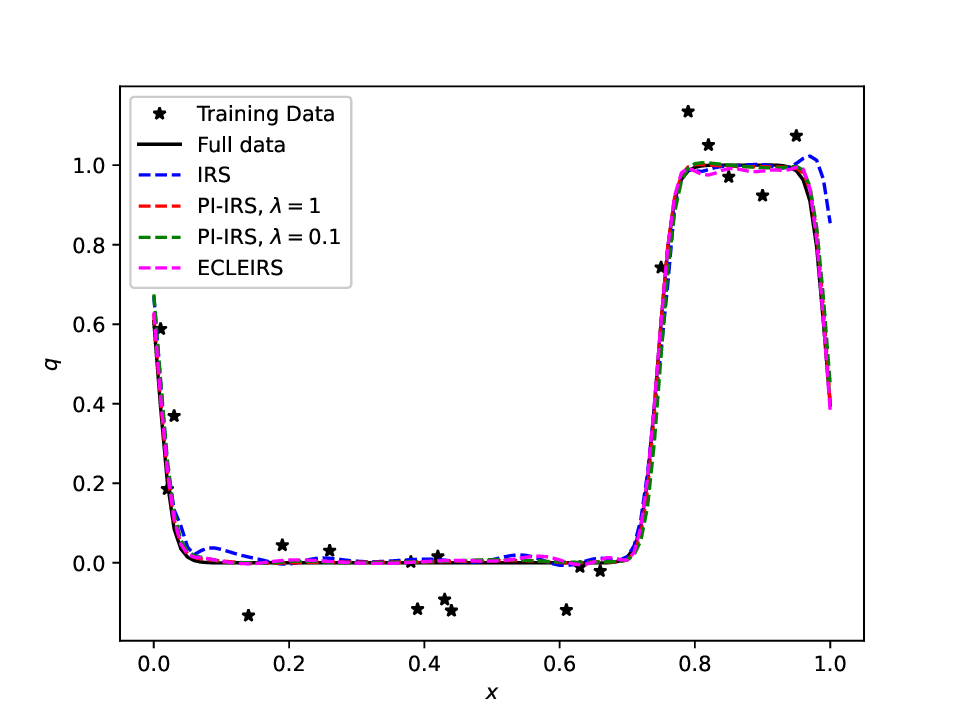}}
    \subfigure[\label{fig:ErrorNoiseAdvec}]{\includegraphics[width=0.49\linewidth, trim={0.5cm 0cm 1.5cm 1.0cm},clip]{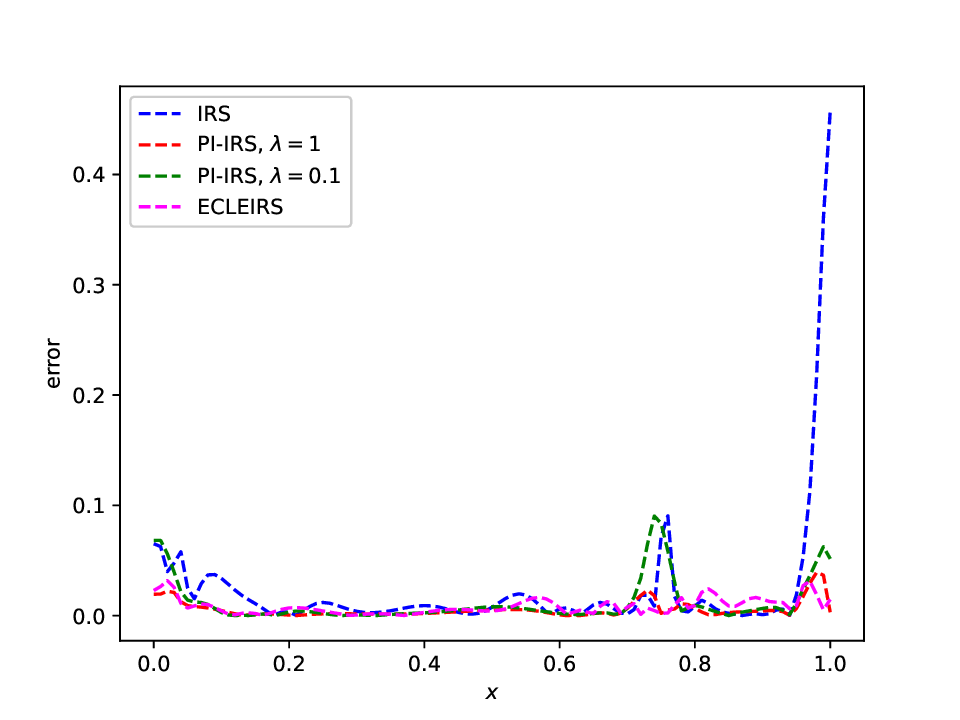}}
    \caption{1-D advection problem: (a) Solution and (b) error for $\bm{\mu} = [1.0, \; 0.95, \;  0.15]$ with spatiotemporal sparsity of $20\%$ and $\sigma_N = 0.1$ for different reduced state dynamics identification approaches. The predictions by PI-IRS and ECLEIRS are so close to ground truth results that they almost overlap in (a), as also indicated by low error values in (b).}
    \label{fig:SolErrorNoiseAdvec}
\end{figure}

The clean solutions obtained for different reduced state dynamics approaches in the presence of $20\%$ sparsity with added noise with $\sigma_N = 0.1$ are shown in \figref{SolErrorNoiseAdvec}. The results indicate that while IRS provides a clean solution representation from the noisy and sparse data, large oscillations are observed near the wave. While no such oscillations are observed for PI-IRS with $\lambda = 0.1$, we observe higher errors compared to PI-IRS with $\lambda = 1$ and ECLEIRS. The latter two approaches provide similar errors for this noise level. These results indicate that by adding conservation law information in the inverse problem, either through penalization of the optimization objective or through exact enforcement by designing the architecture appropriately, we can obtain suitable solution representation from sparse and noisy data without any prior data or knowledge of clean solution representation. While we show solution signal reconstruction capability of our proposed approach for a time instance, similar results are obtained for other time instances and different system parameters.

\begin{figure}[t]
    \centering
    \subfigure[\label{fig:BoxplotPenalty_advec_sigma0}]{\includegraphics[width=0.49\linewidth, trim={0.5cm 0cm 5cm 2.5cm},clip]{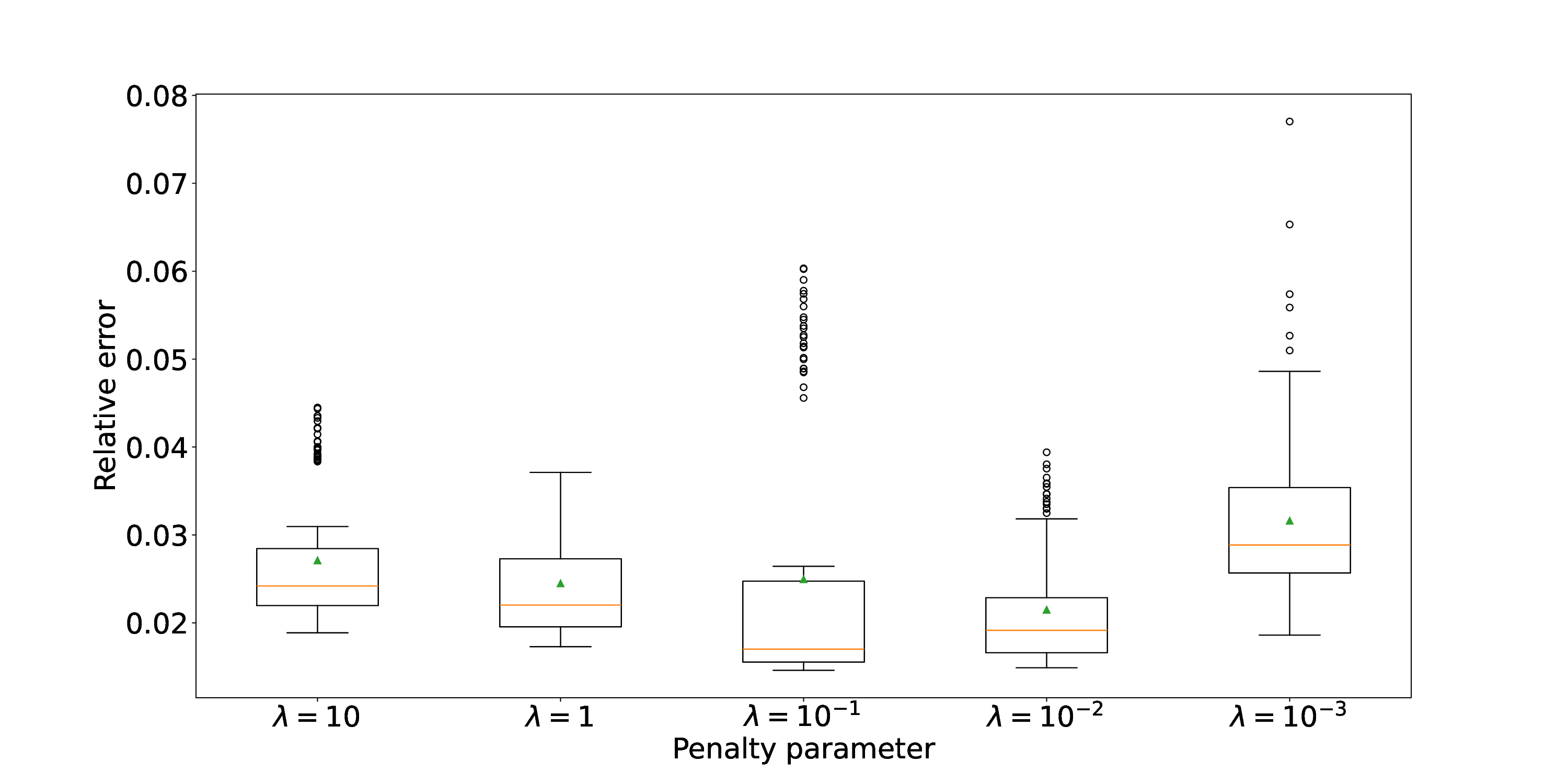}}
    \subfigure[\label{fig:BoxplotPenalty_advec_sigma0p2}]{\includegraphics[width=0.49\linewidth, trim={0.5cm 0cm 5cm 2.5cm},clip]{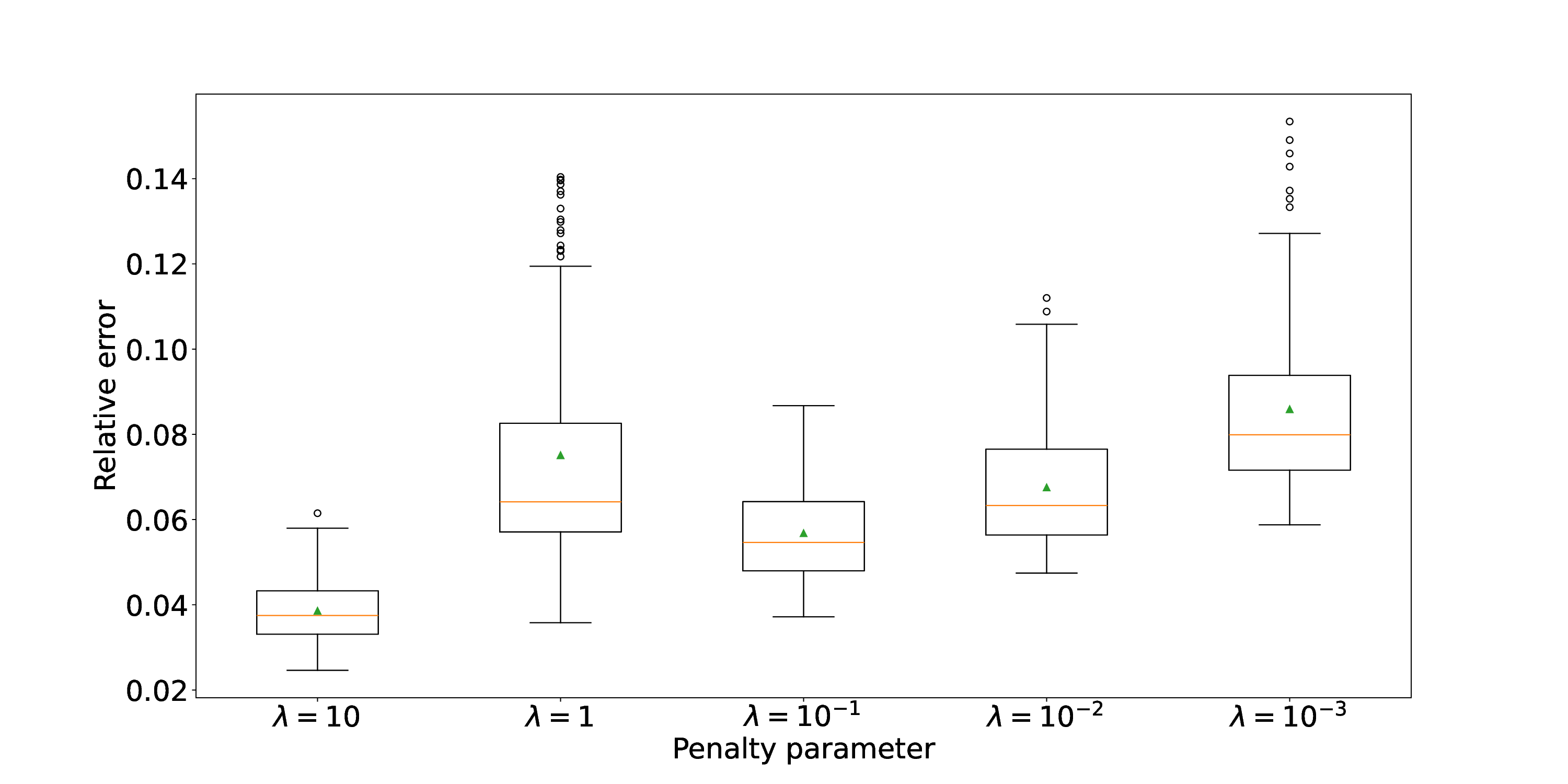}}
    \caption{1-D advection problem: Box plots of relative error (defined in \eref{rel_error_def}) for training data with (a) $\sigma_N = 0$ (no noise) and (b) $\sigma_N = 0.2$ with sparsity of $20\%$ for PI-IRS with different penalty parameter values.}
    \label{fig:BoxplotPenalty_advec}
\end{figure}

PI-IRS and ECLEIRS are suitable for obtaining clean solution representation from sparse and noisy data. However, PI-IRS involves the selection of a penalty parameter, whereas ECLEIRS is free from such a parameter selection. We assess the influence of the penalty parameter on the denoising capability of PI-IRS by comparing the relative error defined as
\begin{equation}
    e_{xt} (\mu) = \frac{\vert \vert q(x,\; t;\; \mu) - q^m(x,\; t;\; \mu) \vert \vert_2}{\vert \vert q(x,\; t;\; \mu) \vert \vert_2},
    \label{eq:rel_error_def}
\end{equation}
where the error is computed in space-time for each parameter individually. These errors are presented as boxplots in \figref{BoxplotPenalty_advec} with the mean and standard deviation computed over the entire training parameter set. The results show different trends of the variation in relative error for two different noise levels. For scenarios with sparse and clean data ($\sigma_N = 0$), we observe that PI-IRS with $\lambda = 10^{-1}$ yields the most accurate results. However, this parameter value exhibits several outliers like those observed for other values of $\lambda$. Conversely, PI-IRS with $\lambda = 10$ provides the lowest errors in the presence of significant noise ($\sigma_N = 0.2$) in the data. The results show that larger $\lambda$ values give better results for this high noise level. This observation can be attributed to the need for more significant penalization of the conservation loss to ensure that physically realistic solution representations are obtained. These results also highlight a major drawback of using the penalty formulation in PI-IRS: the penalty parameter selection is sensitive to noise levels. Similar trends were also observed for different sparsity levels, however, that discussion is not included for brevity. Therefore, the procedure for selecting the ideal penalty parameter is often laborious and involves significant computing time in the offline stage. Conversely, ECLEIRS has the advantage of exactly enforcing the conservation law, and thereby not requiring tuning of any such parameters. 

\begin{figure}
    \centering
    \subfigure[\label{fig:BoxplotCompareModel}]{\includegraphics[width=0.49\linewidth, trim={0.5cm 0cm 5cm 2.5cm},clip]{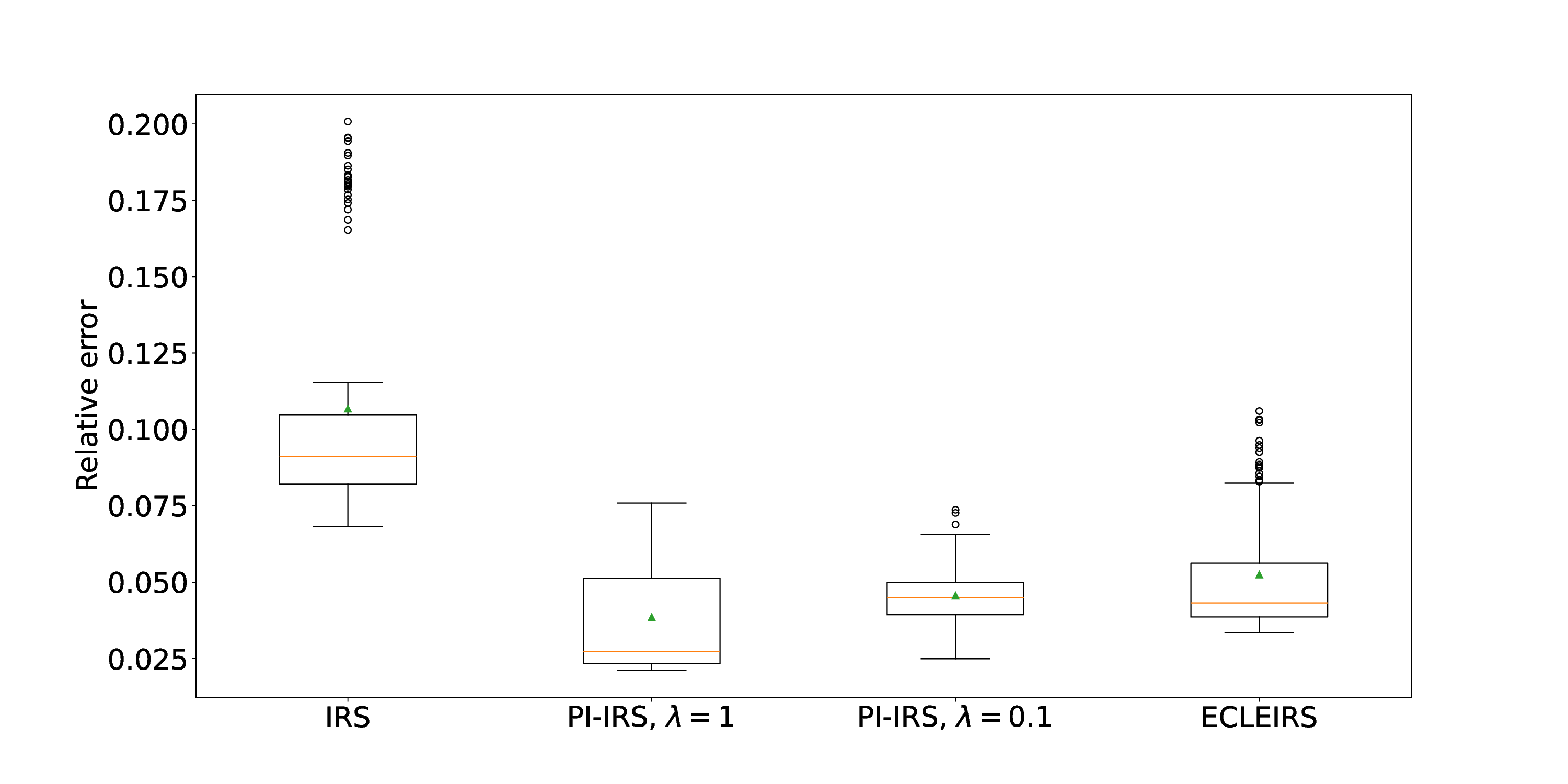}}
    \subfigure[\label{fig:BoxplotEcleir_advec}]{\includegraphics[width=0.49\linewidth, trim={0.5cm 0cm 5cm 2.5cm},clip]{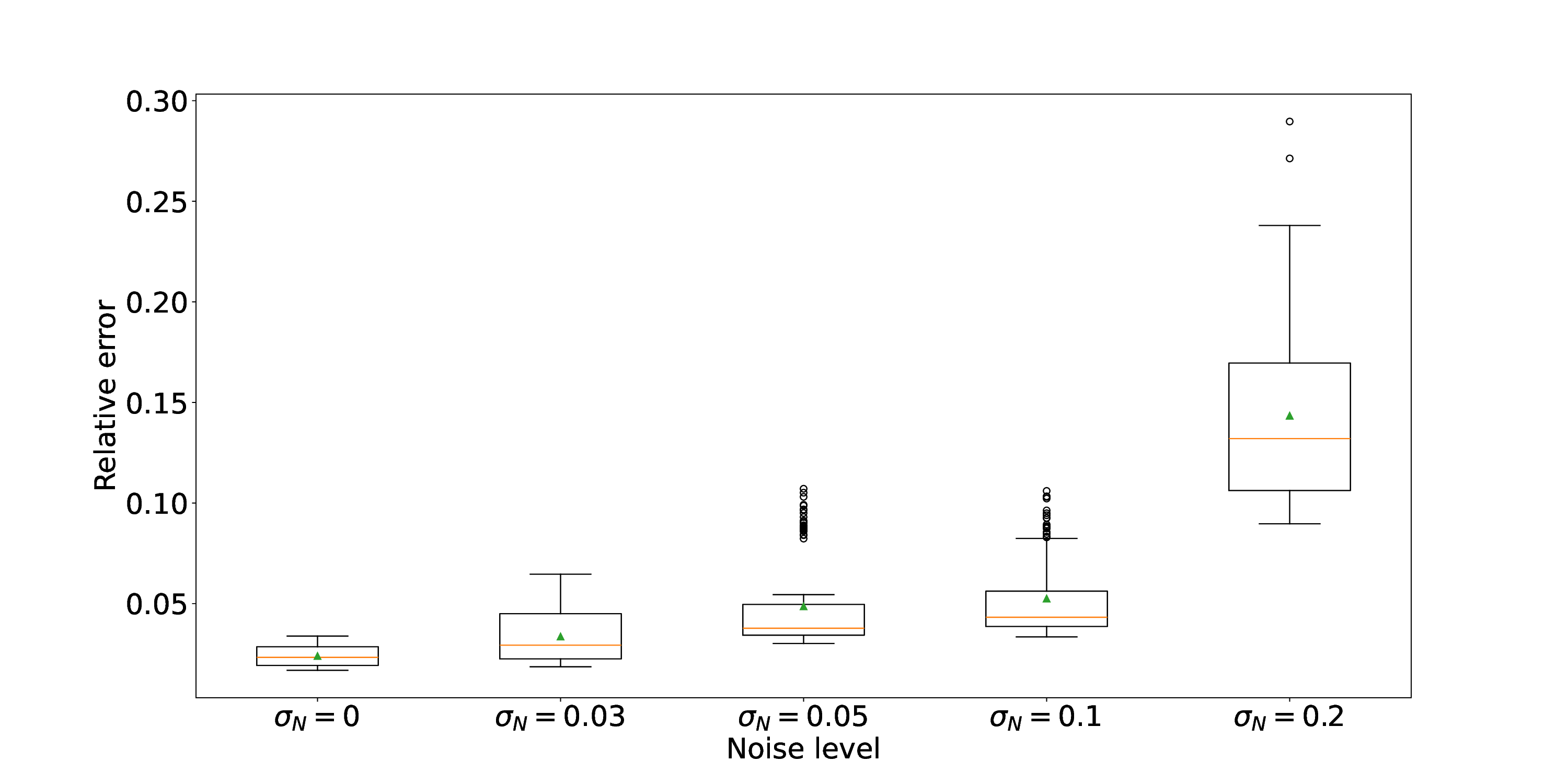}}
    \caption{1-D advection problem: Box plots of relative error (defined in \eref{rel_error_def}) for training data comparing (a) different reduced state dynamics approaches for $\sigma_N = 0.1$ and (b) ECLEIRS for different noise levels with sparsity of $20\%$.}
    \label{fig:BoxplotCompareModel_advec}
\end{figure}

The performance of different reduced state dynamics approaches in providing a clean solution representation from noisy and sparse data is compared in \figref{BoxplotCompareModel_advec}. The results in \figref{BoxplotCompareModel} show that PI-IRS for both penalty parameters and ECLEIRS perform well and provide a low relative error. Conversely, IRS leads to a high error, highlighting the importance of considering conservation laws for this dynamic inverse problem. The results show that while ECLEIRS results are similar to PI-IRS with $\lambda = 10^{-1}$, PI-IRS with $\lambda = 1$ leads to the most accurate results. However, as mentioned earlier, this improved accuracy is at the cost of tuning the penalty parameter for different sparsity and noise levels, whereas no such tuning is needed for ECLEIRS. The performance of ECLEIRS for different noise levels is shown in \figref{BoxplotEcleir_advec}. This  performance does not deteriorate for $\sigma_N \leq 0.1$, while at greater noise levels, we start seeing a more significant increase in error. 

\begin{figure}
    \centering
    \subfigure[\label{fig:BoxplotEcleir_rhotcompare}]{\includegraphics[width=0.49\linewidth, trim={0.5cm 0cm 5cm 2.5cm},clip]{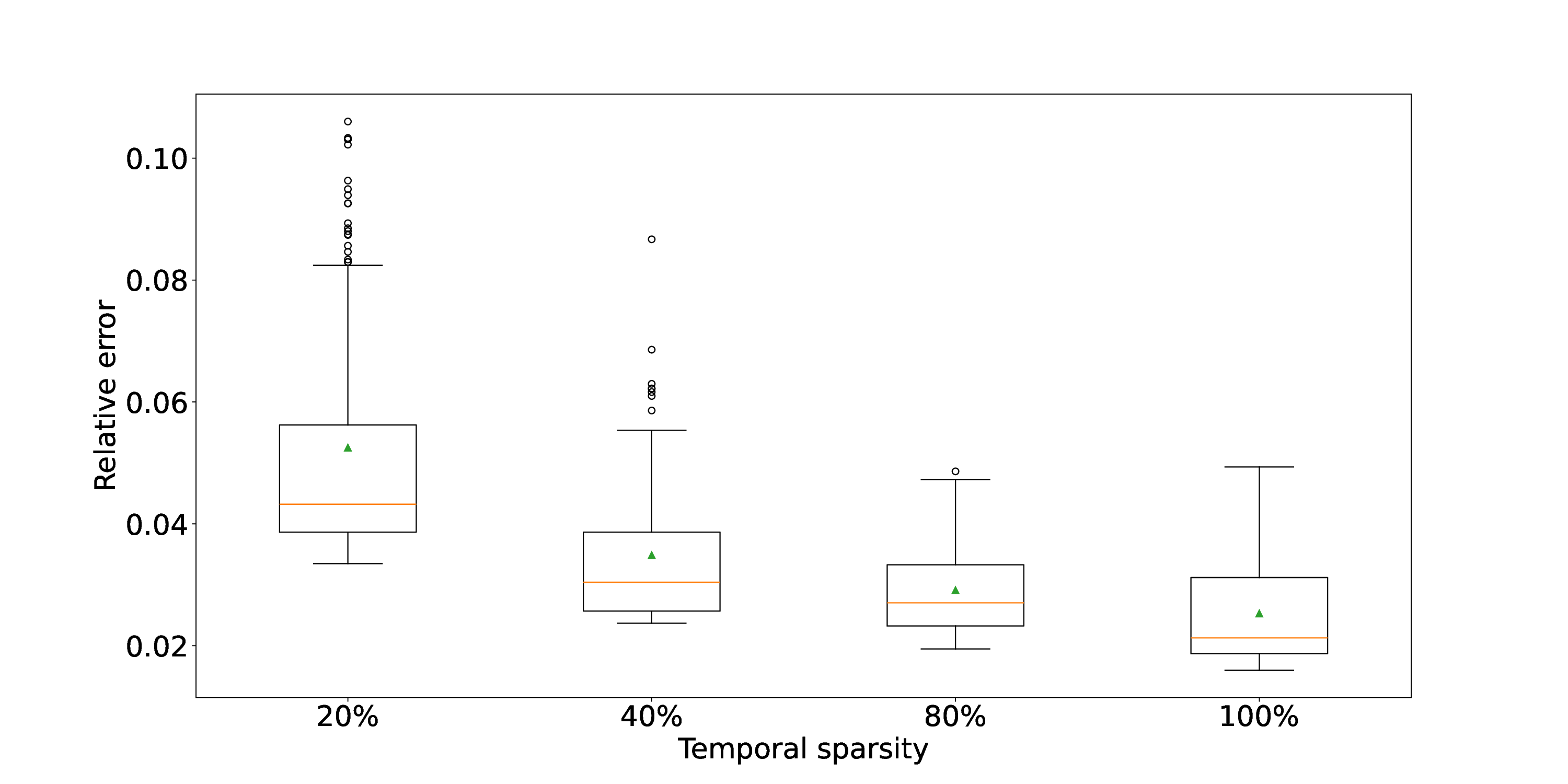}}
    \subfigure[\label{fig:BoxplotEcleir_rhoxcompare}]{\includegraphics[width=0.49\linewidth, trim={0.5cm 0cm 5cm 2.5cm},clip]{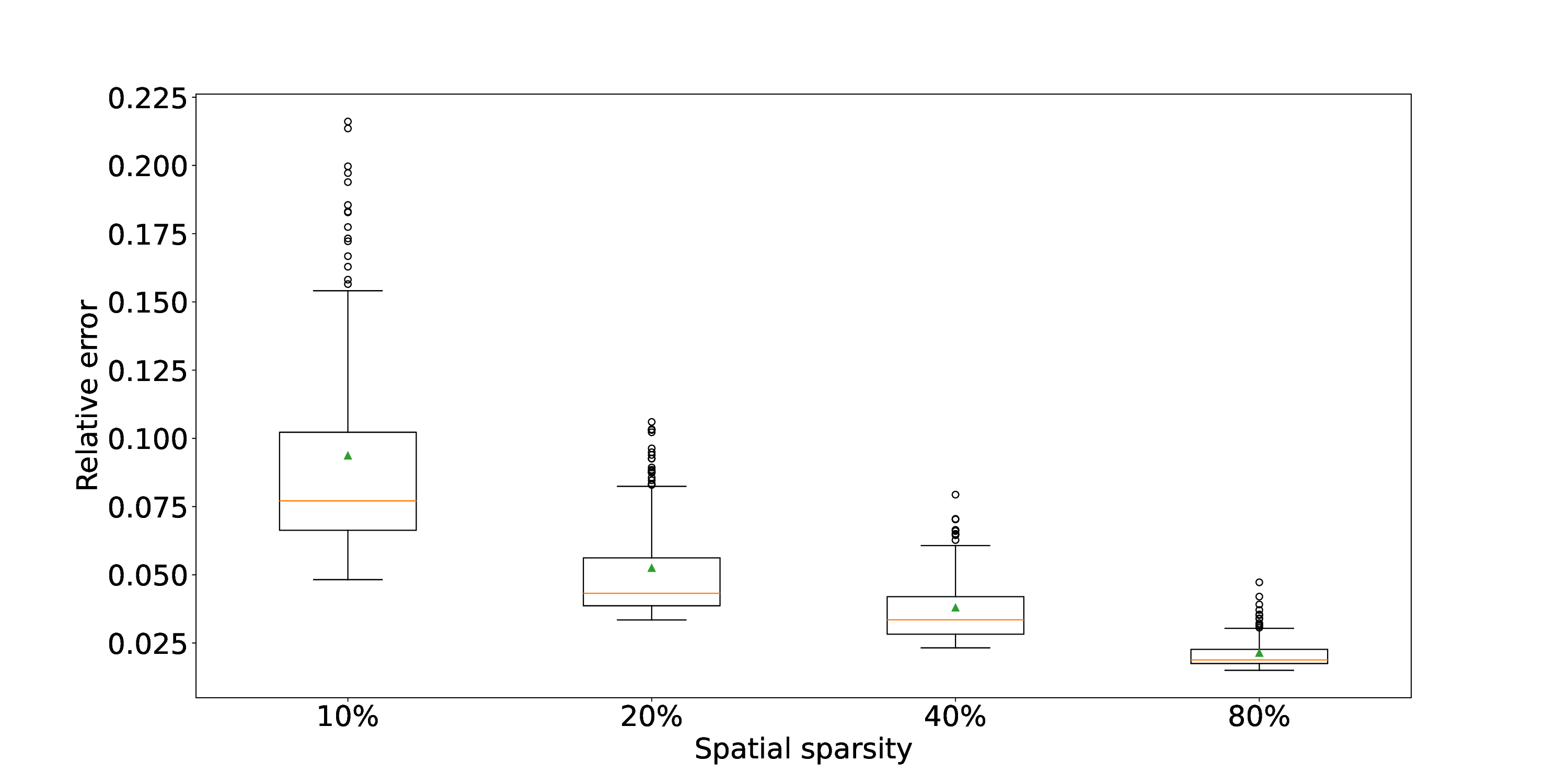}}
    \caption{1-D advection problem: Box plots of relative error (defined in \eref{rel_error_def}) using ECLEIRS for (a) fixed spatial sparsity of $20\%$ and varying temporal sparsity and (b) fixed temporal sparsity of $10\%$ and varying spatial sparsity for data with fixed noise level of $\sigma_N = 0.1$.}
    \label{fig:BoxplotEcelire_rhocomp}
\end{figure}

We also assess the performance of ECLEIRS for varying spatial and temporal sparsity to understand the behavior of different approaches for learning a clean solution representation. Box plots showing the relative errors for different temporal and spatial sparsity levels are shown in \figref{BoxplotEcelire_rhocomp}. The results show that increasing the temporal sparsity level, that is having more temporal data, reduces the error in obtaining the clean solution representation. Similarly, increasing the spatial sparsity, that is having more spatial data, reduces the error in obtaining the solution representation of the clean signal. This behavior highlights a tradeoff between spatial and temporal sparsity. Therefore, we can have a higher temporal sampling frequency to approximate a clean solution signal with high accuracy for scenarios where only a few spatial sensor locations are available to collect noisy data. Similarly, if the time scales for physics evolution are small and unresolved by temporal collection frequency, then accurate solution representations can be obtained by increasing the spatial sampling locations.

\subsubsection{Dynamics prediction for the validation dataset}

In the previous section, we assessed the performance of different reduced state dynamics approaches in obtaining a clean solution signal representation from sparse and noisy data. In this section, we assess the capability of the identified reduced state dynamics in predicting the solution dynamics for parameters that were not included in the model training procedure. These validation parameters are both within the parameter space of learned mapping $\mathcal{D}_{\mu}$ to test the ability of the learned models in the interpolation regime and also parameters that are marginally outside the learning parameter space to assess the performance of the learned models in the extrapolation regime.

\begin{figure}[ht!]
    \centering
    \subfigure[\label{fig:TotalError_kind_advec}]{\includegraphics[width=\textwidth, trim={4cm 7.5cm 7cm 9cm},clip]{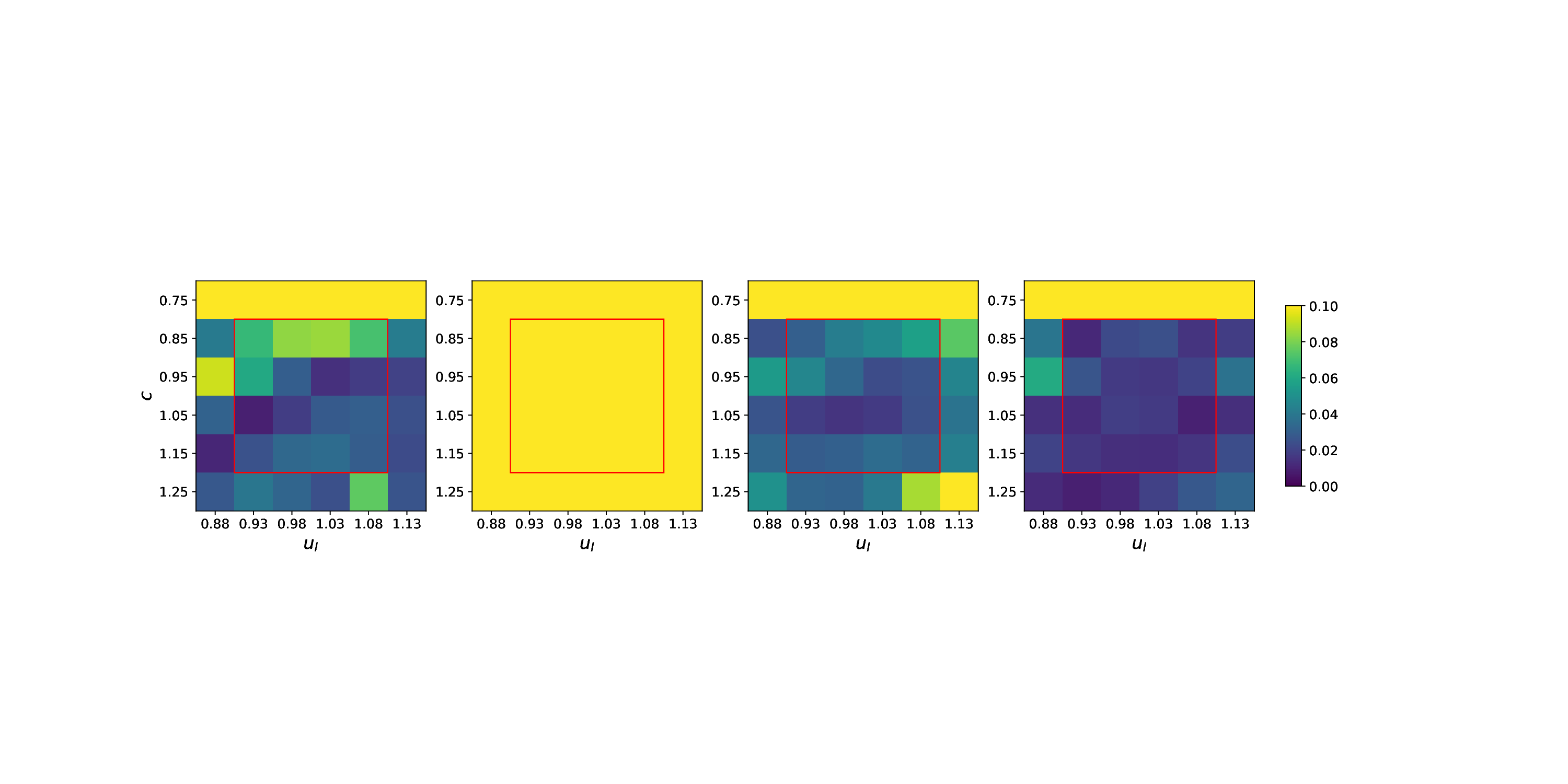}}
    \subfigure[\label{fig:TotalError_jind_advec}]{\includegraphics[width=\textwidth, trim={4cm 7.5cm 7cm 9cm},clip]{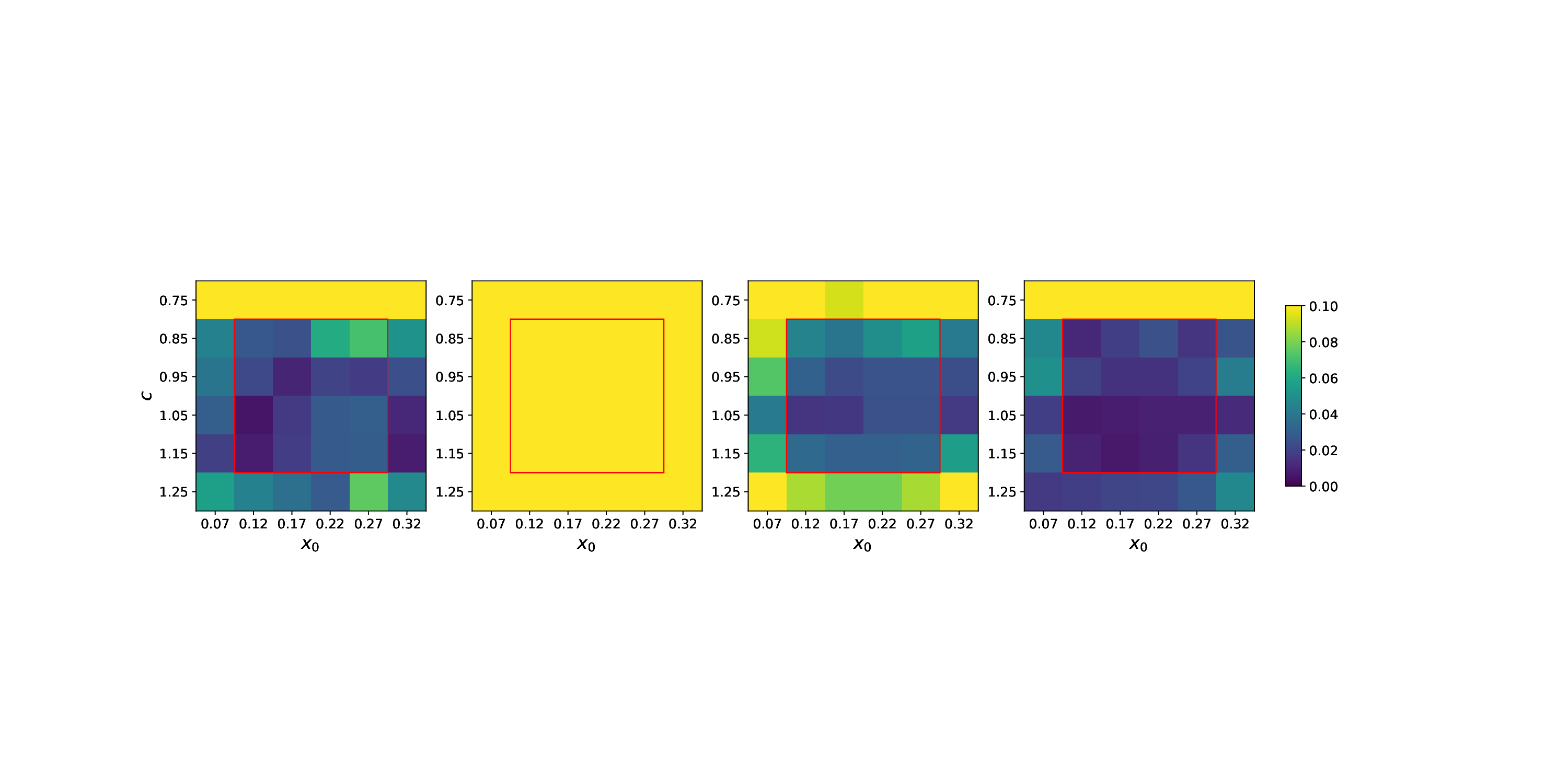}}
    \subfigure[\label{fig:TotalError_iind_advec}]{\includegraphics[width=\textwidth, trim={4cm 7.5cm 7cm 9cm},clip]{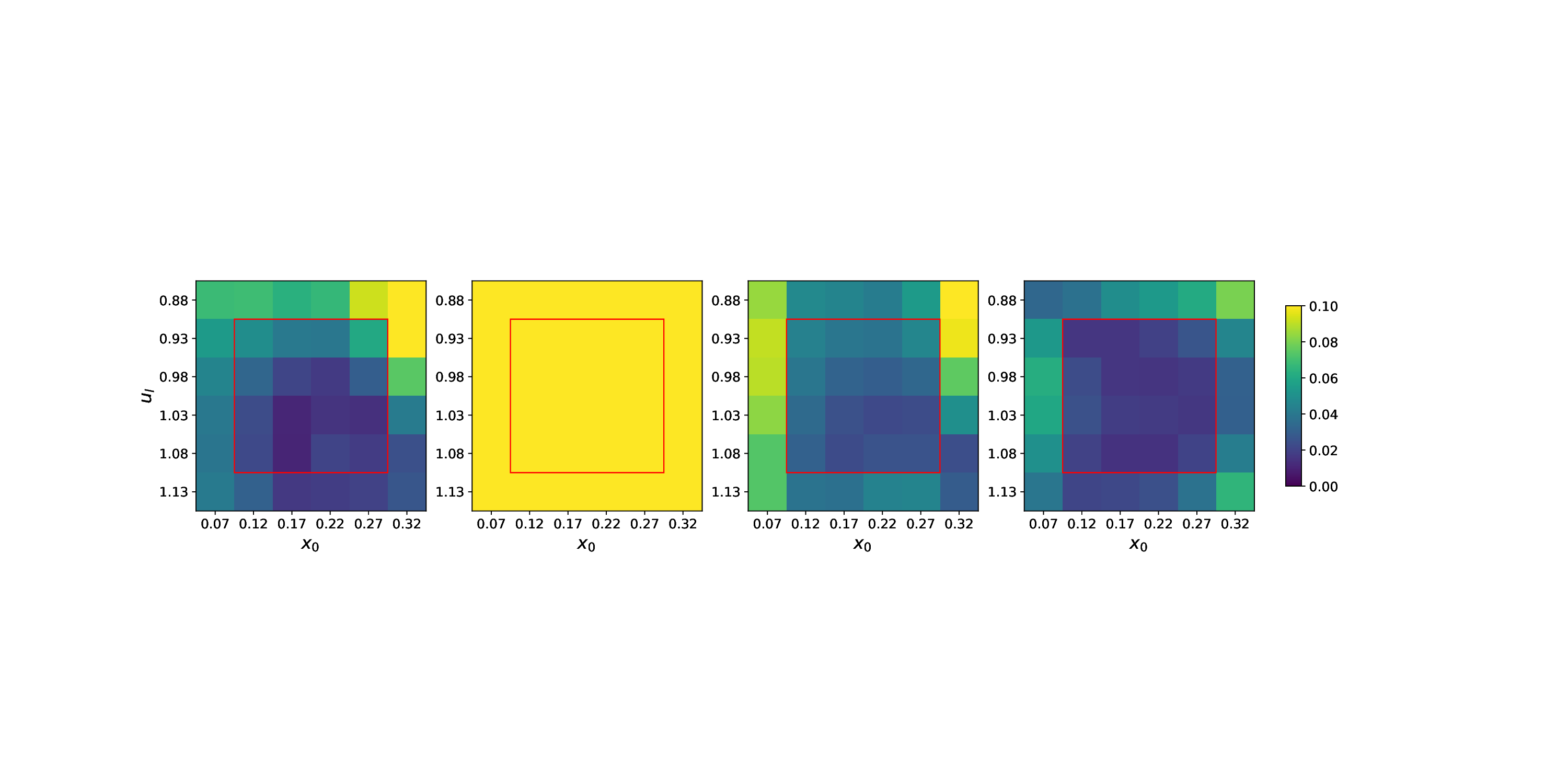}}
    \caption{1-D advection problem: Relative error (defined in \eref{rel_error_def}) in space-time for (a) $x_0 = 0.22$, (b) $u_l = 1.03$ and (c) $c = 0.85$. Different columns denote different reduced state dynamics approaches trained on high-resolution clean data: IRS (1st column), PI-IRS with $\lambda = 1 \times 10^{-1}$ (2nd column), PI-IRS with $\lambda = 1 \times 10^{-4}$ (3rd column) and ECLEIRS (4th column). The parameters inside the red box are within the interpolation space and the those outside are in the extrapolation space.}
    \label{fig:SpaceTime_CompareModels}
\end{figure}

We first assess the performance of various reduced state dynamics approaches for predicting the solution for different parameters in the validation dataset for the scenario where clean and full spatiotemporal data was available for the model training. The relative error in space-time for different parameter combinations and different reduced state dynamics approaches are compared in \figref{SpaceTime_CompareModels}. The results indicate that in the presence of full resolution data without noise, IRS, PI-IRS with $\lambda = 10^{-4}$ and ECLEIRS perform well, whereas PI-IRS with $\lambda = 10^{-1}$ do not give accurate results. This observation highlights the strong dependence of PI-IRS performance on the selection of penalty parameters. While a larger value of the penalty parameter was important for sparse reconstruction and denoising the data, these values do not work well when high-resolution clean data is available. Among all the tested models, ECLEIRS performs best with the lowest error even in parameter extrapolation regions. 

\begin{figure}[ht!]
    \centering
    \subfigure[\label{fig:TotalError_kind_consverror}]{\includegraphics[width=\textwidth, trim={4cm 7.5cm 7cm 9cm},clip]{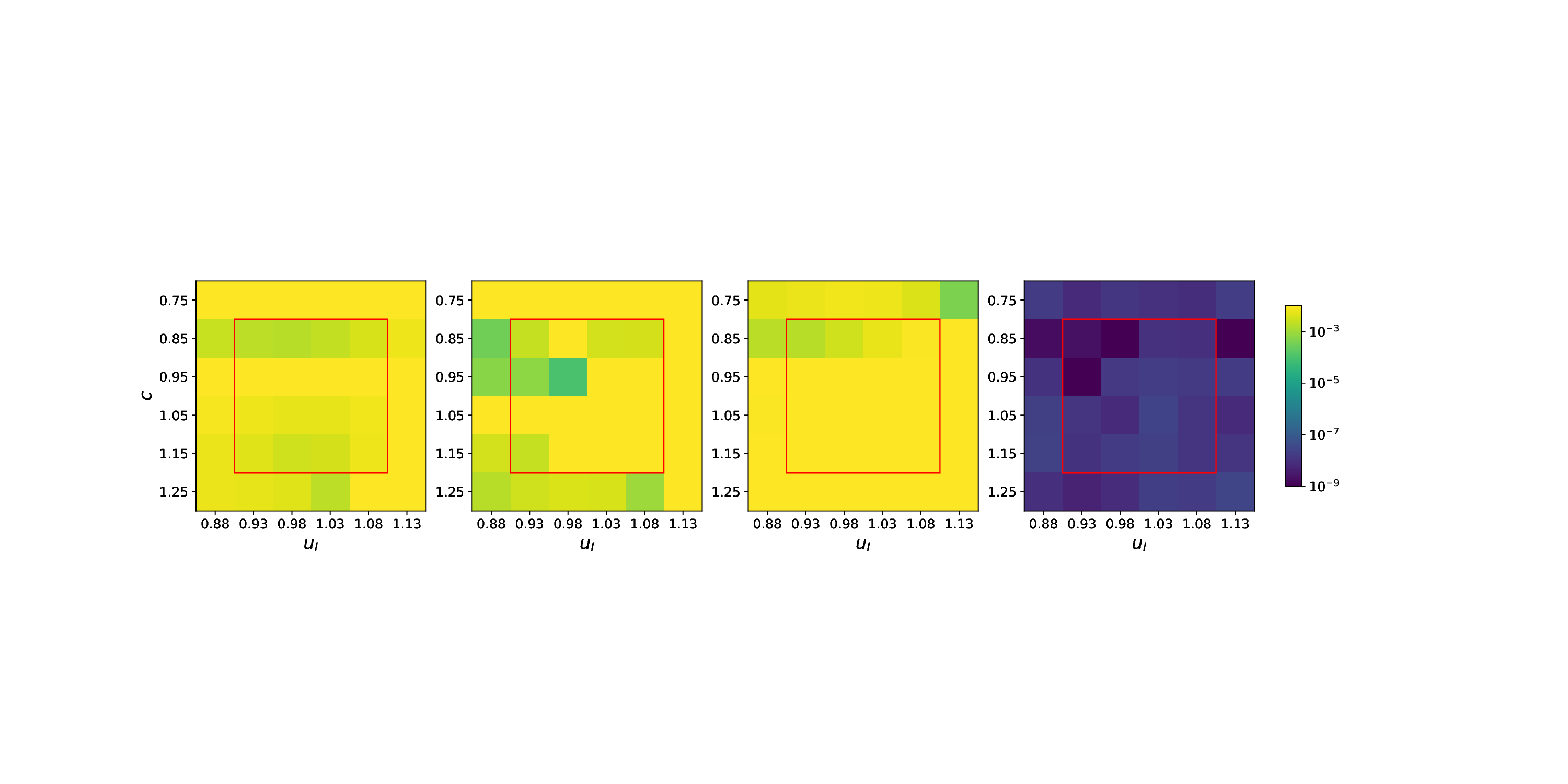}}
    \subfigure[\label{fig:TotalError_jind_consverror}]{\includegraphics[width=\textwidth, trim={4cm 7.5cm 7cm 9cm},clip]{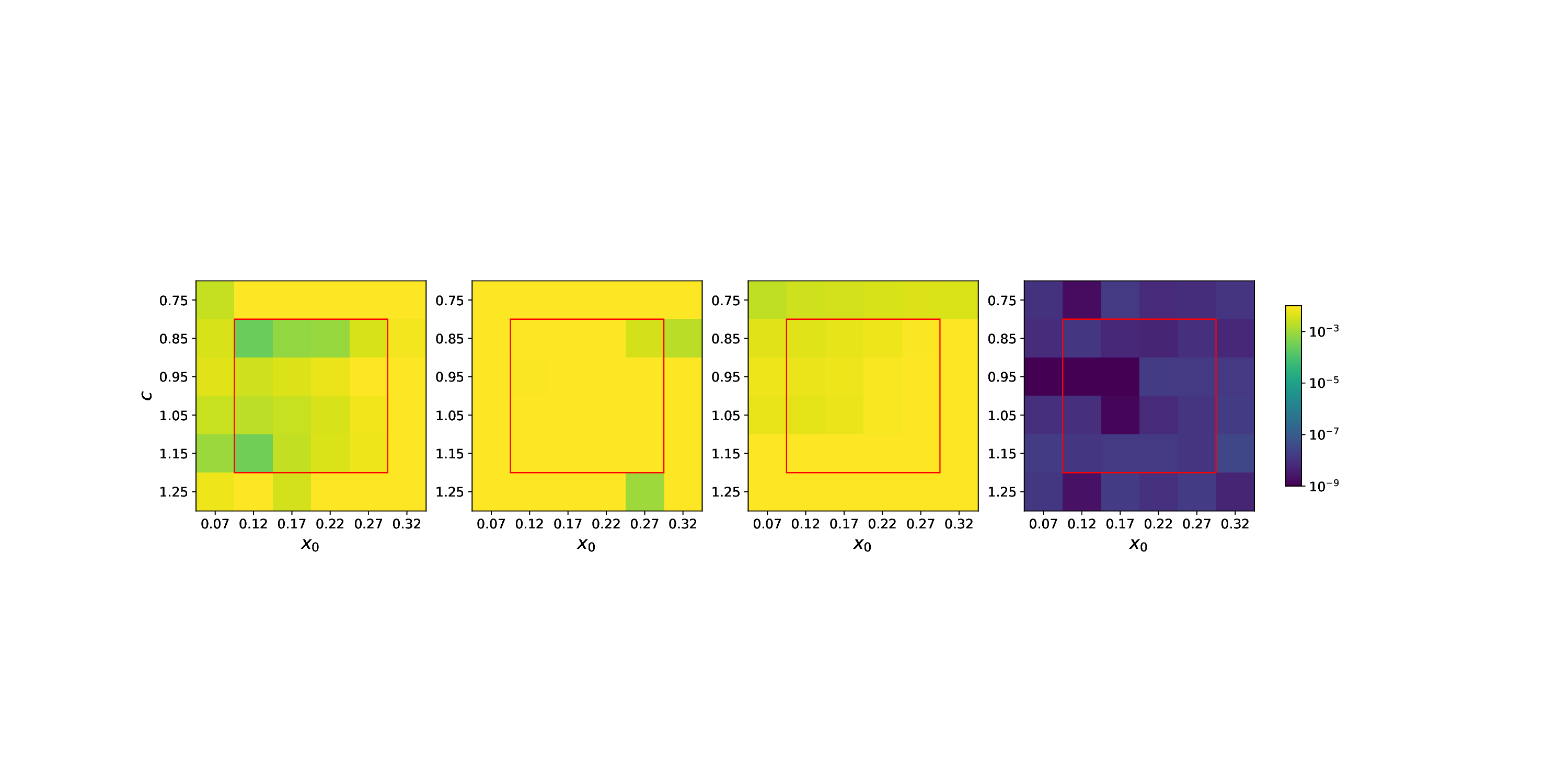}}
    \subfigure[\label{fig:TotalError_iind_consverror}]{\includegraphics[width=\textwidth, trim={4cm 7.5cm 7cm 9cm},clip]{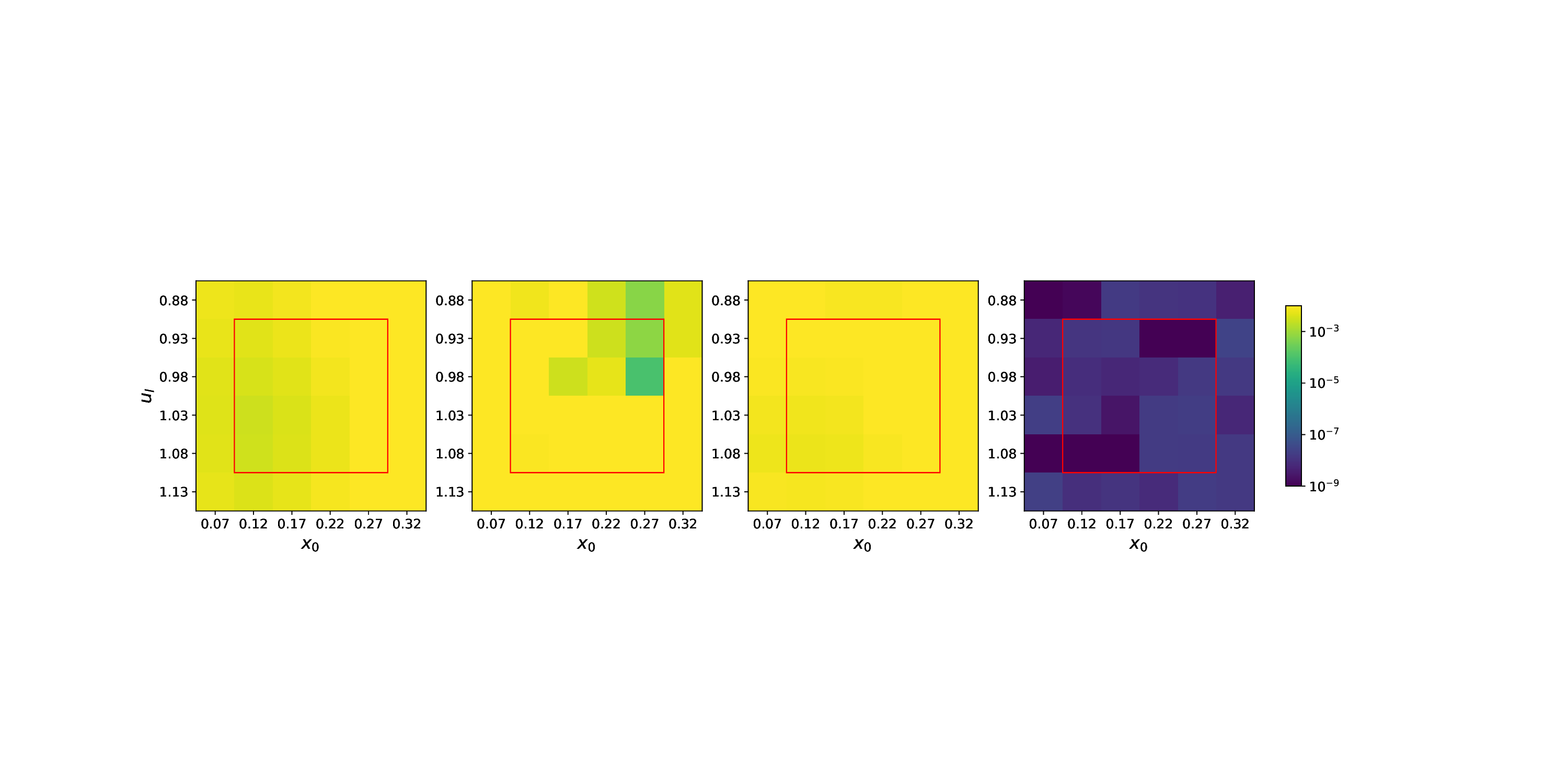}}
    \caption{1-D advection problem: Mean conservation error (defined in \eref{consv_error}) in space-time for (a) $x_0 = 0.22$, (b) $u_l = 1.03$ and (c) $c = 0.85$. Different columns denote different reduced state dynamics approaches trained on high-resolution clean data: IRS (1st column), PI-IRS with $\lambda = 1 \times 10^{-4}$ (2nd column), PI-IRS with $\lambda = 1 \times 10^{-6}$ (3rd column) and ECLEIRS (4th column). The parameters inside the red box are within the interpolation space and the those outside are in the extrapolation space.}
    \label{fig:SpaceTime_CompareModels_consverror}
\end{figure}

In addition to the assessment of the error in approximating the solution, we also assess the mean conservation error defined as
\begin{equation}
    e_{c} (\pmb{\mu}) = \Bigg\vert \frac{1}{n_x n_t} \sum_{i=1}^{n_x} \sum_{j=1}^{n_t}  \frac{\p q (x_i, t_j; \pmb{\mu})}{\p t} + \frac{\p c q (x_i, t_j; \pmb{\mu})}{\p x} \Bigg\vert,
    \label{eq:consv_error}
\end{equation}
where $\vert \cdot \vert$ indicates the absolute value. These results for the model trained on full resolution clean data are shown in \figref{SpaceTime_CompareModels_consverror}. We observe that both IRS and PI-IRS results have a high conservation error, even for large values of penalty parameters. This behavior is expected, even with the penalty formulation, as PI-IRS does not guarantee conservation outside the training data. On the contrary, ECLEIRS satisfies conservation law exactly by construction, and therefore, it exhibits a low conservation error even for parameters outside the training dataset. The conservation error for ECLEIRS presented here is approximately $\mathcal{O}(10^{-8})$ due to single precision operations used for training the model. We expect the conservation error to have even lower values if double-precision arithmetic is used. These results indicate that ECLEIRS is superior in accuracy and satisfying the conservation law.

\begin{figure}[t!]
    \centering
    \subfigure[\label{fig:BoxplotCompareModel_testing}]{\includegraphics[width=0.49\linewidth, trim={0.5cm 0cm 5cm 2.5cm},clip]{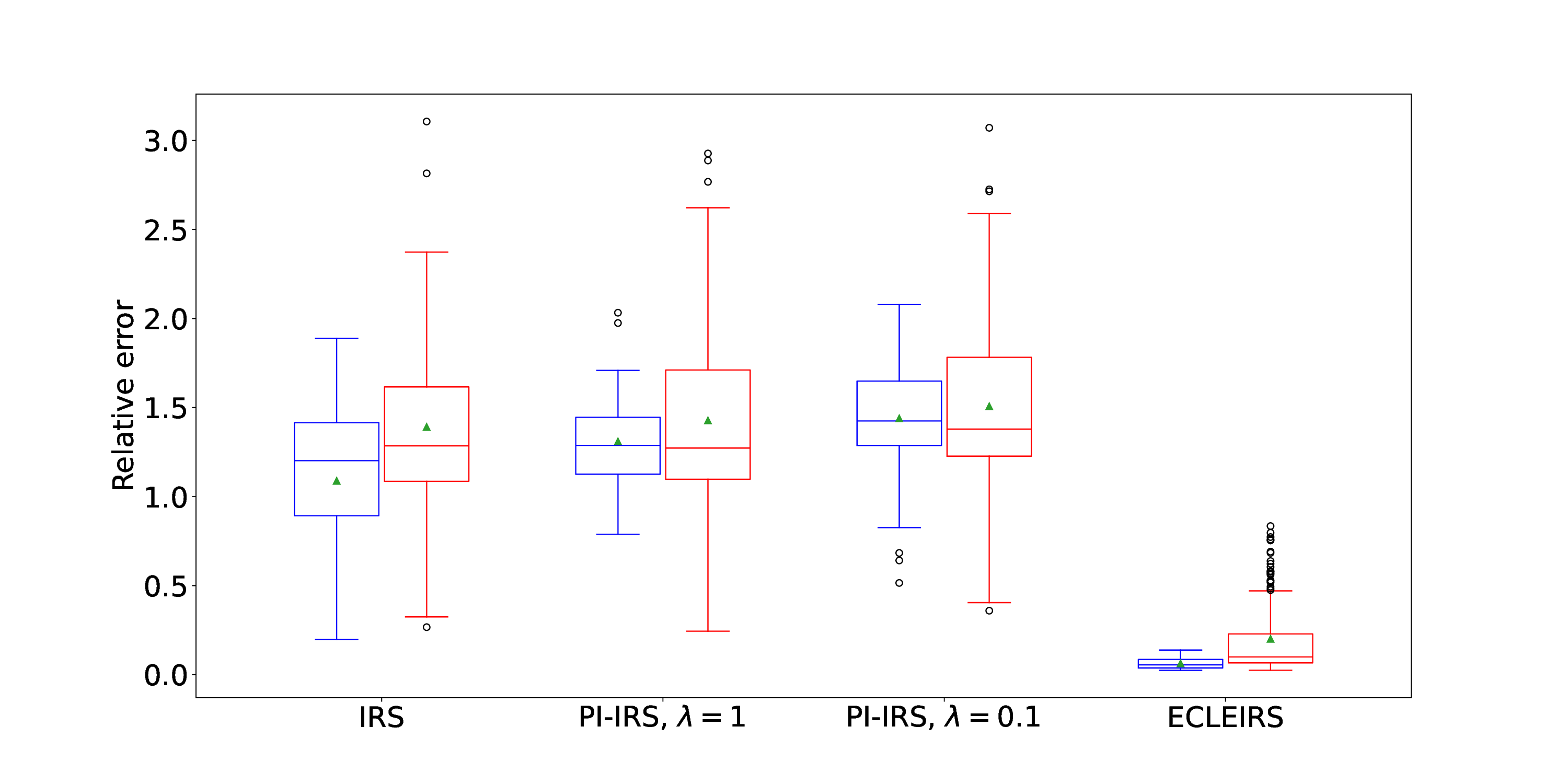}}
    \subfigure[\label{fig:BoxplotCompareModel_testing_2}]{\includegraphics[width=0.49\linewidth, trim={0.5cm 0cm 5cm 2.5cm},clip]{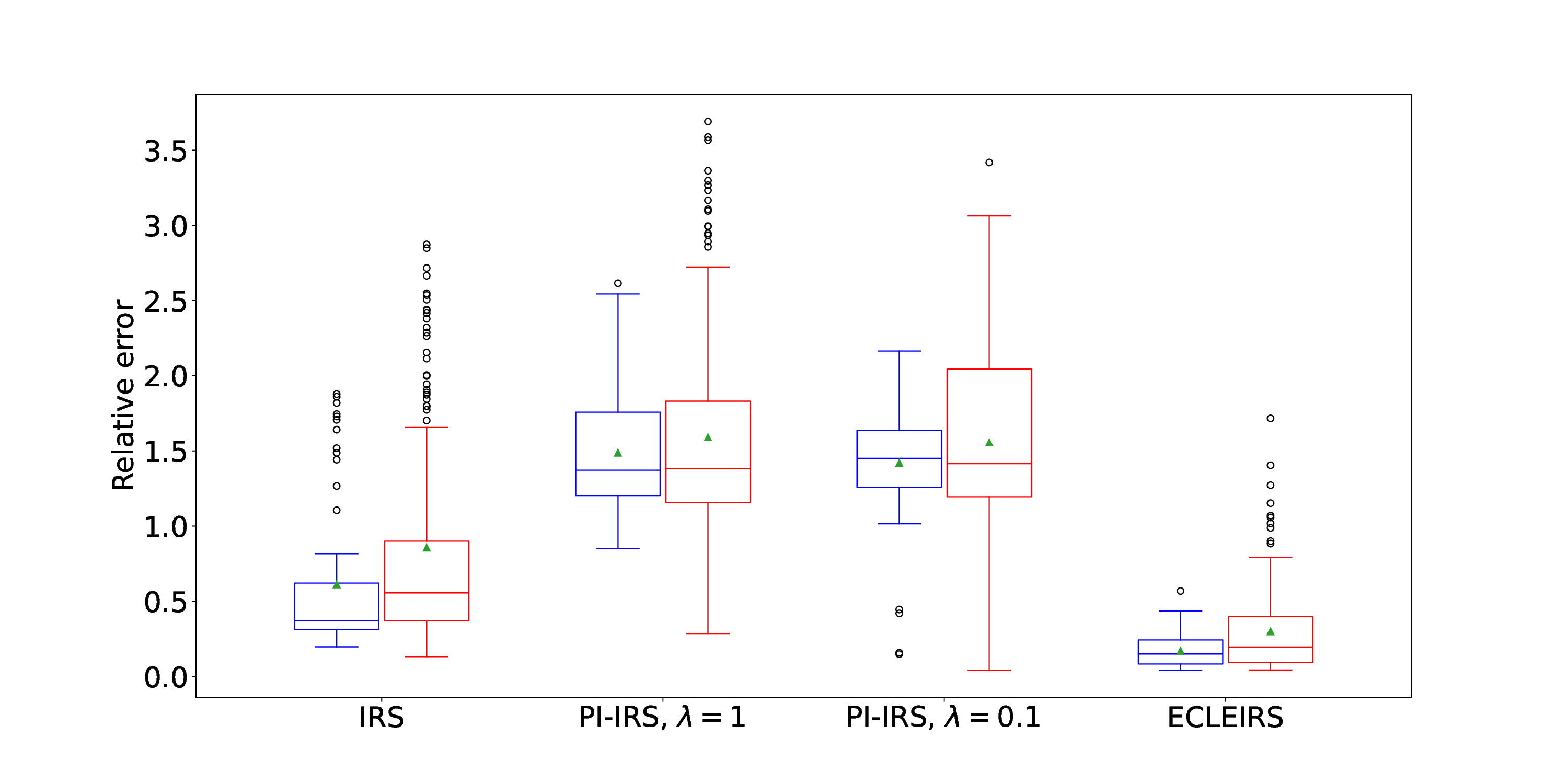}}
    \caption{1-D Advection problem: Box plots of relative errors (defined in \eref{rel_error_def}) for the validation dataset comparing different reduced state dynamics approaches trained using $20\%$ spatiotemporal sparse data with different noise levels (a) $\sigma_N = 0.03$ and (b) $\sigma_N = 0.1$. Blue box plots are for parameters in the training parameter space $\mathcal{D}^{\mu}$, while the red box plots are for parameters outside the training parameter space.}
    \label{fig:BoxplotCompareModel_advec_testing}
\end{figure}
\begin{figure}[t!]
    \centering
    \subfigure[\label{fig:BoxplotEcleirs_testing}]{\includegraphics[width=0.49\linewidth, trim={0.5cm 0cm 5cm 2.5cm},clip]{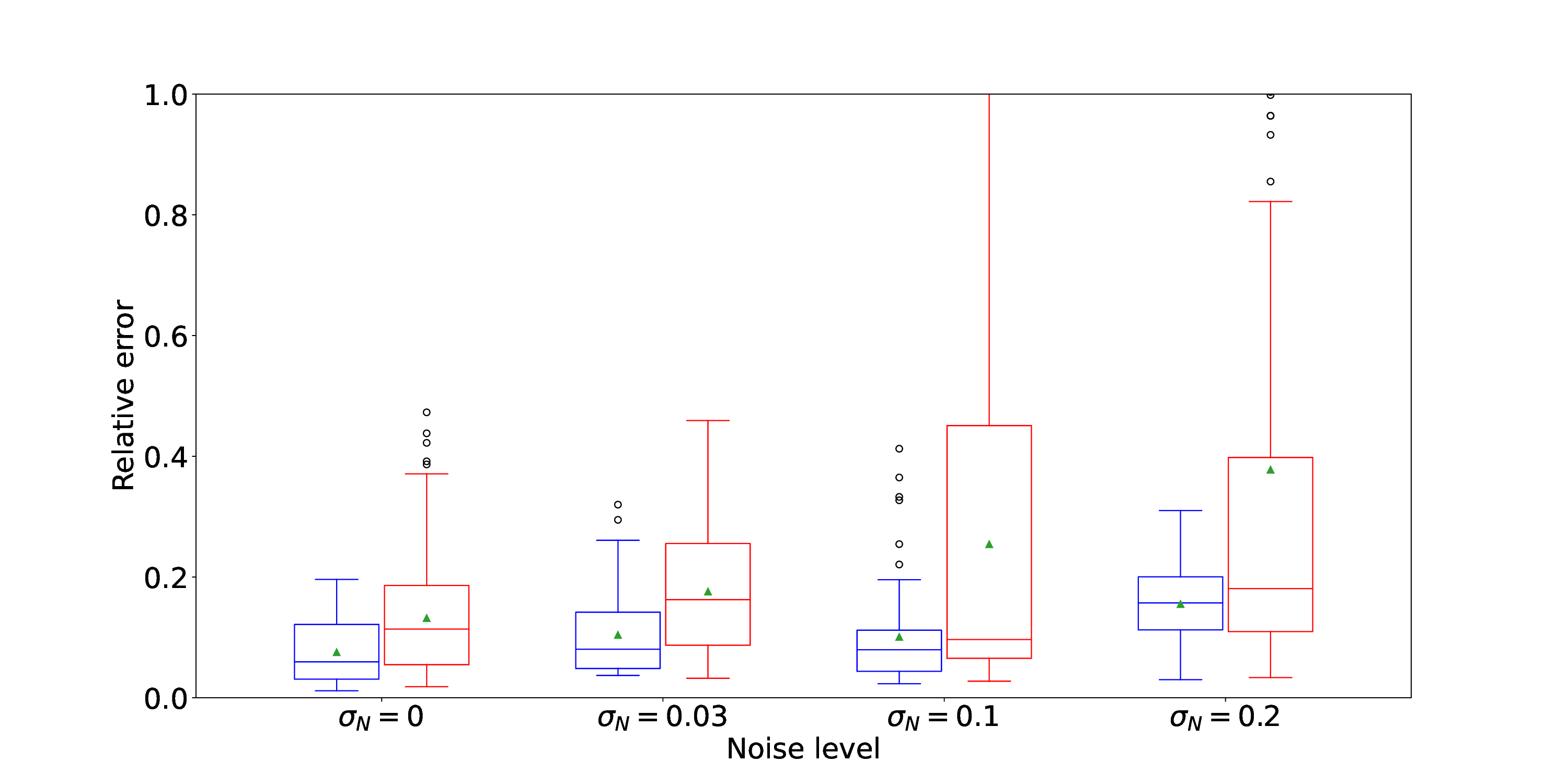}}
    \subfigure[\label{fig:BoxplotEcleirs_testing_consv}]{\includegraphics[width=0.49\linewidth, trim={0.5cm 0cm 5cm 2.0cm},clip]{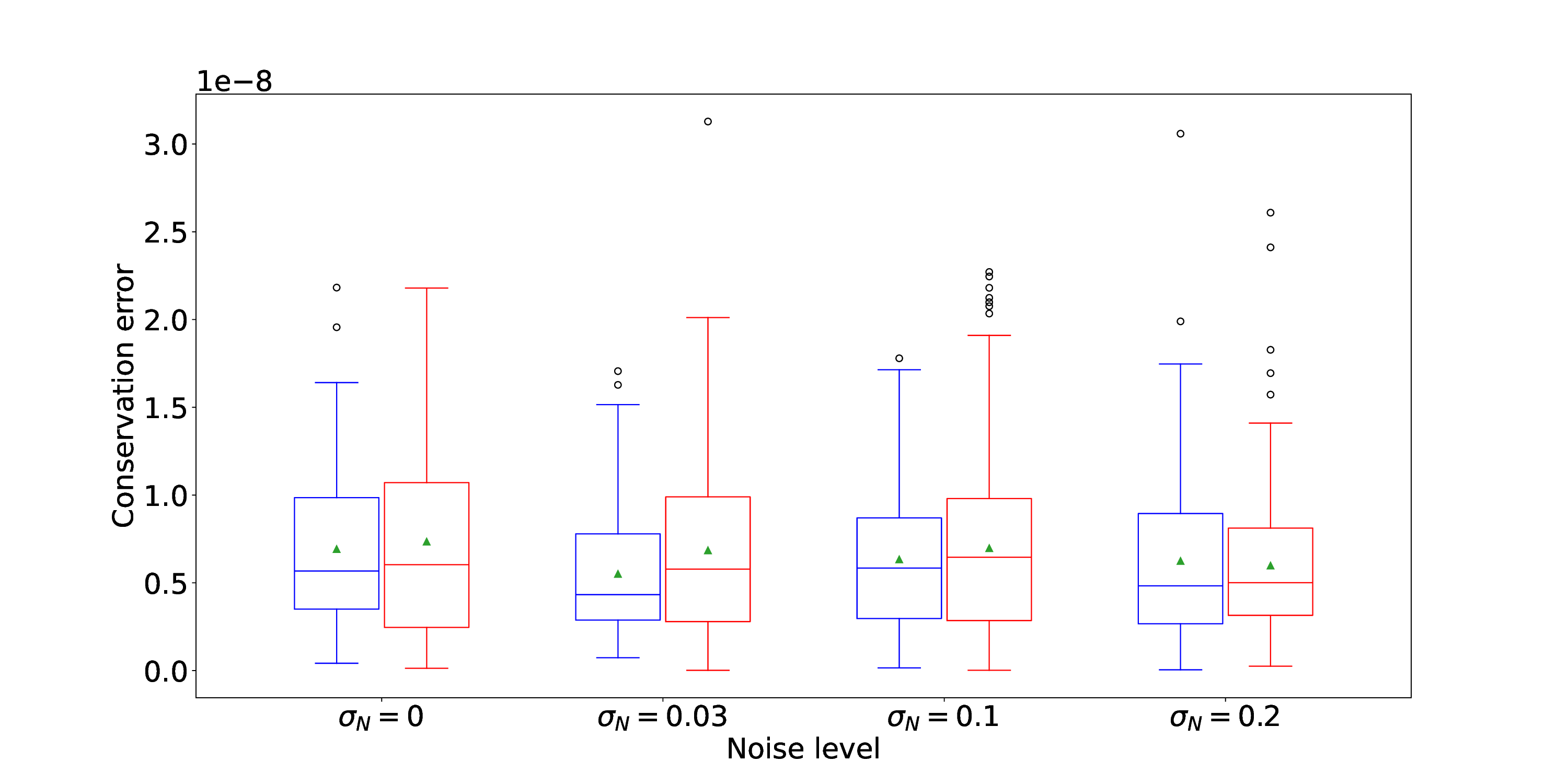}}
    \caption{1-D Advection problem: Box plots of (a) relative error (defined in \eref{rel_error_def}) and (b) mean conservation error (defined in \eref{consv_error}) for ECLEIRS comparing different noise level for $20\%$ spatial sparsity and $100\%$ temporal sparse training data. Blue box plots are for parameters in the learning parameter space $\mathcal{D}^{\mu}$, while the red box plots are for parameters outside the learning parameter space.}
    \label{fig:BoxplotEcleirs_advec_testing}
\end{figure}

With the previous results, we have explored the performance of reduced state dynamics approaches when clean data is available at full spatial and temporal resolution. However, as in several situations, we may only have access to sparse spatiotemporal noisy data. Therefore, it is important to assess if the reduced state dynamics learned from sparse and noisy data extends well to the validation dataset consisting of parameters not included in the model learning process. Various reduced state dynamics approaches considered in this article are compared in \figref{BoxplotCompareModel_advec_testing} to highlight the differences in their performance for the validation dataset. We separately analyze results for parameters within the learning parameters space, referred to as interpolation parameters, and those outside this space, referred to as extrapolation parameters. The results show high errors for IRS and PI-IRS for penalty parameters in both interpolation and extrapolation parameter space. Note that these penalty parameters are the same ones selected in the previous section for assessing the sparse reconstruction and denoising capability of PI-IRS. These results indicate that while PI-IRS excels at denoising from sparse data, the reduced state dynamics identified by this approach do not extend well for unseen parameters not included in the learning process. Conversely, ECLEIRS excels not only for denoising and sparse reconstruction but also performs well for dynamics prediction for unseen parameters, as indicated by low errors observed in the validation dataset. This observation holds for both noise levels and also for interpolation and extrapolation parameter spaces. The error for extrapolation parameter space is higher than that for interpolation parameter space for all the approaches. However, these errors appear to degrade much more slowly for ECLEIRS compared to the other reduced state dynamics approaches. We carefully examine the performance of ECLEIRS for different noise levels on spatially sparse data by analyzing the errors shown in \figref{BoxplotEcleirs_advec_testing}. The results show that the median and standard deviation of the errors do not change significantly with an increase in noise level for the interpolation parameters. However, there is a slight increase in mean error with increasing noise, which is expected as the sparse reconstruction and denoising ability slowly decreases for extremely sparse and noisy data. The faster deterioration of results for extrapolation parameters with increasing noise levels is also expected. However, as shown earlier, this degradation is much slower than other approaches. Furthermore, we observe numerous outliers exhibiting massive errors at higher noise levels. Overall, ECLEIRS is more robust than IRS or PI-IRS. The analysis of mean conservation error for different noise levels shows low conservation error. This error is insensitive to the noise levels and sparsity of the training data for both interpolation and extrapolation parameters. This observation is expected as conservation is exactly enforced by construction in ECLEIRS formulation. This attribute offers a meaningful advantage over constrained optimization-type approaches like PI-IRS, which can only ensure conservation during training and lacks generalization to unseen parameters. 

\subsection{1-D Burgers problem}

The second numerical experiment is conducted for the parameterized 1-D Burgers equation, which is expressed in conservative form as 
\begin{equation}
    \frac{\p q}{\p t} + \frac{\p}{\p x} \Big(\frac{1}{2} c q^2\Big) = 0,
    \label{eq:1DBurger}
\end{equation}
where $c$ is the advective parameter used to parameterize the wave speed. The commonly used Burgers equation uses $c = 1$. In addition to advection term parameterization, we also parameterize the initial conditions of the problem as
\begin{equation}
    q(x,0) = q_0 \sin{(2\pi \omega x)} \quad \forall \quad x \in \Omega,
\end{equation}
where $q_0$ and $\omega$ are the other two system parameters and $\Omega = [0,1]$ is the problem domain. Therefore, this numerical experiment consists of parameter vector $\pmb{\mu} = [c, \; q_0,\; \omega]$ sampled from the parameter space $\mathcal{D}^{\mu} \in \mathbb{R}^3$. The data is generated by solving \eref{1DBurger} using finite-volume based WENO-JS spatial discretization scheme \cite{Jiang1996} along with a third order TVD Runge-Kutta temporal integration scheme \cite{Shu1988}, which results in a stable formulation. The data is generated for the CFL number of $0.5$ using $201$ grid points in the spatial direction. The details of the parameter space of the data used for learning and validating the model are shown in \tabref{Burg_dataset}. Note that the validation dataset involves a completely different set of parameters than the training data. Some parameter combinations in the validation lie in the convex hull of the training parameter space $\mathcal{D}^{\mu}$, while several others are outside this parameter space in the extrapolation regime. This selection of parameter space allows us to assess the performance of the identified reduced state dynamics for both interpolation and slight extrapolation from the parameter space. The training dataset is further sparsely sampled in space and time to assess the performance of the proposed approaches in the presence of sparse data. In addition to sparse sampling, Gaussian noise with zero mean and standard deviation of $\sigma_N$ is added to the sparsely sampled data. This synthetic sampling of and noise addition to the high-fidelity data allows assessment of the proposed approach for scenarios where fully resolved data may not be available.

\begin{table}[t]
    \centering
        \caption{Details of the training and validation dataset for 1-D Burgers problem. The learning dataset is further randomly divided between training ($75 \%$) and testing set ($25 \%$) to ensure that the model is not overfitted.}
    \begin{tabular}{|c|c|c|}
    \hline
         \textbf{Dataset Name} & \textbf{Number of data points} & \textbf{Parameter values} \\
         & ($n_x \times n_t$) & ($c \times q_0 \times x_{in}$) \\
    \hline
         Learning & $201 \times 390$ & $c \in \{0.8, 0.9, 1.0, 1.1, 1.2\}$ \\
        & & $q_0 \in \{0.8, 0.9, 1.0, 1.1, 1.2\}$ \\
        & & $\omega \in \{0.8, 0.9, 1.0, 1.1, 1.2\}$ \\
         \hline
         Validation & $201 \times 390$ & $c \in \{0.75, 0.85, 0.95, 1.05, 1.15, 1.25\}$ \\
        & & $q_0 \in \{0.75, 0.85, 0.95, 1.05, 1.15, 1.25\}$ \\
        & & $\omega \in \{0.75, 0.85, 0.95, 1.05, 1.15, 1.25\}$ \\
        \hline
    \end{tabular}
    \label{tab:Burg_dataset}
\end{table}

\subsubsection{Denoising and sparse reconstruction capability on the learning dataset}

\begin{figure}[t]
    \centering
    \subfigure[\label{fig:SolutionNoiseBurgers1D}]{\includegraphics[width=0.49\linewidth, trim={0.0cm 0cm 1.5cm 1.0cm},clip]{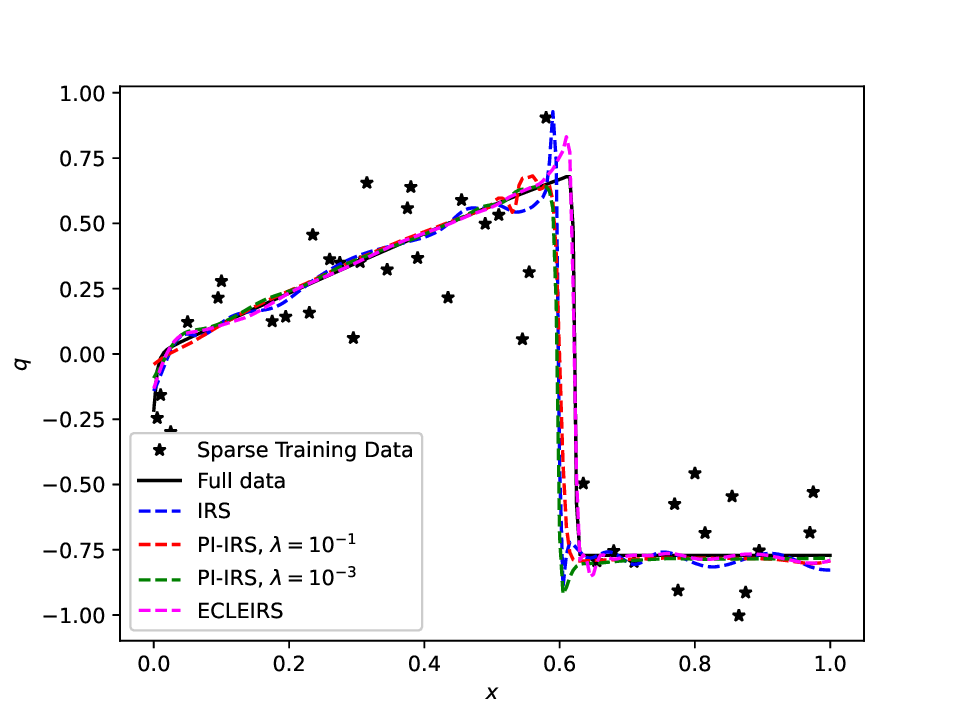}}
    \subfigure[\label{fig:ErrorNoiseBurgers1D}]{\includegraphics[width=0.49\linewidth, trim={0.0cm 0cm 1.5cm 1.0cm},clip]{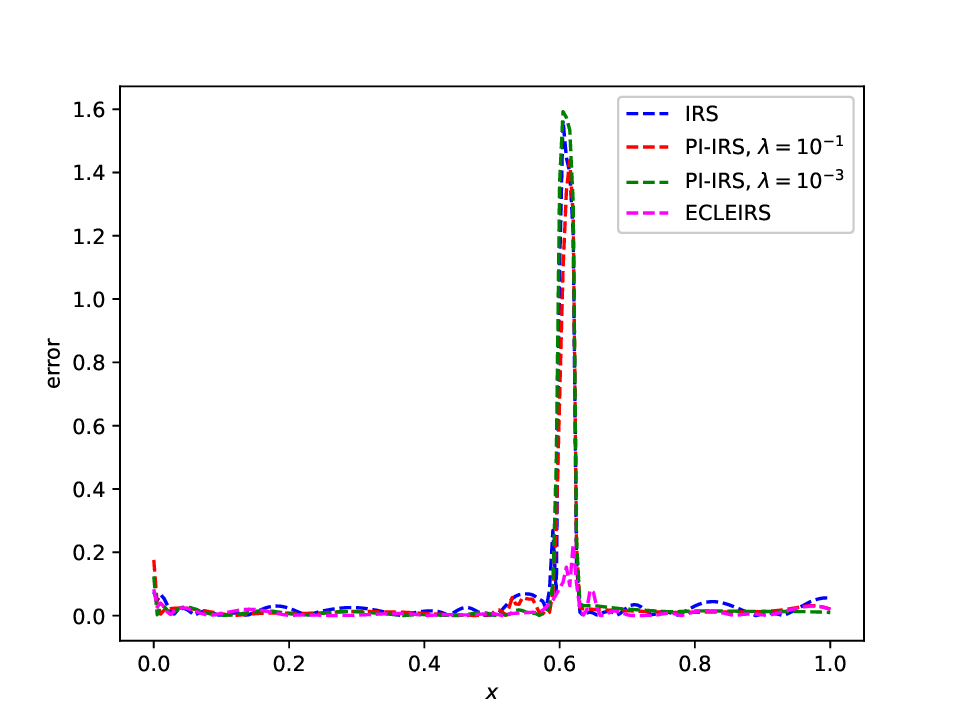}}
    \caption{1-D Burgers problem: (a) Predicted solution and (b) error for $\bm{\mu} = [1.0, \; 0.8, \;  0.8]$ with spatiotemporal sparsity of $20\%$ and $\sigma_N = 0.2$ for different reduced state dynamics identification approaches.}
    \label{fig:SolErrorNoiseBurgers1D}
\end{figure}

Similar to the 1-D advection problem, we first assess the ability of the proposed solution representation approaches to reconstruct the clean solution signal from a dataset comprising sparse and noisy parametric data. The performance of different solution representation approaches in obtaining a clean solution representation from the sparse and noisy signal is shown in \figref{SolErrorNoiseBurgers1D}. The results indicate how well different solution representation approaches work in identifying the clean solution signal from sparse and noisy data. Among these approaches, PI-IRS with lower $\lambda$ performs better compared to PI-IRS with higher $\lambda$ and IRS, as these latter approaches exhibit large oscillations near the shock. Despite having an overshoot of prediction near the shock, ECLEIRS appears to provide the most accurate representation of the signal by capturing the correct location of the shock. This observation is reinforced by looking at the error plots, which show much higher errors for IRS and PI-IRS at different penalty parameters compared to ECLEIRS. 

\begin{figure}
    \centering
    \subfigure[\label{fig:BoxplotPenalty_Burgers1D_sigma0}]{\includegraphics[width=0.49\linewidth, trim={0.5cm 0cm 5cm 2.5cm},clip]{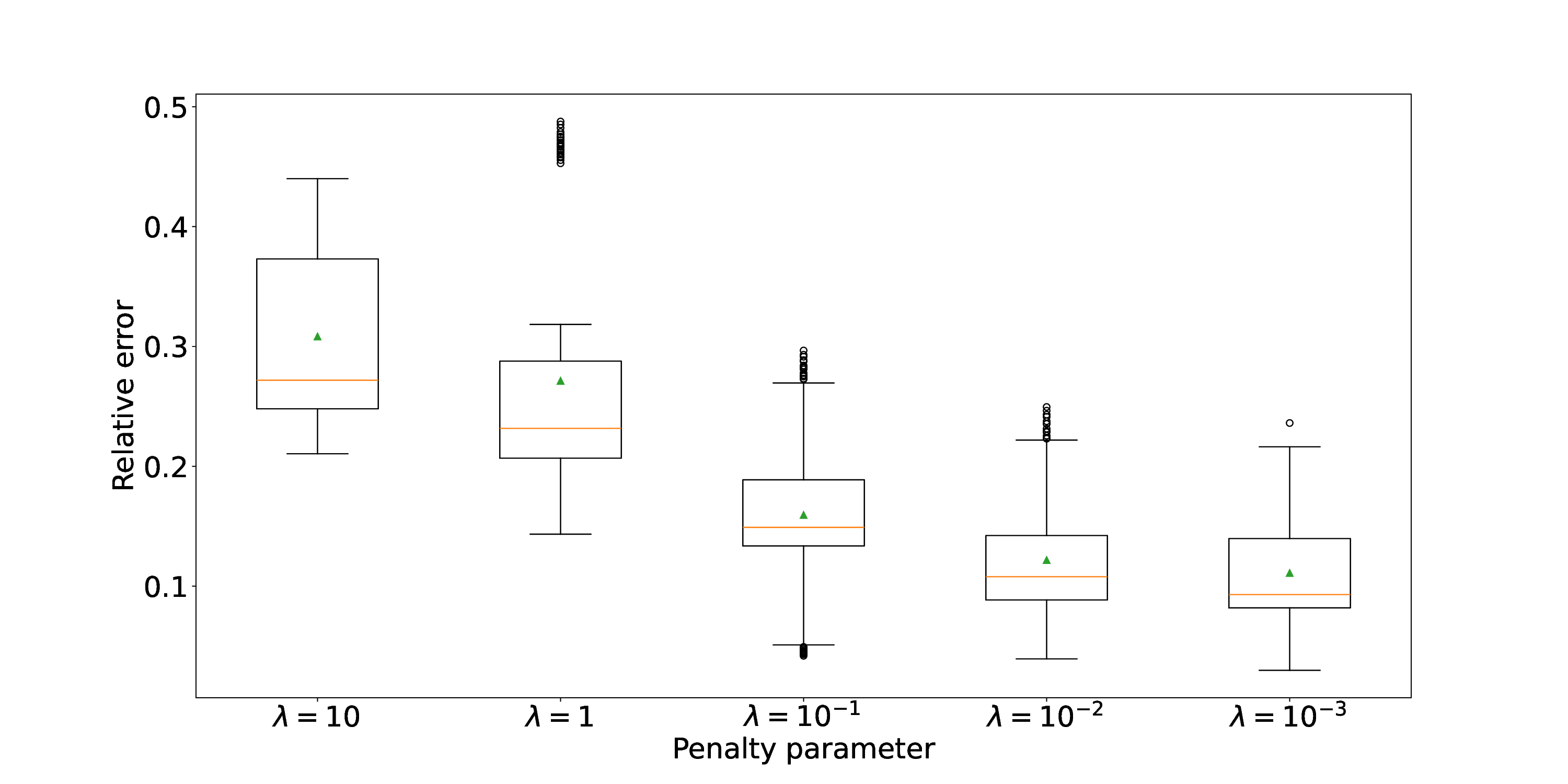}}
    \subfigure[\label{fig:BoxplotPenalty_Burgers1D_sigma0p4}]{\includegraphics[width=0.49\linewidth, trim={0.5cm 0cm 5cm 2.5cm},clip]{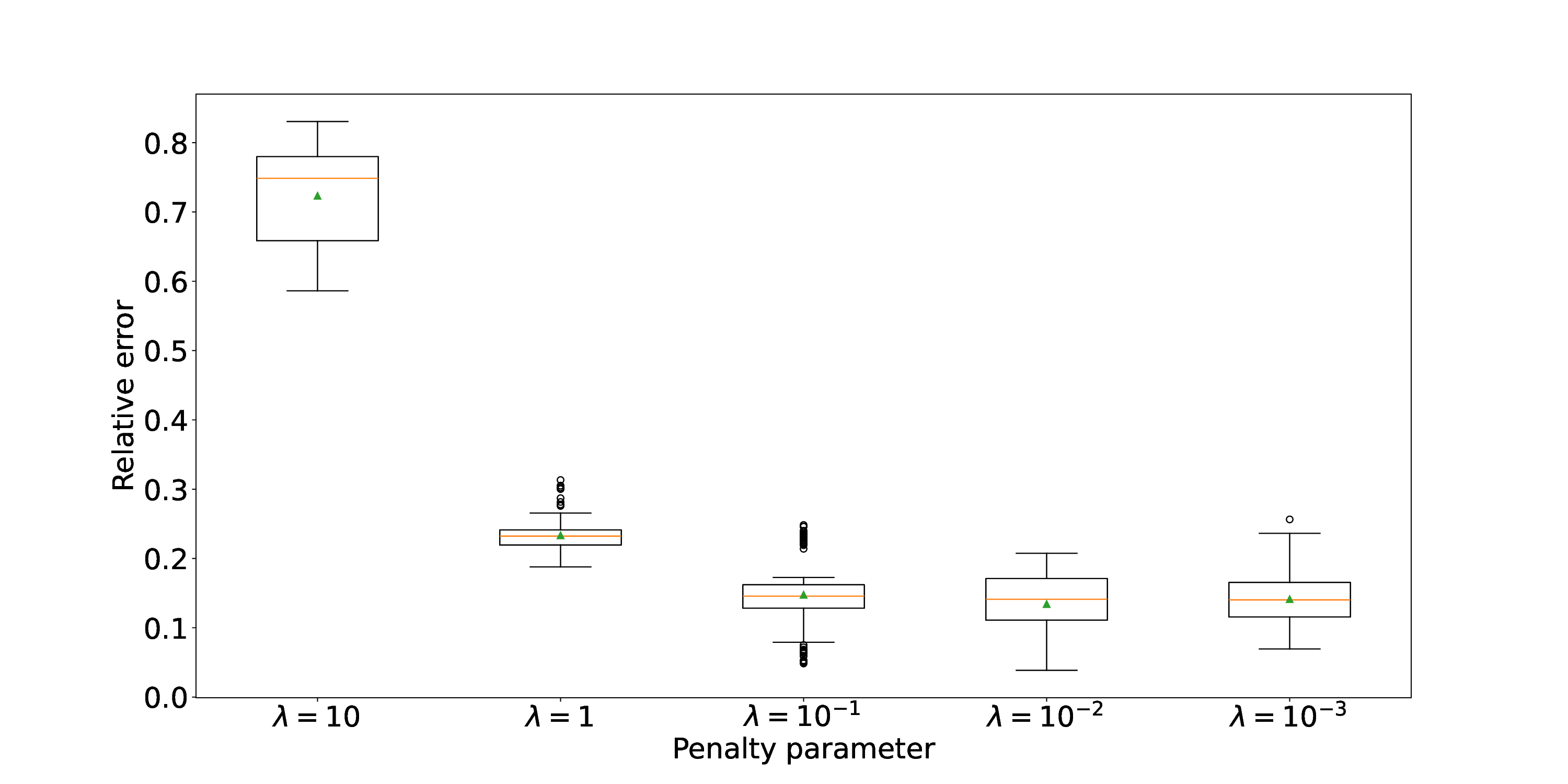}}
    \caption{1-D Burgers problem: Box plots of relative error (defined in \eref{rel_error_def}) for training data with (a) $\sigma_N = 0$ (no noise) and (b) $\sigma_N = 0.4$ with sparsity of $20\%$ for PI-IRS with different penalty parameter values.}
    \label{fig:BoxplotPenalty_Burgers1D}
\end{figure}

While using PI-IRS, it is important to assess the sensitivity of solution representation error to penalty parameters. We assess this sensitivity by comparing the box plots of relative errors defined in \eref{rel_error_def} and plotted in \figref{BoxplotPenalty_Burgers1D}. The box plots indicate how the relative errors vary for different parameter combinations in the learning dataset. The results indicate that error decreases as the penalty parameter decreases, especially when data has no noise. At higher noise levels, the mean error appears to asymptote while the standard deviation increases for $\lambda < 10^{-1}$. We expect the mean error to increase with a further decrease in penalty parameter as we know that $\lambda \to 0$ will result in the same results as IRS, which exhibits higher error. These results also show that the ``ideal" value of $\lambda$ is different from those observed for the 1-D advection problem and highlights the sensitivity of PI-IRS on penalty parameters. This attribute of PI-IRS serves as a major drawback despite the ability of physics regularization to better identify solution representation from sparse and noisy data.

\begin{figure}[t]
    \centering
    \subfigure[\label{fig:BoxplotModelCompare_Burgers1D_sigma0}]{\includegraphics[width=0.49\linewidth, trim={0.5cm 1.5cm 5cm 2.5cm},clip]{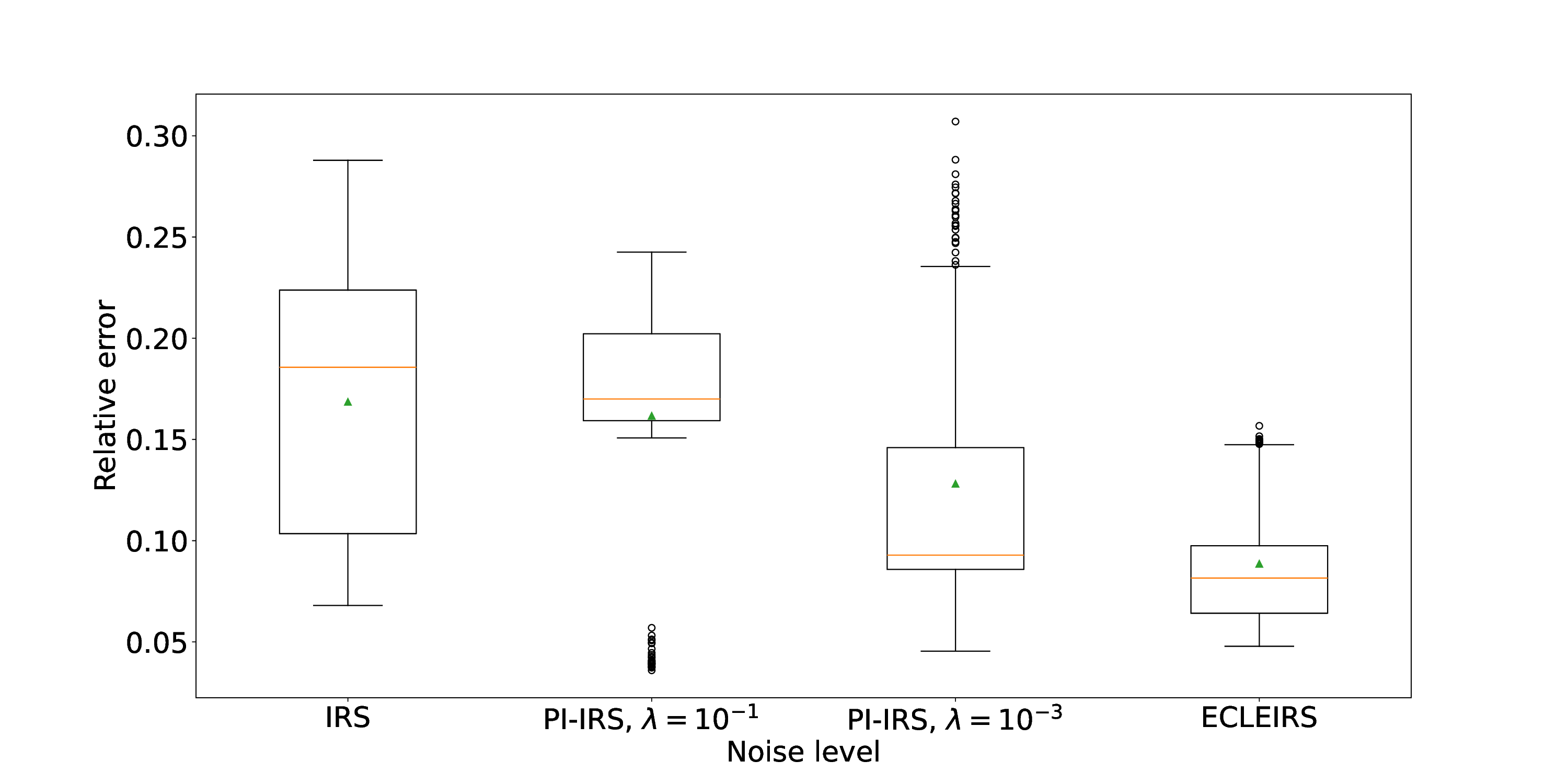}}
    \subfigure[\label{fig:BoxplotModelCompare_Burgers1D_sigma0p4}]{\includegraphics[width=0.49\linewidth, trim={0.5cm 1.5cm 5cm 2.5cm},clip]{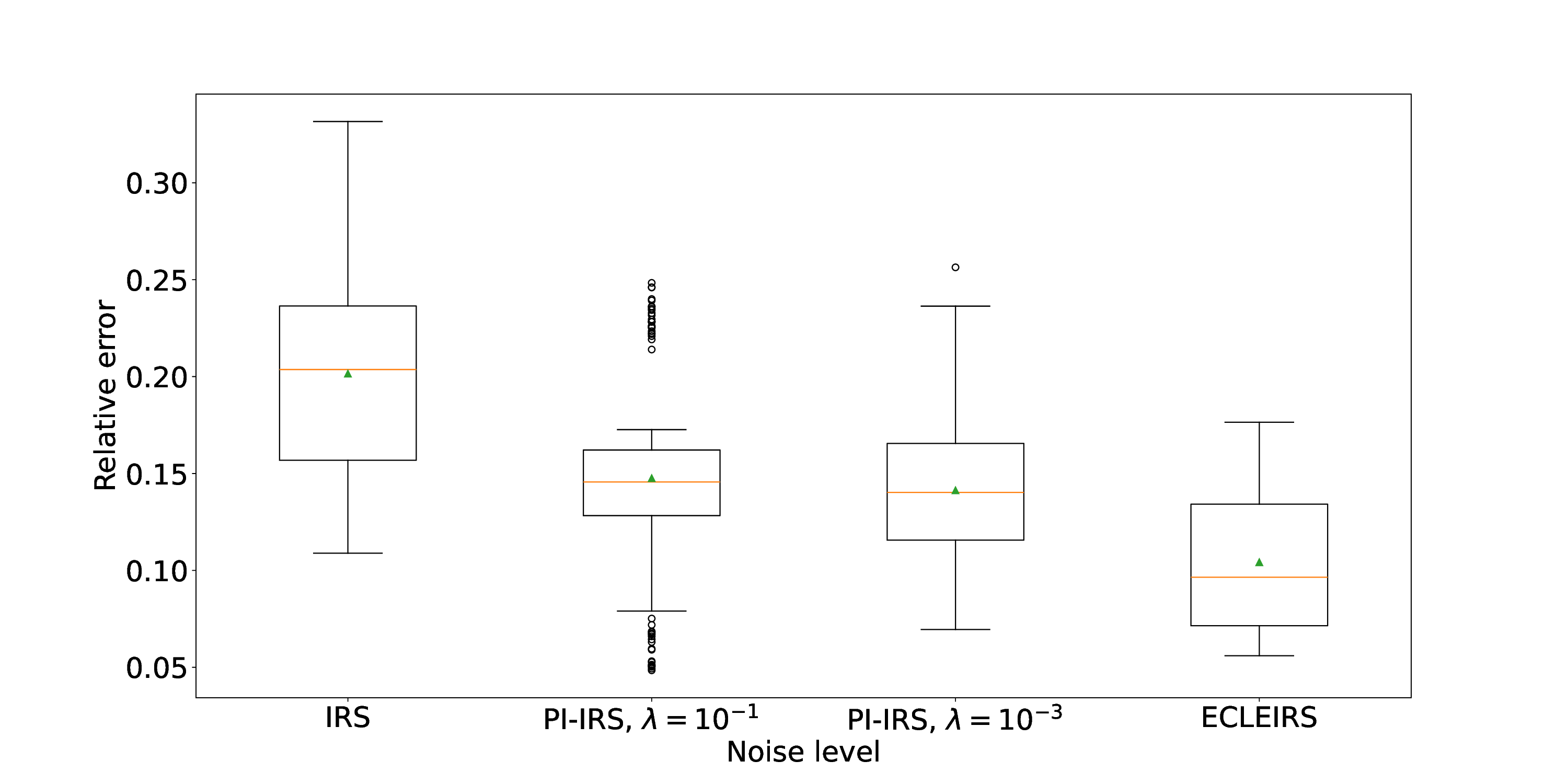}}
    \caption{1-D Burgers problem: Box plots of relative error (defined in \eref{rel_error_def}) for training data with $20\%$ spatiotemporal sparsity and added noise level of (a) $\sigma_N = 0.2$ and (b) $\sigma_N = 0.4$ comparing different solution representation approaches.}
    \label{fig:BoxplotModelCompare_Burgers1D}
\end{figure}

The relative errors for different solution representation approaches are compared in \figref{BoxplotModelCompare_Burgers1D} for different noise levels. These results indicate IRS results in the highest errors for both noise levels. While PI-IRS works better than IRS for both noise, the performance improvement depends on the selection of penalty parameters. Conversely, ECLEIRS offers a parameter-free approach that also provides the lowest errors for different noise levels. 

\begin{figure}[t]
    \centering
    \subfigure[\label{fig:BoxplotNoiseCompare_Burgers1D_rho0p2}]{\includegraphics[width=0.49\linewidth, trim={0.5cm 0cm 5cm 2.5cm},clip]{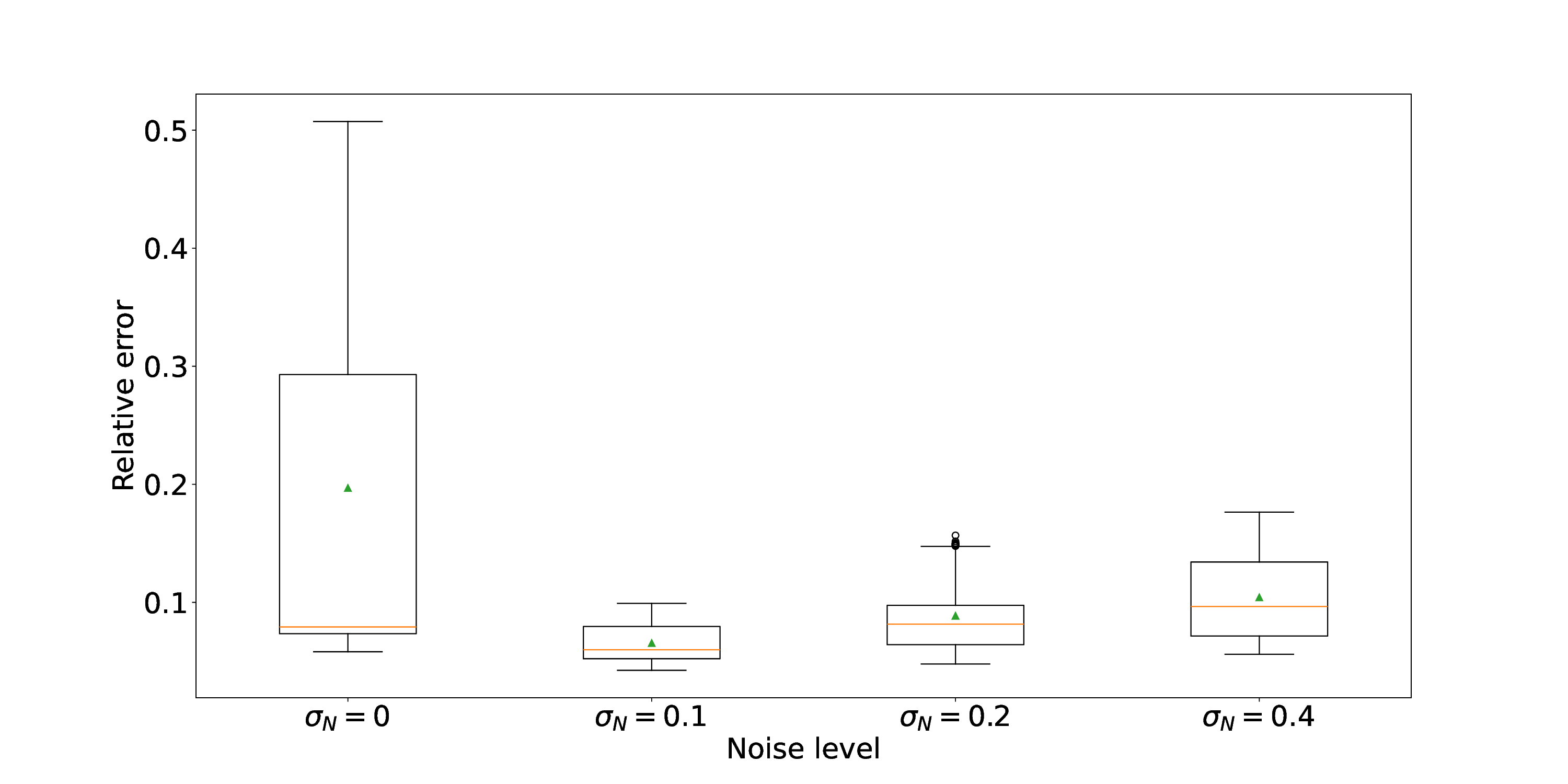}}
    \subfigure[\label{fig:BoxplotNoiseCompare_Burgers1D_rho1p0}]{\includegraphics[width=0.49\linewidth, trim={0.5cm 0cm 5cm 2.5cm},clip]{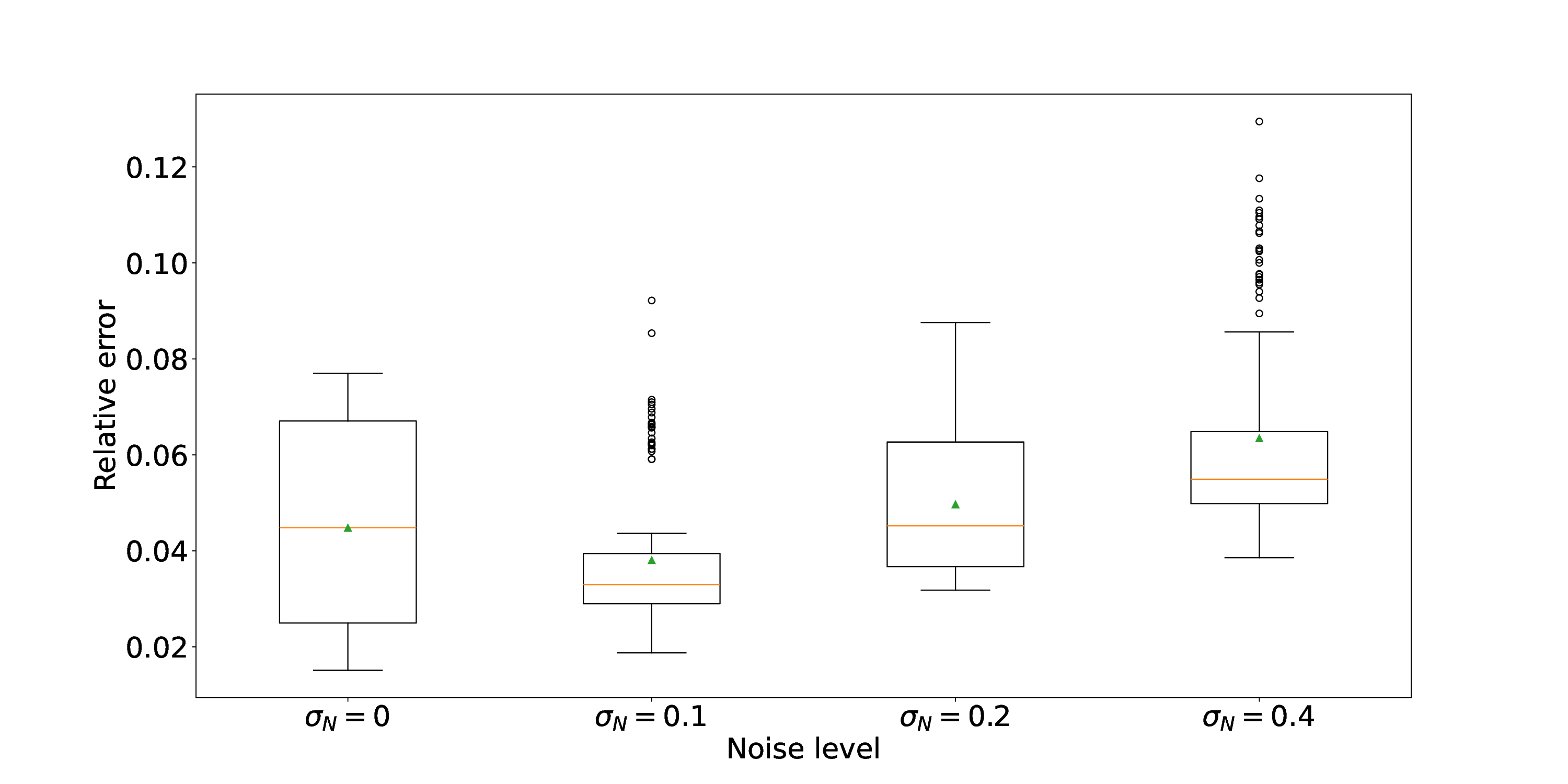}}
    \caption{1-D Burgers problem: Box plots of relative error (defined in \eref{rel_error_def}) for ECLEIRS prediction for training data with different noise levels and spatial sparsity of (a) $20\%$ and (b) $40\%$ while keeping temporal sparsity of $20\%$.}
    \label{fig:BoxplotNoiseCompare_Burgers1D}
\end{figure}

\begin{figure}[t!]
    \centering
    \subfigure[\label{fig:TotalError_kind_burg}]{\includegraphics[width=\textwidth, trim={4cm 7.5cm 7cm 9cm},clip]{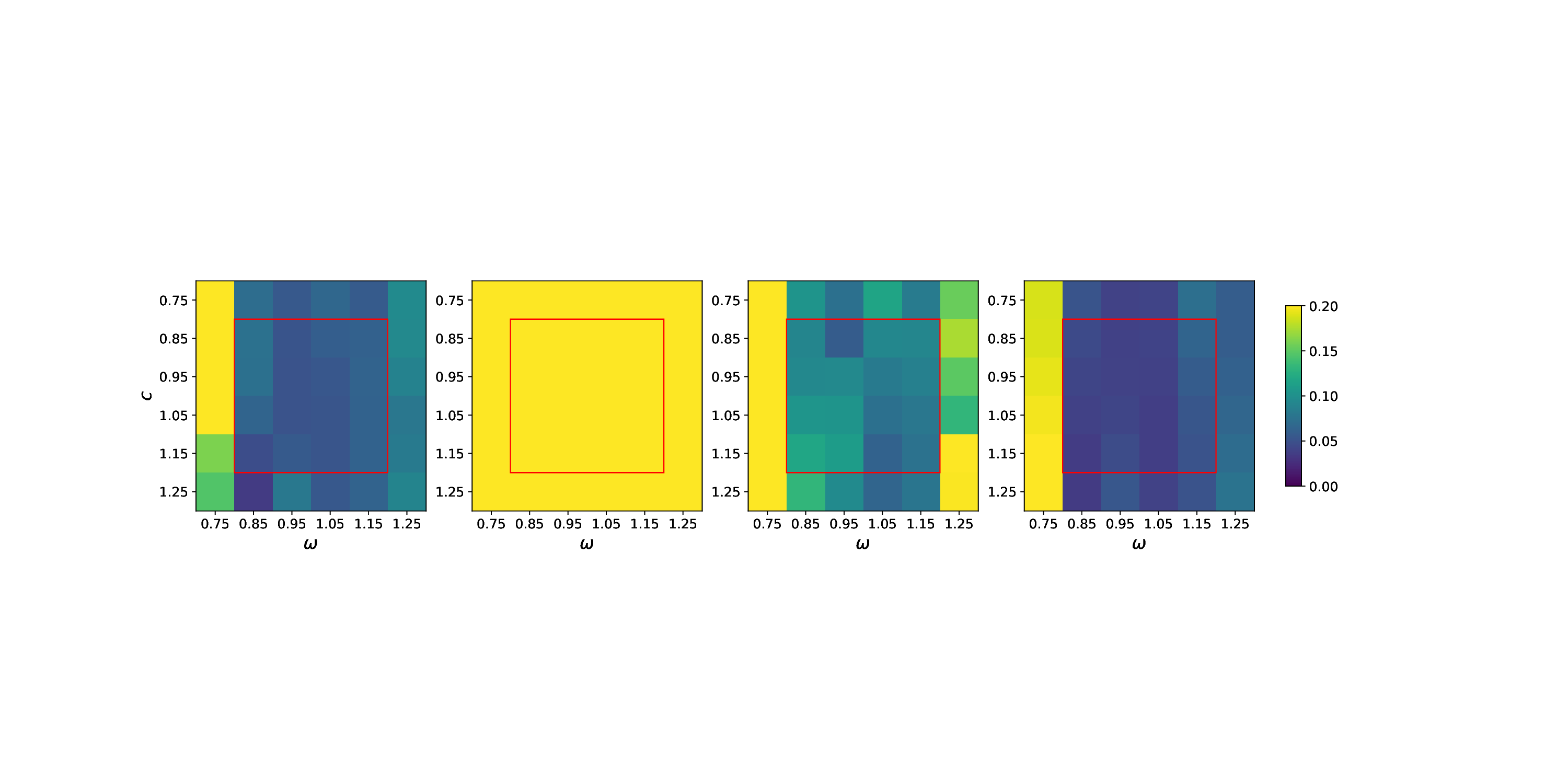}}
    \subfigure[\label{fig:TotalError_jind_burg}]{\includegraphics[width=\textwidth, trim={4cm 7.5cm 7cm 9cm},clip]{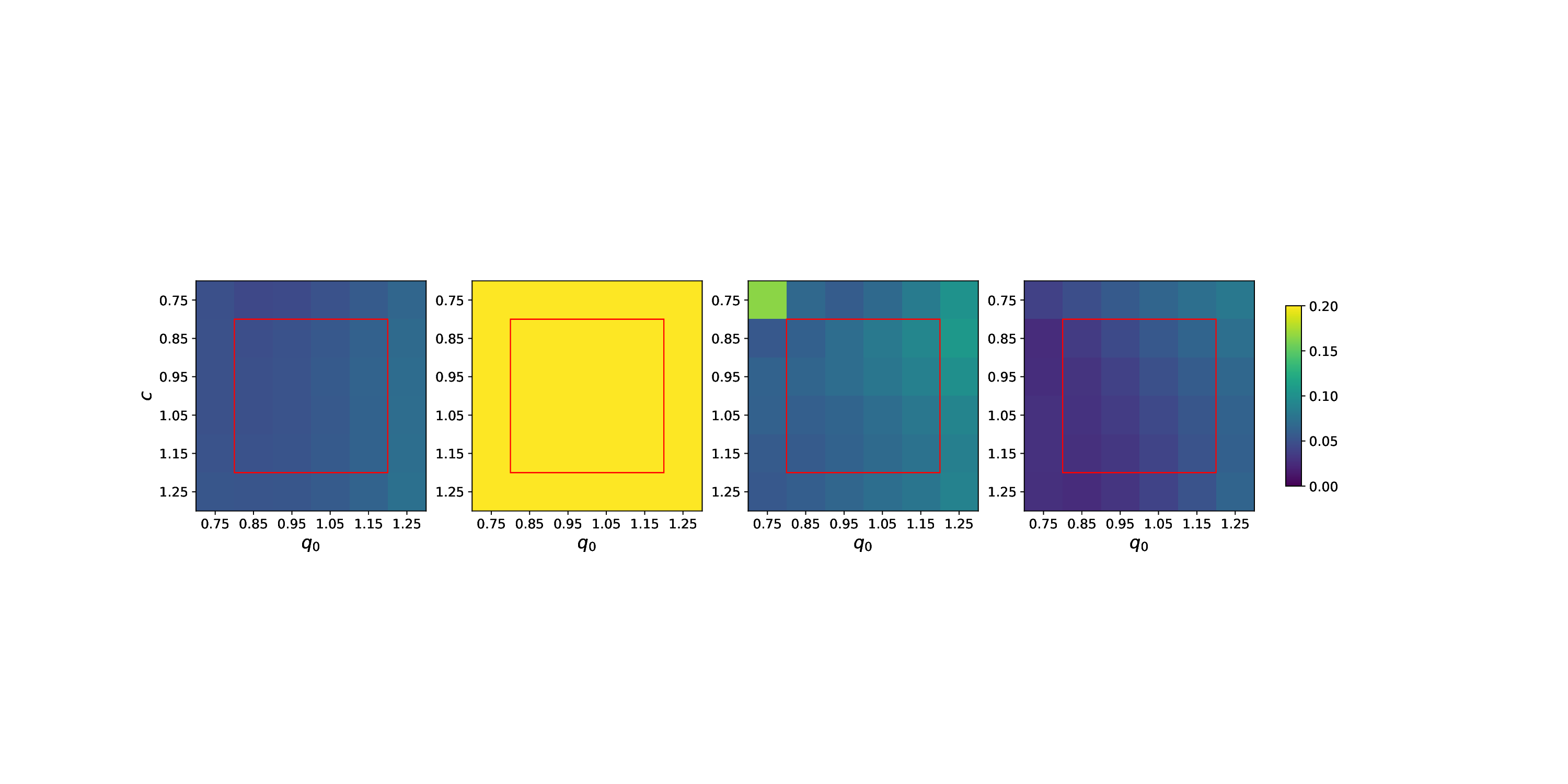}}
    \subfigure[\label{fig:TotalError_iind_burg}]{\includegraphics[width=\textwidth, trim={4cm 7.5cm 7cm 9cm},clip]{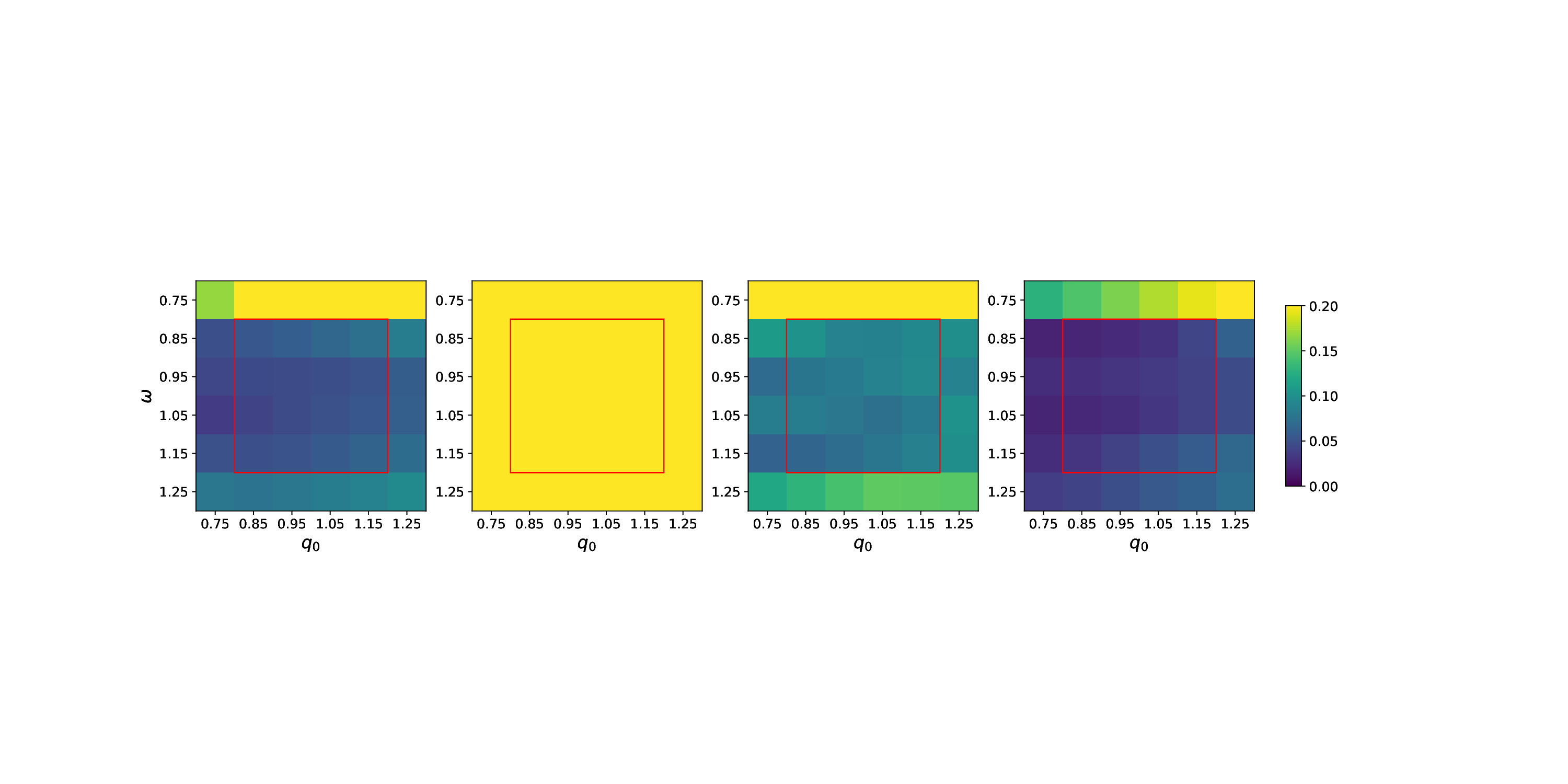}}
    \caption{1-D Burgers problem: Relative error in space-time (defined in \eref{rel_error_def}) for (a) $q_0 = 1.15$, (b) $\omega = 1.15$ and (c) $c = 0.95$. Different columns denote different reduced state dynamics approaches trained on high-resolution clean data: IRS (1st column), PI-IRS with $\lambda = 1 \times 10^{-1}$ (2nd column), PI-IRS with $\lambda = 1 \times 10^{-3}$ (3rd column) and ECLEIRS (4th column). The parameters inside the red box are within the interpolation space and the those outside are in the extrapolation space.}
    \label{fig:SpaceTime_CompareModels_Burgers}
\end{figure}

While ECLEIRS provides more accurate results than PI-IRS and IRS, assessing the variability of results for different noise and sparsity levels is important. The results showing the sensitivity of ECLEIRS on different noise levels and two spatial sparsity levels are shown in \figref{BoxplotNoiseCompare_Burgers1D}. The results indicate a large variance of relative errors despite low median errors for ECLEIRS trained on data with $20\%$ spatial sparsity and no added noise. This result indicates an anamalous behavior of ECLEIRS while predicting strong shocks for several parameters and time values in the absence of noise. These errors are typically visible as overshoots and undershoots near the shock location, as also seen in \figref{SolErrorNoiseBurgers1D}. This behavior could be attributed to the ability of the architecture to respect exact conservation under limited data scenarios while predicting a strong shock, as PI-IRS with penalty type enforcement does not suffer from this issue. These errors can be reduced by hyperparameter tuning of the neural network architecture. We keep the hyperparameters the same across the entire test case to ensure a fair comparison between different reduced state dynamics approaches and different sparse/noise levels in training data. Furthermore, these issues also appear to resolve at higher noise levels as we observe lower errors. For both spatial sparsity values, we observe that even a lower noise level of $\sigma_N = 0.1$ appears to resolve the issue with high error and provides the lowest errors for ECLEIRS. As expected, this error increases with the noise level, as also observed for the 1-D advection problem. The results also show that the anomalous behavior is less severe at $40\%$ temporal sparsity, and therefore, it appears to be an artifact of sparse spatial sampling. At $40\%$ spatial sparsity, an increase in error with noise level is observed except for the zero noise scenario, which agrees with the results observed for the 1-D advection problem.

\subsubsection{Dynamics prediction for the validation dataset}

\begin{figure}[t!]
    \centering
    \subfigure[\label{fig:TotalError_kind_consverror_burg}]{\includegraphics[width=\textwidth, trim={4cm 7.5cm 7cm 9cm},clip]{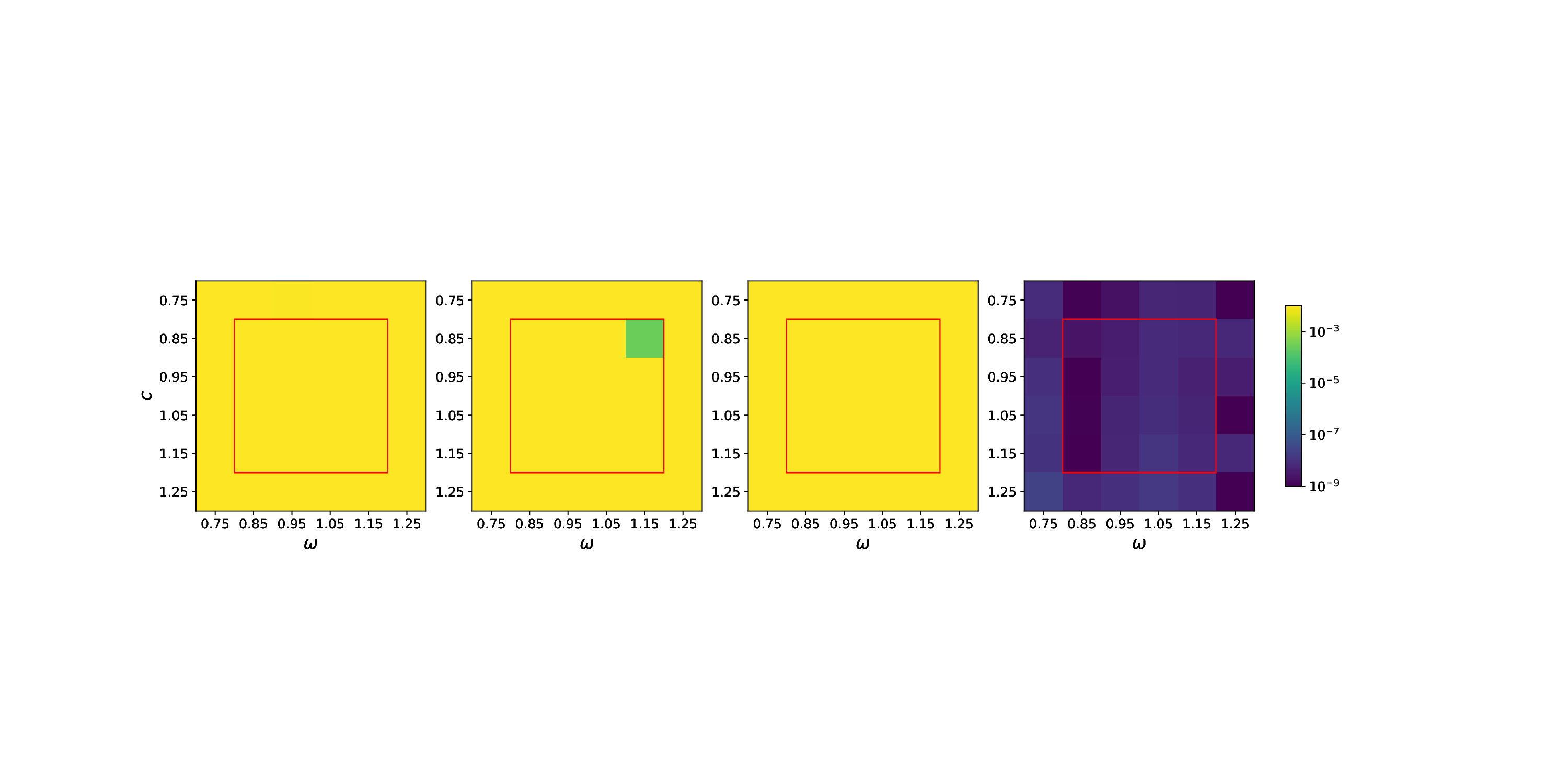}}
    \subfigure[\label{fig:TotalError_jind_consverror_burg}]{\includegraphics[width=\textwidth, trim={4cm 7.5cm 7cm 9cm},clip]{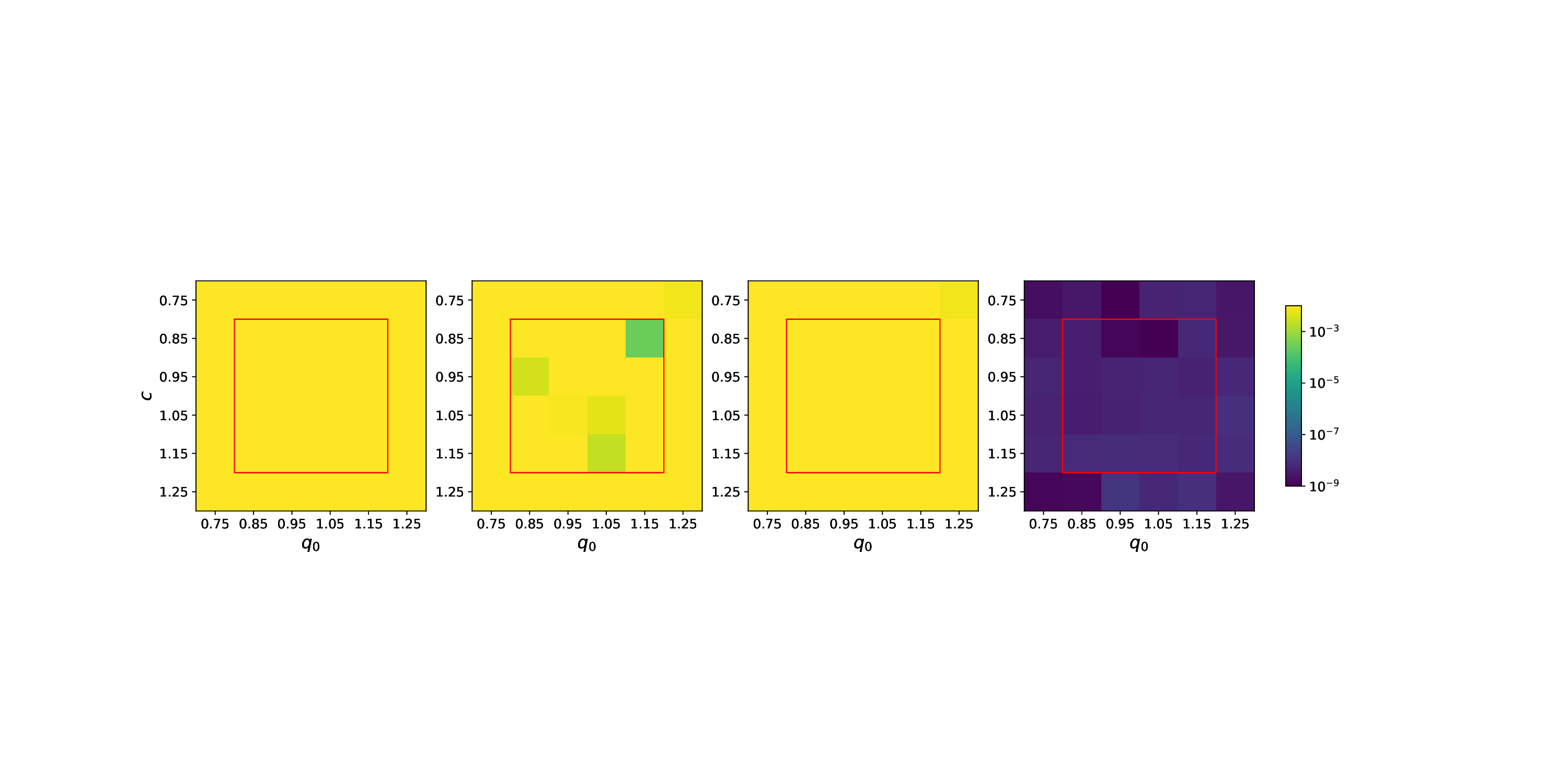}}
    \subfigure[\label{fig:TotalError_iind_consverror_burg}]{\includegraphics[width=\textwidth, trim={4cm 7.5cm 7cm 9cm},clip]{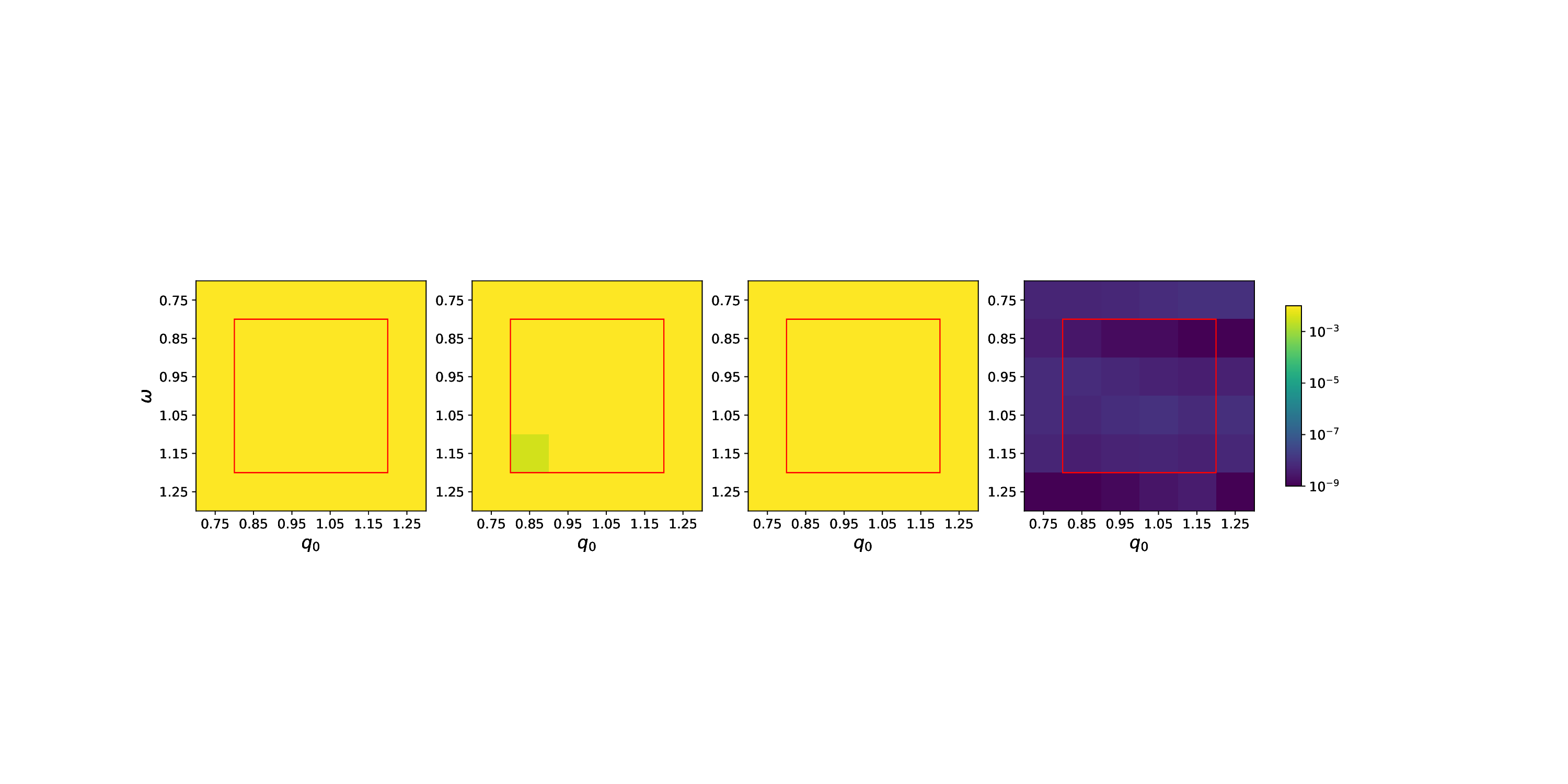}}
    \caption{1-D Burgers problem: Mean conservation error (defined in \eref{consv_error_burgers}) in space-time for (a) $q_0 = 1.15$, (b) $\omega = 1.15$ and (c) $c = 0.95$. Different columns denote different reduced state dynamics approaches trained on high-resolution clean data: IRS (1st column), PI-IRS with $\lambda = 1 \times 10^{-1}$ (2nd column), PI-IRS with $\lambda = 1 \times 10^{-3}$ (3rd column) and ECLEIRS (4th column). The parameters inside the red box are within the interpolation space and the those outside are in the extrapolation space.}
    \label{fig:SpaceTime_CompareModels_Burgers_consverror}
\end{figure}

In this section, we compare the performance of different modeling approaches trained on different levels of sparse and noisy data on parameters that are unseen within the convex hull of the learning parameter space $\mathcal{D}^{\mu}$ and also for those outside this parameter space. The relative error in solution prediction, as defined in \eref{rel_error_def}, obtained for different reduced state dynamics modeling approaches is compared for different system parameters in \figref{SpaceTime_CompareModels_Burgers}. These results are for models learned using full-resolution data without any noise. The results indicate IRS performs well in scenarios where high-resolution clean data is available. On the other hand, the inclusion of physics-informed loss in PI-IRS appears to degrade the results for dynamics prediction as a higher penalty parameter leads to higher errors. We find this interesting because this physics-informed loss was important for solution signal reconstruction and denoising ability in PI-IRS, and provided better results than IRS. Conversely, ECLEIRS provides the most accurate dynamics predictions compared to all the other models. The predictions exhibit lower errors even for several parameters outside the learning parameter space.
    
In addition to the assessment of the error in approximating the solution, we also assess the mean conservation law error for the 1-D Burgers equation defined as
\begin{equation}
    e_{c} (\pmb{\mu}) = \Bigg\vert \frac{1}{n_x n_t} \sum_{i=1}^{n_x} \sum_{j=1}^{n_t} \frac{\p q (x_i, t_j; \pmb{\mu})}{\p t} + \frac{\p }{\p x} \Big( \frac{1}{2} c q^2 (x_i, t_j; \pmb{\mu}) \Big) \Bigg\vert,
    \label{eq:consv_error_burgers}
\end{equation}
where $\vert \cdot \vert$ indicates the absolute value. The mean conservation error for different reduced state dynamics approaches for parameters in the validation set is shown in \figref{SpaceTime_CompareModels_Burgers_consverror}. Similar to 1-D advection case, high errors are observed for IRS and PI-IRS at different penalty parameters. Conversely, ECLEIRS provides low errors for parameters that are within the learning parameter space and also for those that are outside this space.

\begin{figure}[t]
    \centering
    \subfigure[\label{fig:BoxplotModelCompare_burg_noise_low}]{\includegraphics[width=0.49\linewidth, trim={0.5cm 0cm 5cm 2.5cm},clip]{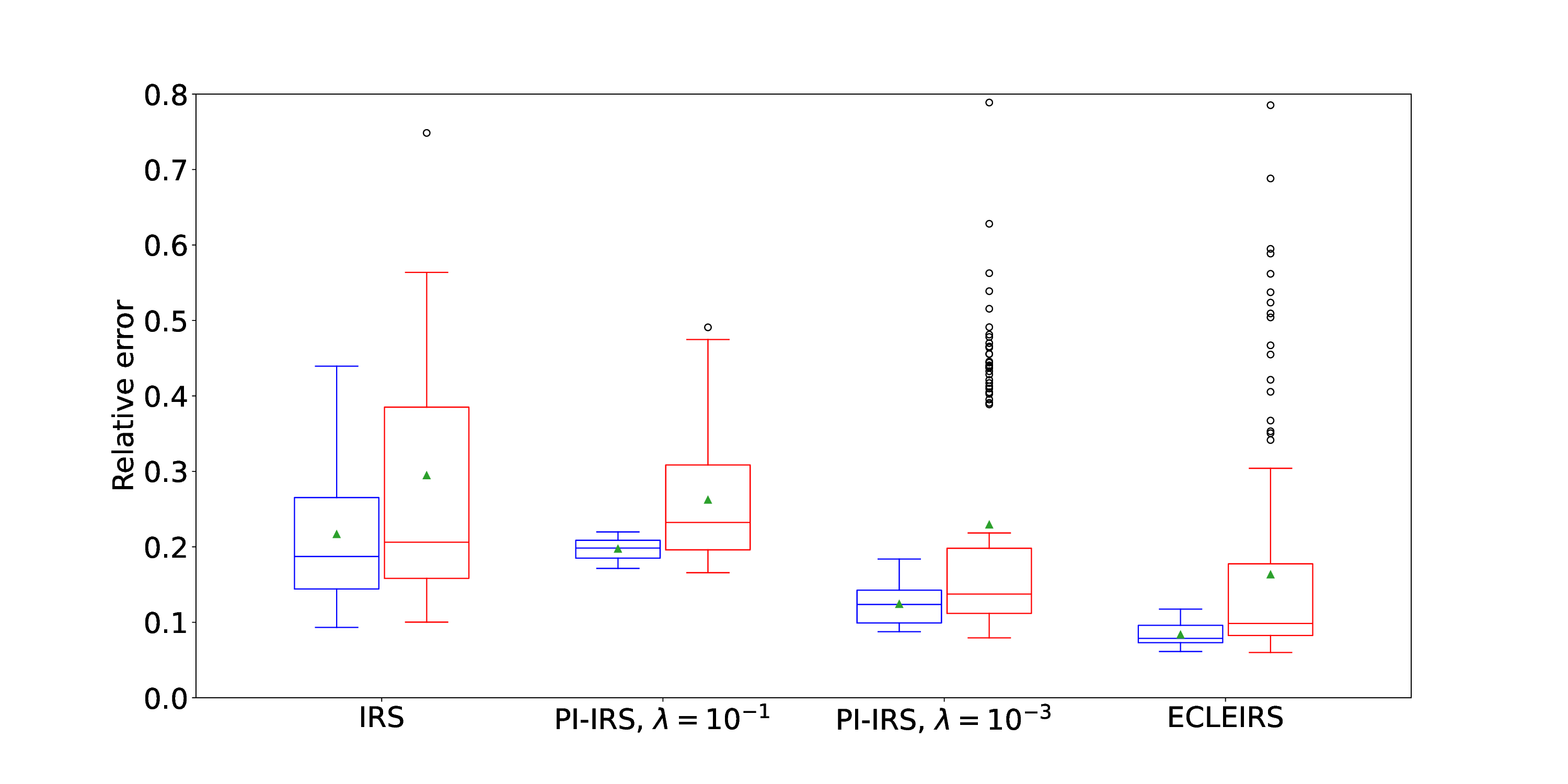}}
    \subfigure[\label{fig:BoxplotModelCompare_burg_noise_high}]{\includegraphics[width=0.49\linewidth, trim={0.5cm 0cm 5cm 2.5cm},clip]{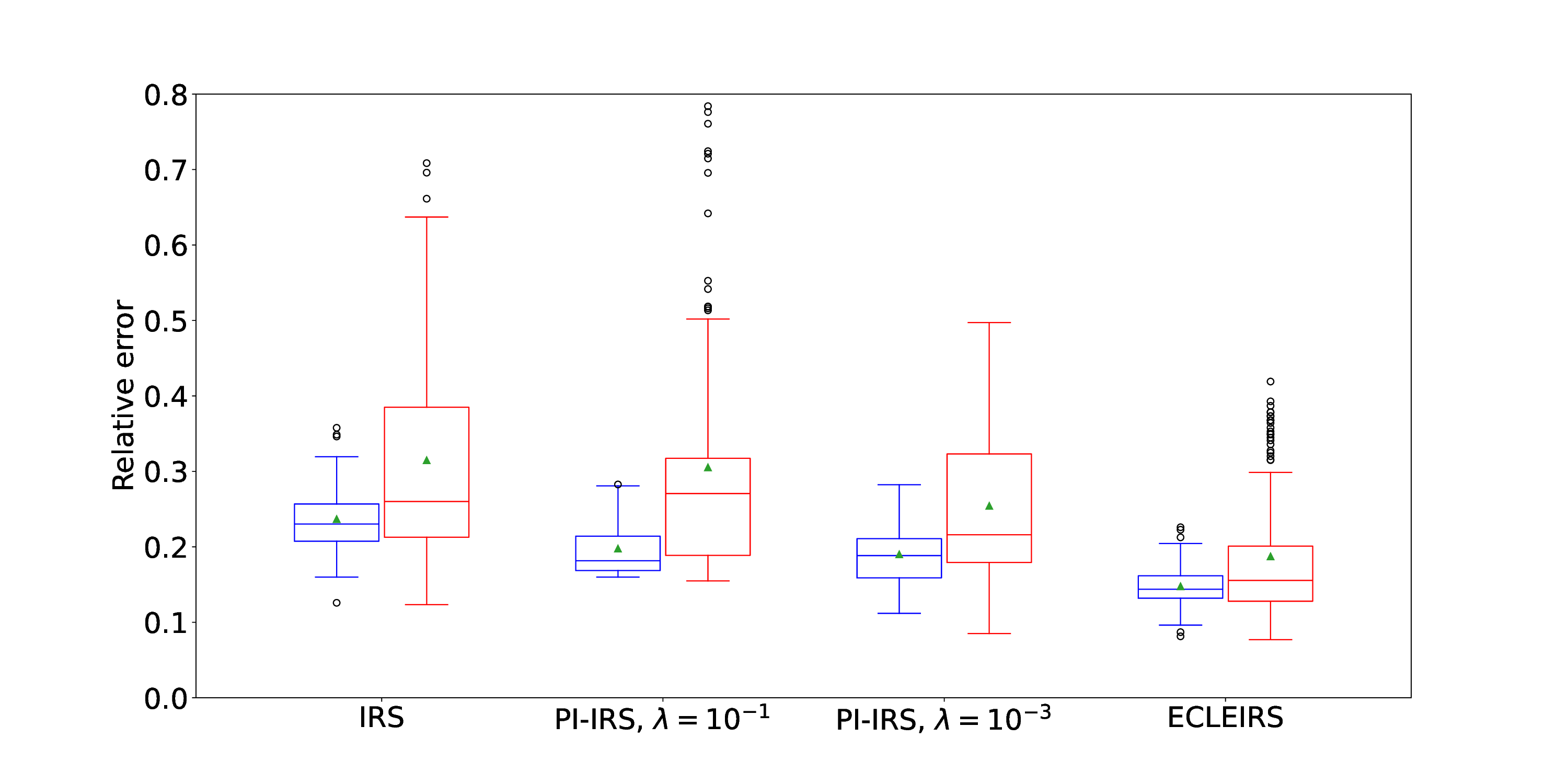}}
    \caption{1-D Burgers problem: Box plots for relative error (defined in \eref{rel_error_def}) comparing different approaches (a) $\sigma_N = 0.1$ and (b) $\sigma_N = 0.2$ for spatiotemporal sparsity of $20\%$. Blue box plots are for parameters in the training parameter space $\mathcal{D}^{\mu}$, while the red box plots are for parameters outside the training parameter space.}
    \label{fig:BoxplotModelCompare_burg}
\end{figure}

\begin{figure}[t]
    \centering
    \subfigure[\label{fig:BoxplotNoiseComp_burg_error}]{\includegraphics[width=0.49\linewidth, trim={0.5cm 0cm 5cm 2.0cm},clip]{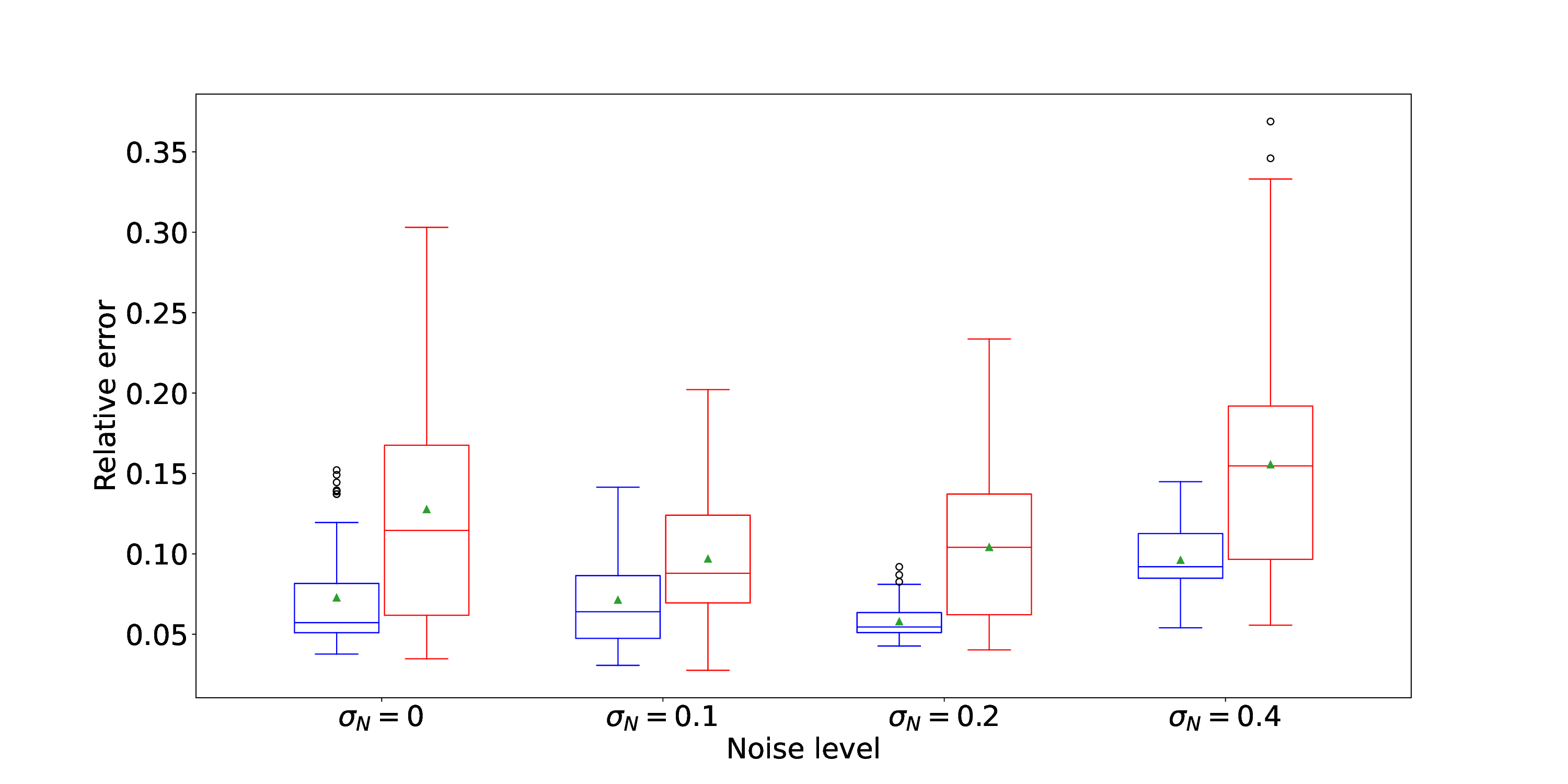}}
    \subfigure[\label{fig:BoxplotNoiseComp_burg_consverror}]{\includegraphics[width=0.49\linewidth, trim={0.5cm 0cm 5cm 2.0cm},clip]{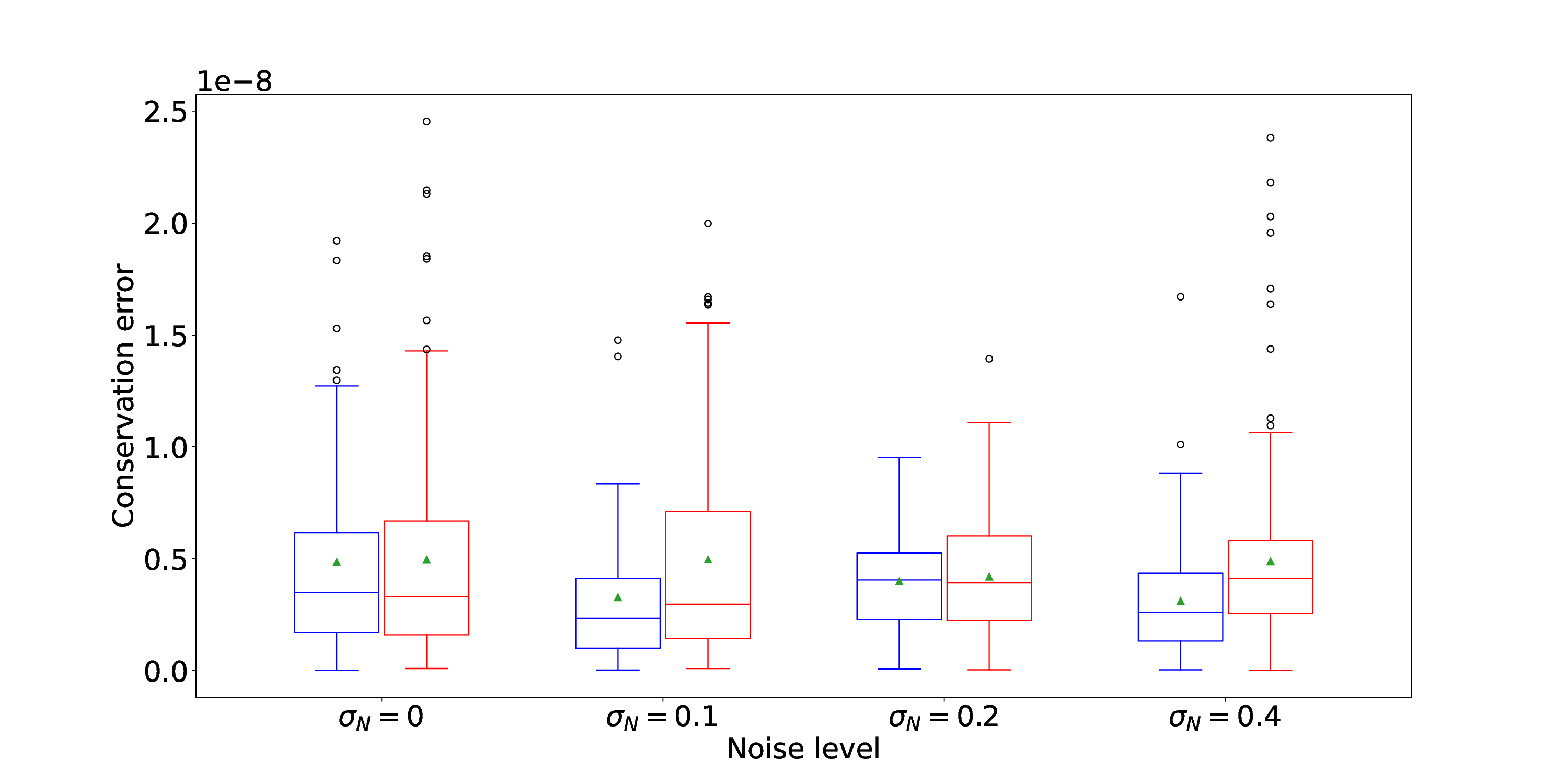}}
    \caption{1-D Burgers problem: Box plots of (a) relative error (defined in \eref{rel_error_def}) and (b) mean conservation error (defined in \eref{consv_error_burgers}) for ECLEIRS comparing different noise level for $40\%$ spatial sparsity and $20\%$ temporal sparse training data. Blue box plots are for parameters in the learning parameter space $\mathcal{D}^{\mu}$, while the red box plots are for parameters outside the learning parameter space.}
    \label{fig:BoxplotNoiseCompare_burg}
\end{figure}

We examine how the models trained on sparse and noisy data perform for predicting dynamics for parameters outside the learning dataset. A more comprehensive comparison of relative errors for different reduced state dynamics approaches is shown in \figref{BoxplotModelCompare_burg} for training data with $20\%$ spatiotemporal sparsity and different noise levels. The results show that IRS exhibits high mean errors with high standard deviation in interpolation and extrapolation parameters. PI-IRS with $\lambda = 10^{-1}$ also exhibit high mean errors for interpolation and extrapolation parameters. However, the standard deviation in errors is much lower for interpolation parameters. PI-IRS with $\lambda = 10^{-3}$ exhibits lower errors than IRS and PI-IRS with higher penalty parameters, while we see significant outliers for both noise levels. ECLEIRS appears to yield the best prediction for both noise levels while having the lowest mean errors and a small standard deviation of the error for interpolation parameters. At the low noise level, the results for extrapolation parameters are comparable for PI-IRS with $\lambda = 10^{-3}$ and ECLEIRS, as the latter exhibits a smaller mean but the former exhibits a small standard deviation. However, for large noise levels, ECLEIRS provide lower means and standard deviation of errors for both interpolation and extrapolation parameters. We observe a general trend that all reduced state dynamics models exhibit lower errors for parameters in the interpolation regime, while these errors increase for parameters in the extrapolation regime. Despite this trend, we observe that the degradation of errors in the extrapolation parameter regime is lowest for ECLEIRS, especially at higher noise levels. The variation of errors for ECLEIRS for different noise levels is compared in \figref{BoxplotNoiseCompare_burg}. Largely, the results indicate a slow increase in mean errors with noise level. Unlike the 1-D advection problem, there is no clear trend for this problem. The results also show that errors for interpolation parameters are lower compared to those for extrapolation parameters, which follows the trend observed in other tests in this article. Meanwhile, the conservation error for ECLEIRS is low and is invariant to the selection of the parameter space as both interpolation and extrapolation parameters yield similar errors. These errors also do not change with different sparsity or noise levels.

\subsection{2-D Euler equation} 

In the third numerical experiment, we consider parameterized 2-D Euler equations, which are expressed in the conservative form as
\begin{equation}
    \frac{\p \pmb{q}}{\p t} + \nabla \cdot \bm{f} (\bm{q}) = 0,
\end{equation}
where $\pmb{q} = [\rho, \; \rho \bm{u}, \; \rho E]^T$ and $\pmb{f} (\pmb{q}) = [\rho \pmb{u},\;  \rho \bm{u} \otimes \bm{u} + p \bm{I},\; (E + p) \bm{u}]$ are the solution and flux vectors respectively. We consider an ideal gas equation of state with $\gamma =1.4$, and a 2-D formulation related to the classical Sod Shock tube over time $t\in[0,2]$. Specifically, define the domain $\Omega\coloneqq [0,1]\times[0,1]$ and material interface at corner $\bm{\mu} = [x^0_1, x^0_2]$. We then partition our domain $\Omega = \Omega_1\cup\Omega_2$, where $\Omega_1 \coloneqq \{ \pmb{x}\textnormal{ : } x_1\leq x_1^0\text{ and } x_2\leq x_2^0\}$ and $\Omega_2 \coloneqq \{ \pmb{x}\textnormal{ : } x_1> x_1^0\text{ or } x_2> x_2^0\} = \Omega\backslash\Omega_1$. We consider zero initial velocity, while density and pressure values of $(\rho, p)\coloneqq (0.125,0.1)$ for $\pmb{x}\in\Omega_1$ and $(\rho, p)\coloneqq (1,1)$ for $\pmb{x}\in\Omega_2$ were chosen. We impose zero-flux boundary conditions via zero velocity in the normal component and reflected boundary conditions for density and energy; along with the initial pressure-density discontinuity, this results in the formation and evolution of complex interacting shocks. The data is generated using a finite volume discretization with Rusanov flux function and $150$ grid points in each spatial direction, and 3rd-order SSP Runge-Kutta time integration with timestep chosen to be $0.5$ CFL. The solution at every other timestep is stored to be used for training reduced state dynamics models. In our learned models, we enforce the continuity equation
\begin{equation}
    \frac{\p \rho}{\p t} + \nabla \cdot \rho \pmb{u} = 0
\end{equation}
using ECLEIRS and PI-IRS. Other equations can also be enforced similarly. However, we restrict to enforcing only the continuity equation due to complexity in implementation and lack of consistency, as discussed in Section \ref{sec:ECELIRSFormulation}.

\begin{figure}
    \centering
    \includegraphics[width=0.4
    \linewidth]{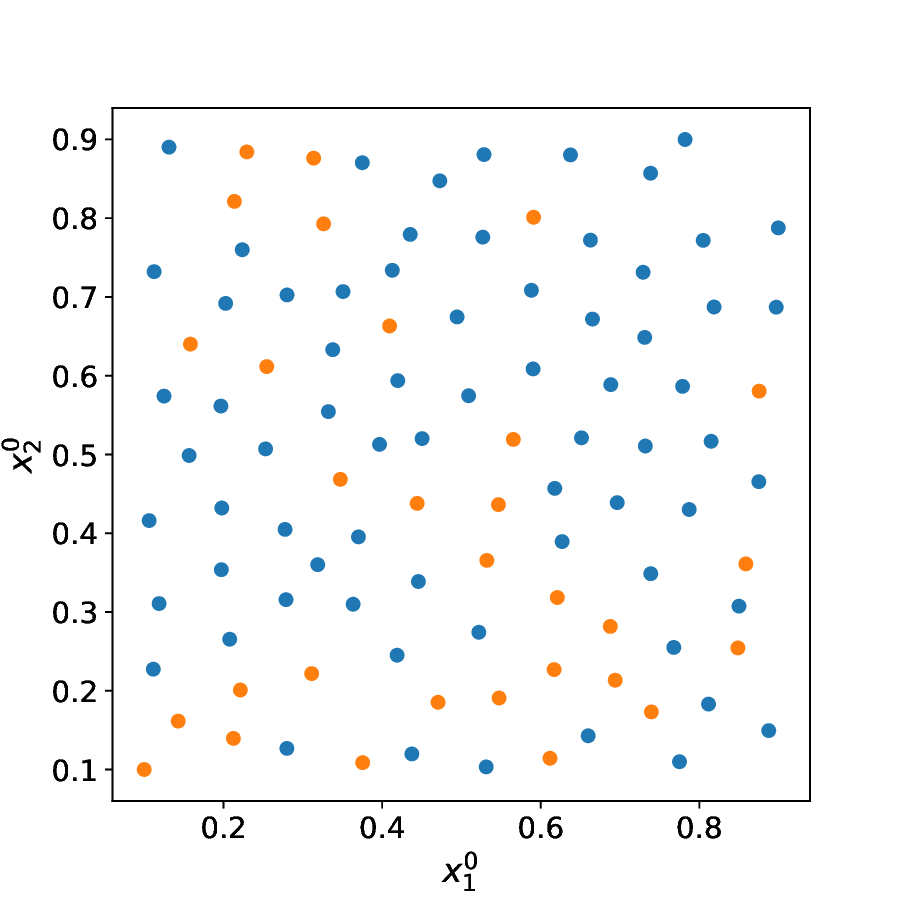}
    \caption{The selection of parameter points used in the learning dataset (blue markers) and validation dataset (orange markers). }
    \label{fig:Euler2D_paramsel}
\end{figure}

The dataset for the parameterized 2-D Euler problem is generated by running 100 simulations with initial density and pressure fields specified by the material discontinuity at $\pmb{\mu}$. The parameter vector $\pmb{\mu}$ is sampled quasi-randomly over $[0.1,0.9]\times[0.1,0.9]$ using a 2-D Halton sequence to facilitate a reasonable covering of the domain without grid imprinting that would likely arise from tensor sampling in $x_1 \times x_2$. Within this dataset, $70\%$ of the data is randomly chosen as the learning dataset to split between training and testing sets and ensure no overfitting during the training process. The data corresponds to the rest $30\%$ of the parameters used as the validation dataset. The distribution of learning and validation parameters can be visualized in \figref{Euler2D_paramsel}. 

\begin{figure}[t]
    \centering
    \subfigure[\label{fig:STnRatio_0p01_Euler}]{\includegraphics[width=0.32\textwidth, trim={0.2cm 0cm 1.0cm 0.5cm},clip]{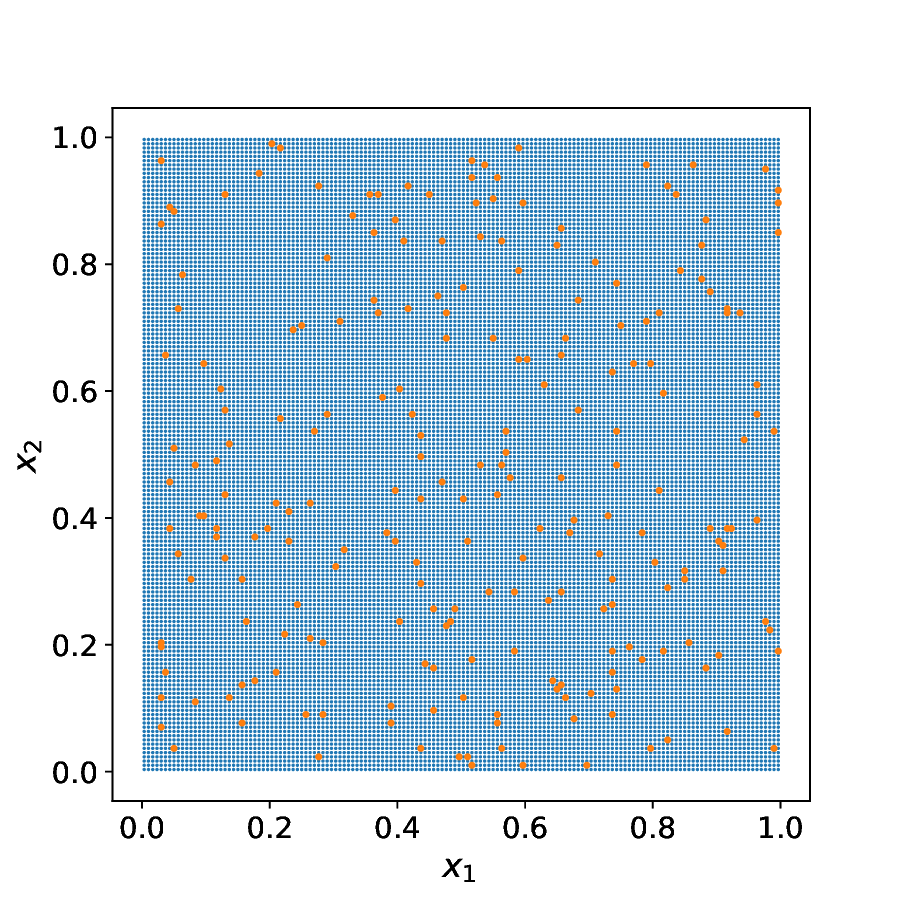}}
    \subfigure[\label{fig:STnRatio_0p05_Euler}]{\includegraphics[width=0.32\textwidth, trim={0.2cm 0cm 1.0cm 0.5cm},clip]{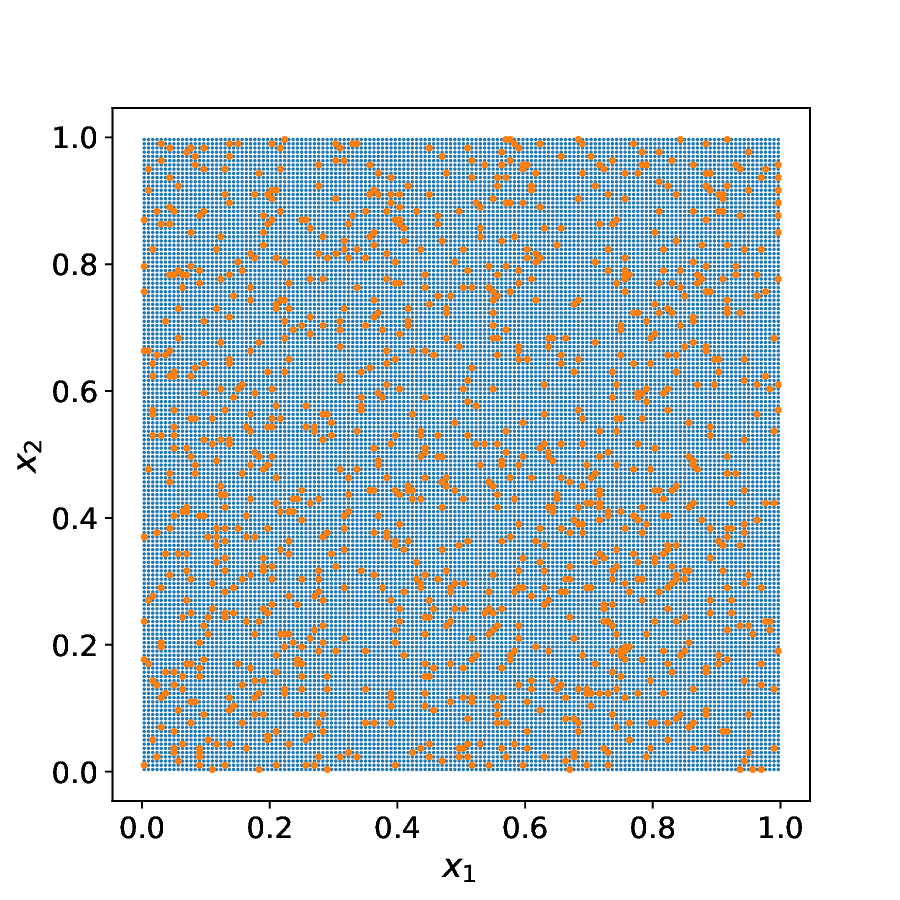}}
    \subfigure[\label{fig:STnRatio_0p2_Euler}]{\includegraphics[width=0.32\textwidth, trim={0.2cm 0cm 1.0cm 0.5cm},clip]{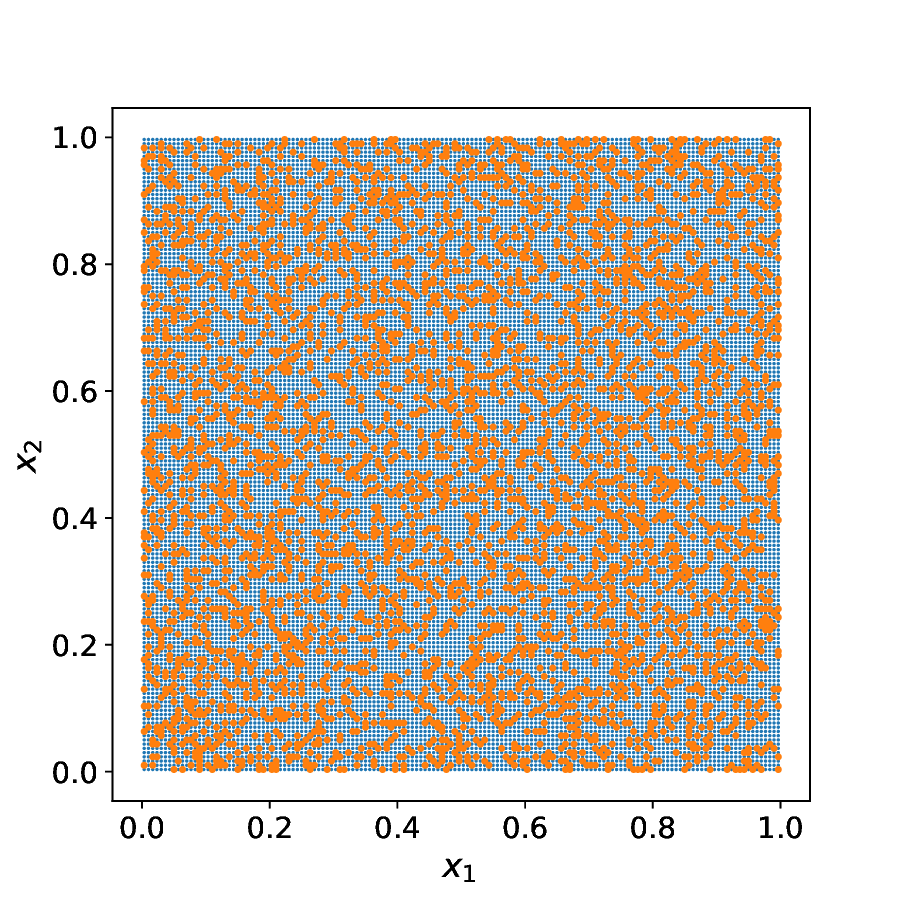}}
    \vspace{-3mm}
    \caption{2-D Euler problem: Spatial locations of data used to learn different reduced state dynamics models for (a) $1\%$, (b) $5\%$ and (e) $20\%$ spatial sparsity. The blue markers denote the spatial resolution of high-fidelity simulation and the orange marker denote the sparse data locations used for learning the model. Note that marker size for sparse points is 5 times greater than the marker size of high-fidelity spatial locations, thereby the selected data is sparser than how it appears here.}
    \label{fig:STnRatio_Euler}
\end{figure}

\subsubsection{Denoising and sparse reconstruction capability on the learning dataset}

For this test case, we have high-spatial resolution data available comprising of $150 \times 150$ spatial grid points and $\approx 400$ temporal grid points. Learning a reduced state dynamics model using such large amounts of data demands huge computational memory and processing power. As often such large amounts of computational resources may not be available, it is valuable to use sparsely sampled spatial and temporal data and identify if the learned reduced state dynamics can accurately predict solutions at full spatial-temporal resolution locations. Some examples of sparse spatial locations used in the learning dataset are shown in \figref{STnRatio_Euler}. 

\begin{figure}[t]
    \centering
    \includegraphics[width=\linewidth]{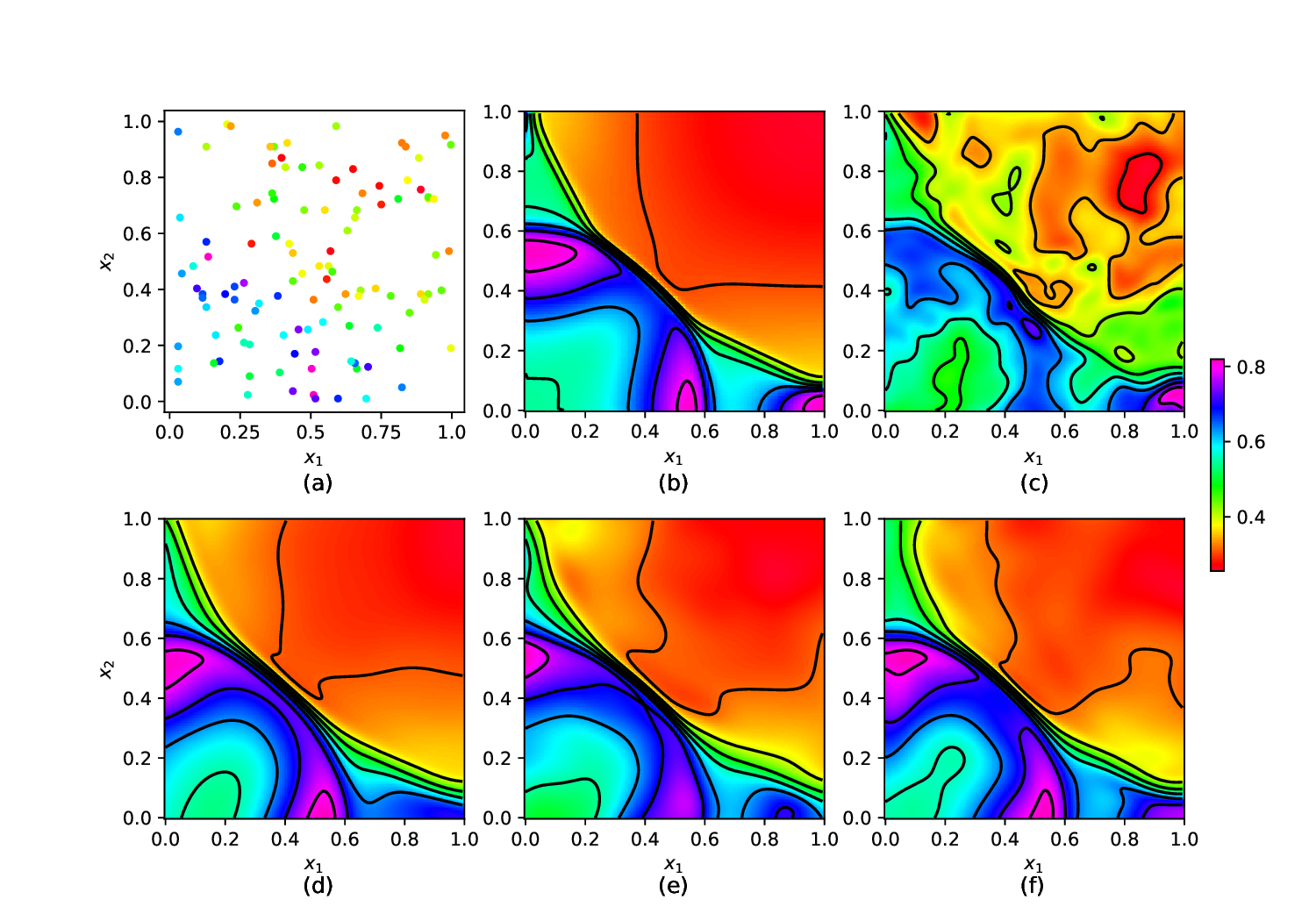}
    \caption{2-D Euler problem: Density at (a) training data based on $0.5 \%$ spatial sparsity, $20\%$ temporal sparsity and added noise of $\sigma_N = 0.1$ extracted from (b) ground truth full resolution solution data without noise. The learned solution representation evaluated at the same spatial locations as groudn truth data for (c) IRS, (d) PI-IRS with $\lambda = 10^{-1}$, (e) PI-IRS with $\lambda = 10^{-3}$ and (f) ECLEIRS. Note that marker size in (a) for sparse points is significantly greater for visibility, thereby the selected data is sparser than how it appears here.}
    \label{fig:Euler2D_sol}
\end{figure}

The ability of different reduced state dynamics approaches for obtaining a clean solution signal from sparse and noisy datasets is shown in \figref{Euler2D_sol}. We observe that the clean solution representation obtained using IRS exhibits spurious structures. On the contrary, both PI-IRS and ECLEIRS provide a much more accurate clean solution representation. There are no significant visible differences between these methods, although PI-IRS with $\lambda = 10^{-3}$ appears to underpredict the magnitude of the shock front. These results highlight the importance of enforcing physical conservation law as a constraint in the inverse problem of identifying clean solutions from corrupted data. We also observe that the addition of physical constraint becomes more important for this 2-D problem and gives accurate results for even sparser spatial data than the other two numerical experiments. 

\begin{figure}[t]
    \centering
    \subfigure[\label{fig:BoxplotPenalty_Euler2D_sigma0p4}]{\includegraphics[width=0.49\linewidth, trim={0.5cm 0cm 5cm 2.5cm},clip]{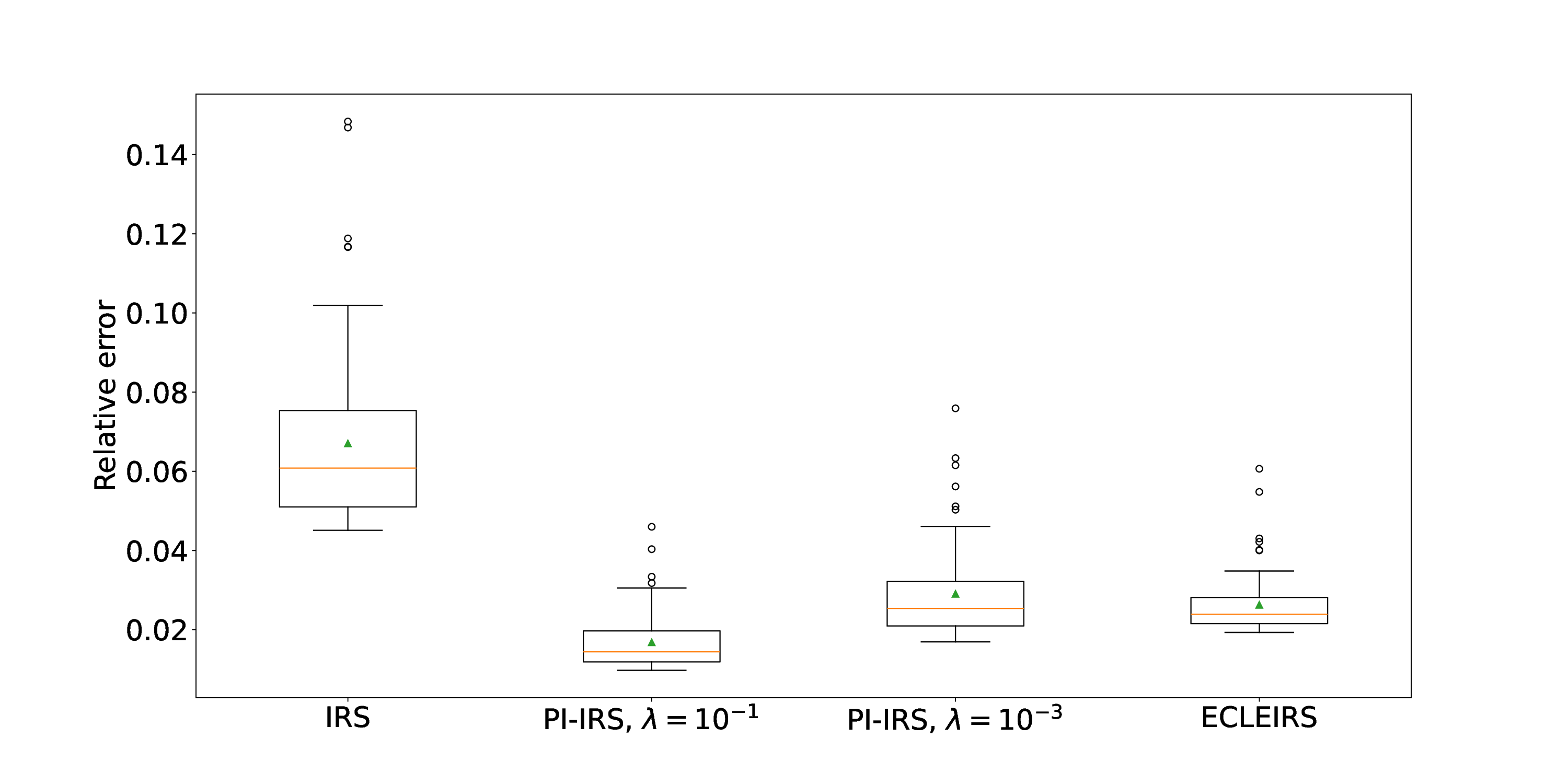}}
\subfigure[\label{fig:BoxplotPenalty_Euler2D_sigma0}]{\includegraphics[width=0.49\linewidth, trim={0.5cm 0cm 5cm 2.5cm},clip]{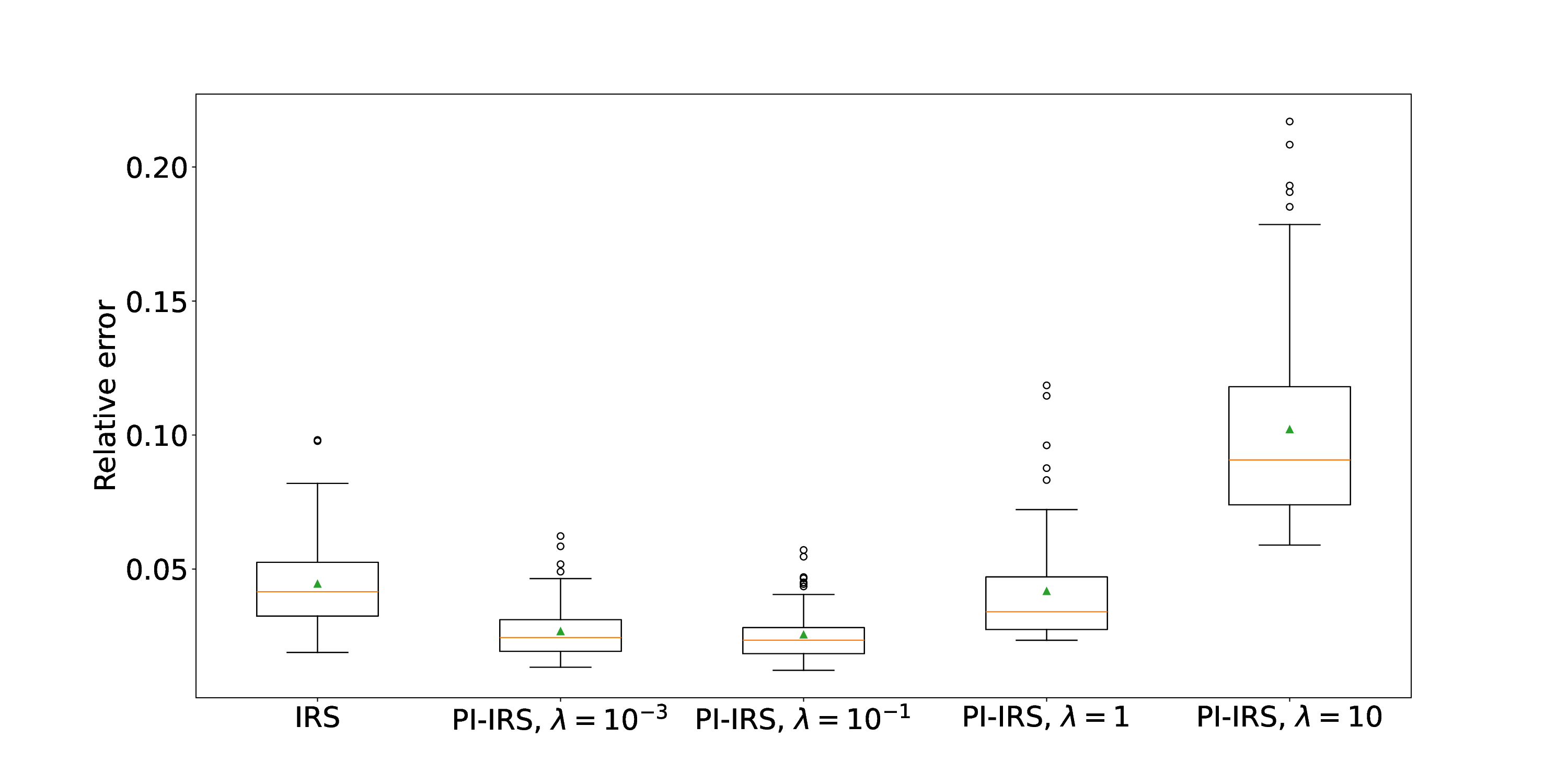}}
    \caption{2-D Euler problem: (a) Box plots of relative error (defined in \eref{rel_error_def}) for training data with $1\%$ spatial sparsity, $40\%$ temporal sparsity and added noise with $\sigma_N = 0.1$ in the learning dataset. (b) Box plots of relative error (defined in \eref{rel_error_def}) comparing PI-IRS at different regularization parameters for $1\%$ spatial sparsity, $1\%$ temporal sparsity and added noise with $\sigma_N = 0.1$ in the learning dataset.}
\label{fig:BoxplotPenaltyCompare_Euler2D}
\end{figure}

We quantify the reconstructive capability of these different reduced state dynamics approaches by assessing the performance for different noise levels, and spatiotemporal sparsity. The performance for different reduced state dynamics approaches is shown in \figref{BoxplotPenaltyCompare_Euler2D}. The results indicate IRS exhibits the highest errors in representing clean solution signals, whereas other approaches, PI-IRS and ECLEIRS perform much better. These results highlight the importance of adding physical constraints in identifying clean solution representation from sparse and noisy data. While PI-IRS with $\lambda = 10^{-1}$ yields the most accurate results, these results are subject to the penalty parameter. The comparison of the effect of penalty parameters on the capability of PI-IRS is illustrated in \figref{BoxplotPenalty_Euler2D_sigma0}. These results indicate that while $\lambda = 10^{-1}$ results in the most accurate results, having a stricter penalty enforcement of the conservation by setting a large value of $\lambda$ can significantly deteriorate the results. In fact, a lower value of $\lambda$ or not adding physical constraint performs better than having a large value of $\lambda$. Conversely, ECLEIRS appears to provide similar levels of accuracy of PI-IRS with the optimal $\lambda$, while not requiring any additional numerical experiments for identifying optimal parameters.

\begin{figure}[t]
    \centering
    \subfigure[\label{fig:Boxplot_2DEuler_SigmaComp}]{\includegraphics[width=0.49\linewidth, trim={0.5cm 0cm 5cm 2.5cm},clip]{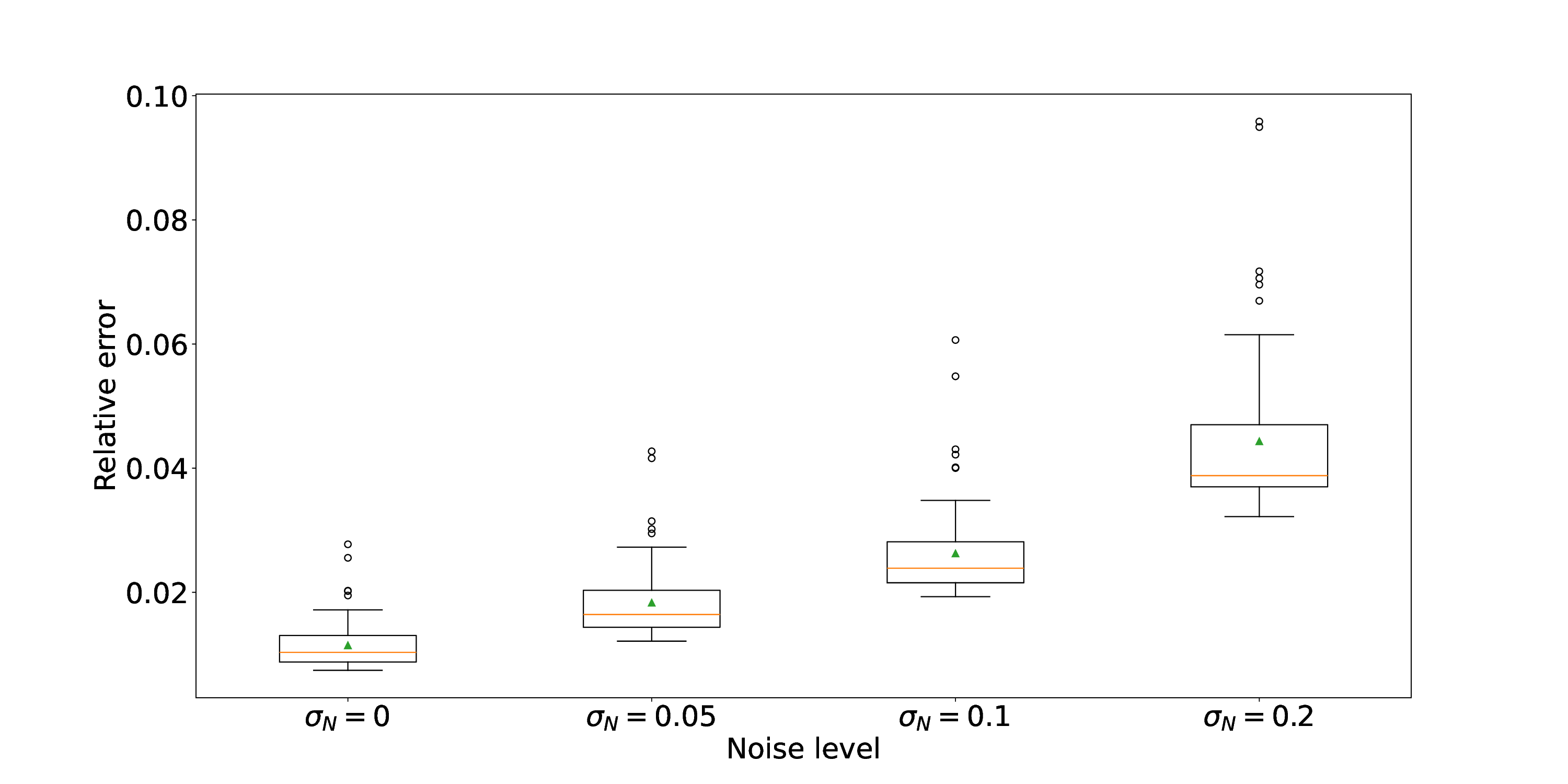}}
\subfigure[\label{fig:Boxplot_2DEuler_RhoxComp}]{\includegraphics[width=0.49\linewidth, trim={0.5cm 0cm 5cm 2.5cm},clip]{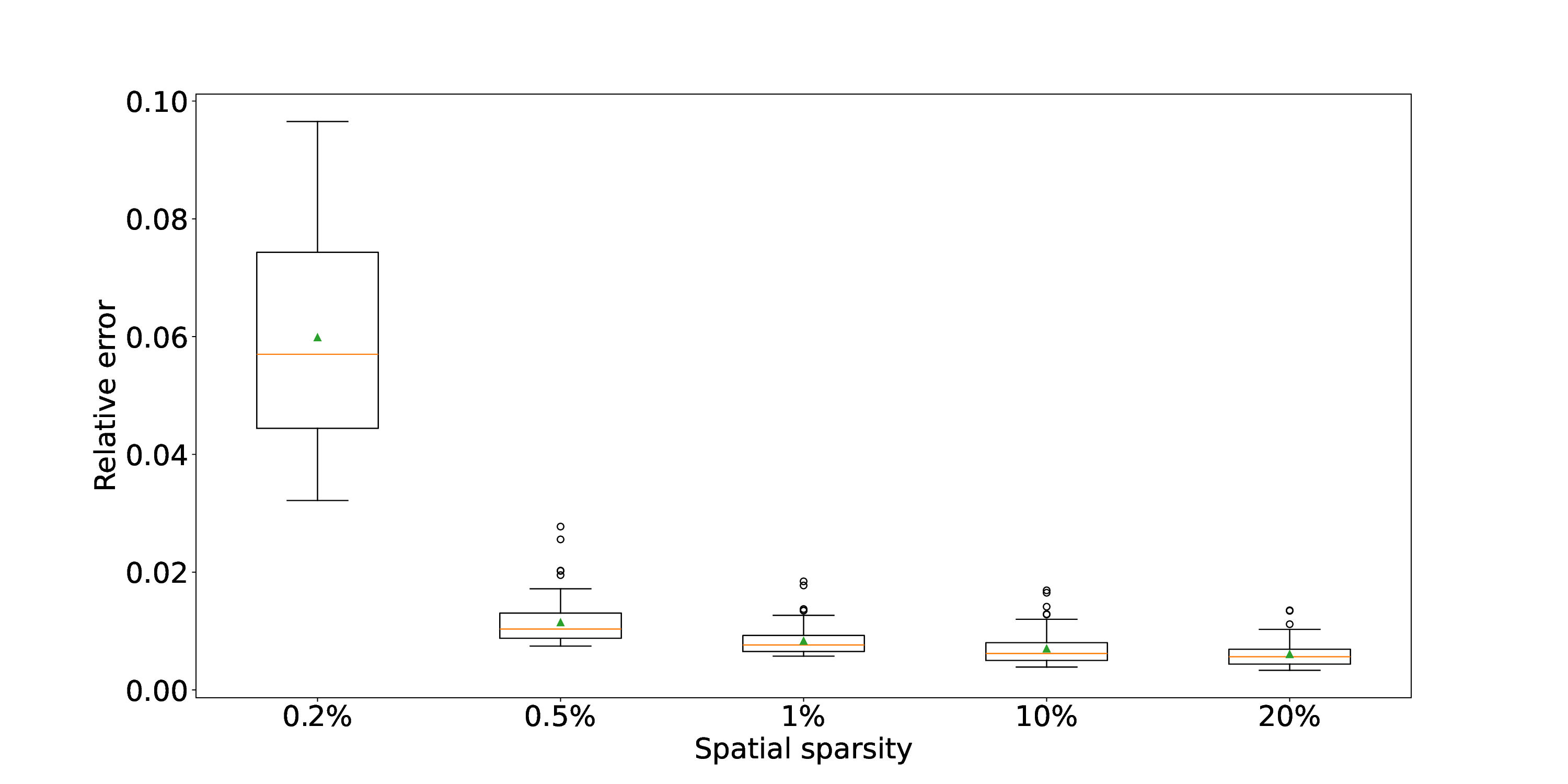}}
    \caption{2-D Euler problem: (a) Box plots for relative error (defined in \eref{rel_error_def}) for ECLEIRS trained on data with $0.5\%$ spatial sparsity and $20\%$ temporal sparsity, and different noise levels in the learning dataset. (b) Box plots of relative error (defined in \eref{rel_error_def}) for ECLEIRS trained on data with different different spatial sparsity, while keeping $20\%$ temporal sparsity and no noise in the learning dataset.}
    \label{fig:Boxplot_Euler2D_Reconstr_ECLEIRS}
\end{figure}

We assess the robustness of ECLEIRS in representing clean solution signals by comparing the models trained over different levels of sparsity and noise data. These results are presented in \figref{Boxplot_Euler2D_Reconstr_ECLEIRS}. The results indicate that ECLEIRS is robust to different levels of noise in the data. The learned solution representation is highly accurate at low levels of noise. While the accuracy decreases with increasing noise levels, even a high noise level of $\sigma_N = 0.2$ results in less than $5\%$ error in prediction. We also assess the performance of ECLEIRS for varying levels of spatial sparsity. The results indicate that the error in representing the clean solution does not grow significantly even for a spatial sparsity of $0.5\%$. Having even more sparse data leads to a rapid increase in error, however, the mean error stays about $5\%$ even for a spatial sparsity of $0.2\%$. Similar behavior was also observed for increasing temporal sparsity, however a detailed discussion is excluded for brevity.

\subsubsection{Dynamics prediction for the validation dataset}

In this section, we evaluate the performance of reduced state dynamics approaches for parameters in the validation dataset depicted in \figref{Euler2D_paramsel}. For these tests, the reduced state dynamics models are learned using sparse and noisy data to assess the ability of these approaches in scenarios with limited availability of good-quality data.

\begin{figure}[t]
    \centering
    \includegraphics[width=
    \linewidth, trim={0.5cm 3.85cm 0cm 5.0cm},clip]{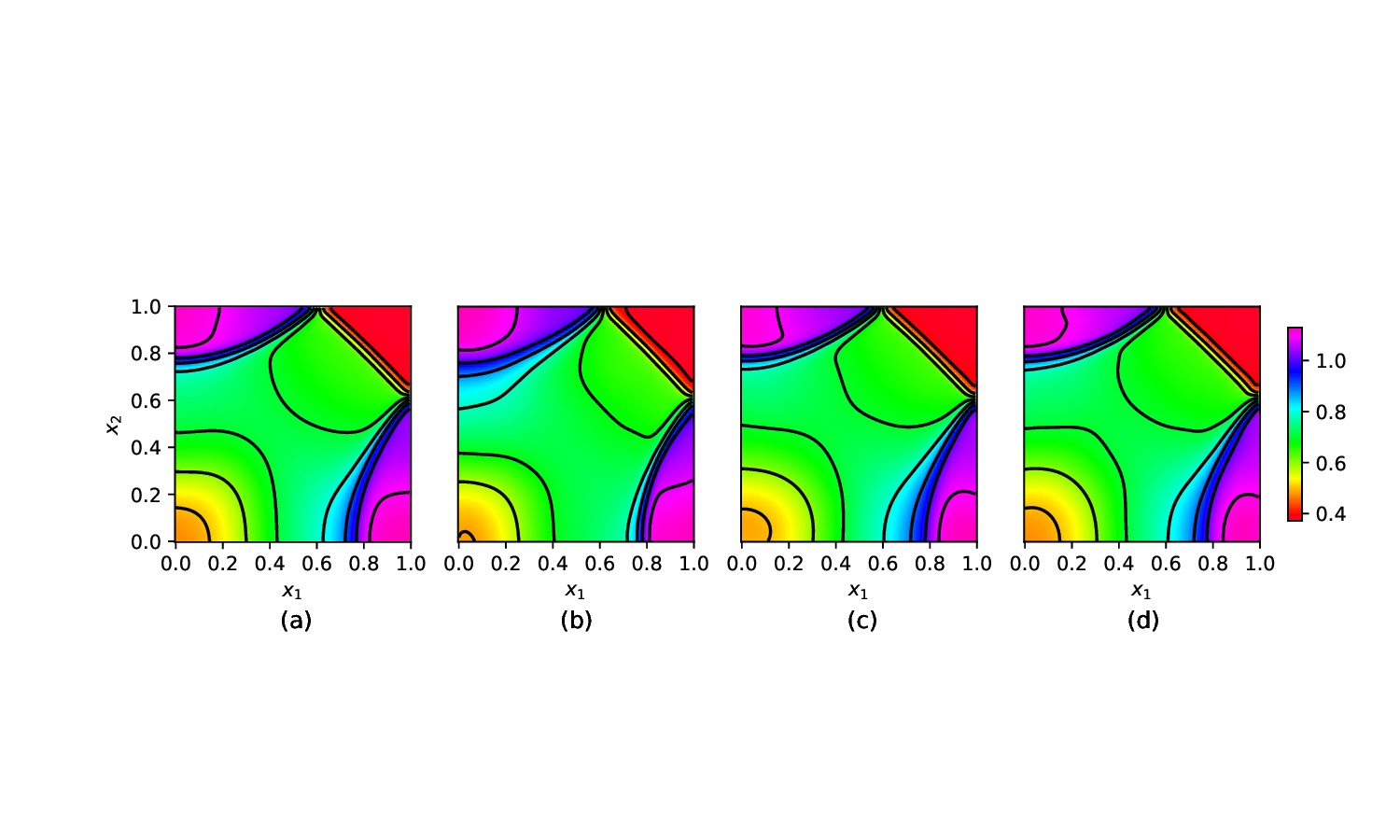}
    \caption{2-D Euler problem: Predicted density at $\pmb{\mu} = [0.34, \; 0.63]$ (in the validation dataset) for (a) ground truth, (b) IRS, (c) PI-IRS with $\lambda = 10^{-1}$ and (d) ECLEIRS. These reduced state dynamics models were trained using data from learning dataset with $20\%$ spatial sparsity, $20\%$ temporal sparsity and no added noise. }
    \label{fig:Euler2D_SolComp_lowNoise}
\end{figure}

\begin{figure}[t]
    \centering
    \includegraphics[width=
    \linewidth, trim={0.5cm 3.7cm 0cm 5.0cm},clip]{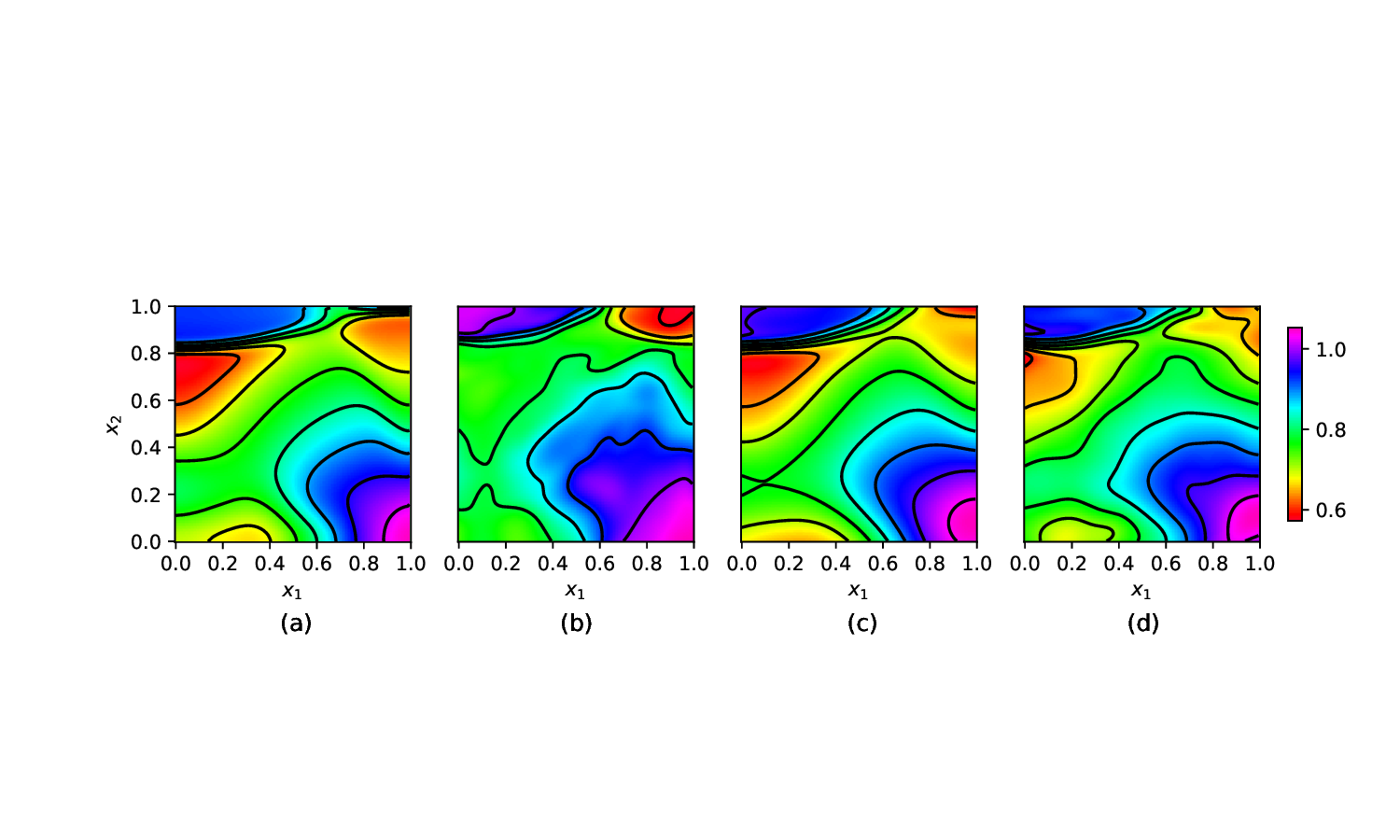}
    \caption{2-D Euler problem: Predicted density at $\pmb{\mu} = [0.527, \; 0.775]$ (in the validation dataset) for (a) ground truth, (b) IRS, (c) PI-IRS with $\lambda = 10^{-1}$ and (d) ECLEIRS. These reduced state dynamics models were trained using data from learning dataset with $1\%$ spatial sparsity, $40\%$ temporal sparsity and added noise with $\sigma_N = 0.1$. }
    \label{fig:Euler2D_SolComp}
\end{figure}

\begin{figure}[t]
    \centering
    \subfigure[\label{fig:Boxplot_2DEuler_ModelComp_testing_1}]{\includegraphics[width=0.49\linewidth, trim={0.5cm 0cm 5cm 2.5cm},clip]{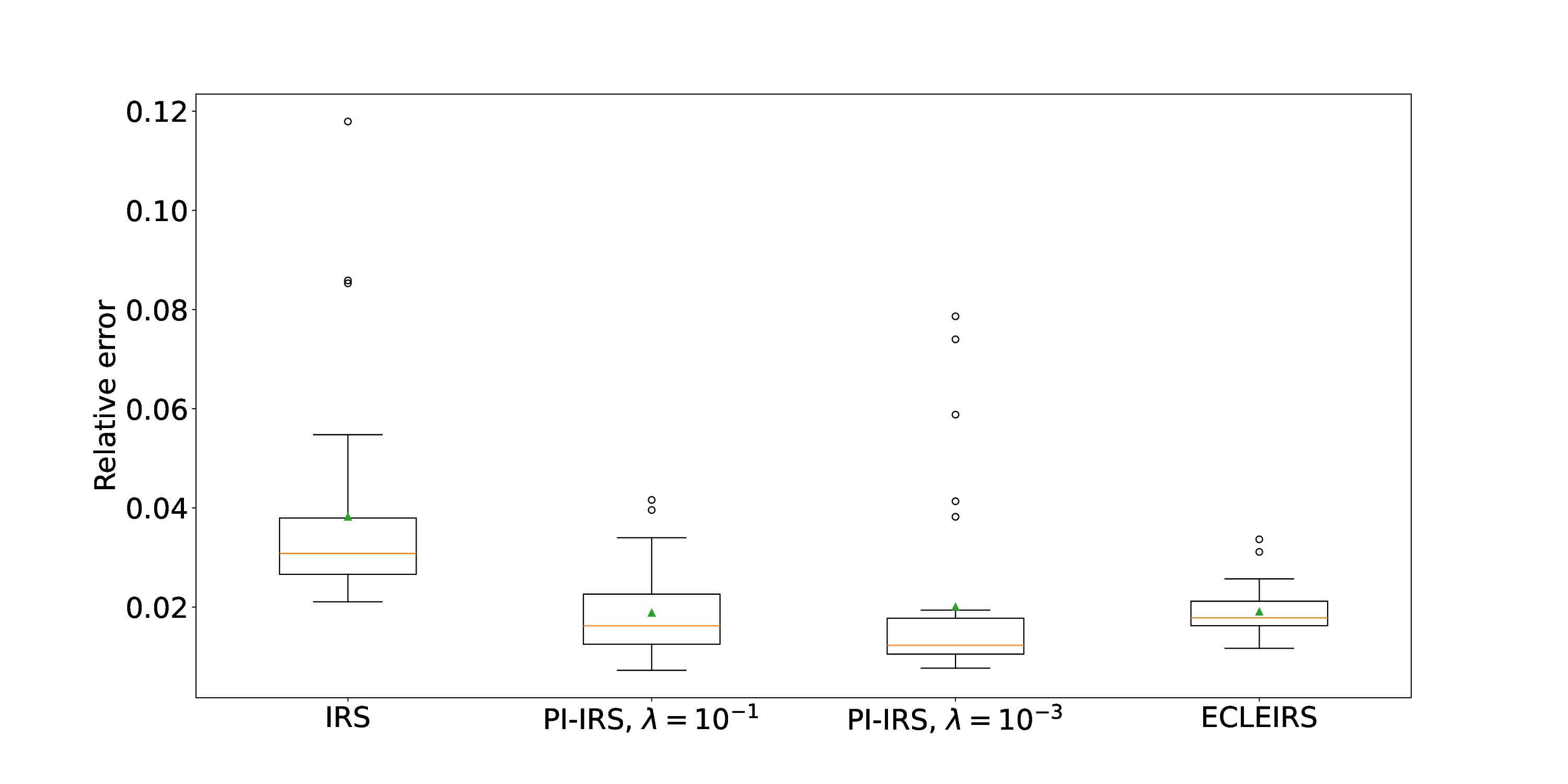}}
    \subfigure[\label{fig:Boxplot_2DEuler_RhoxComp_test}]{\includegraphics[width=0.49\linewidth, trim={0.5cm 0cm 5cm 2.5cm},clip]{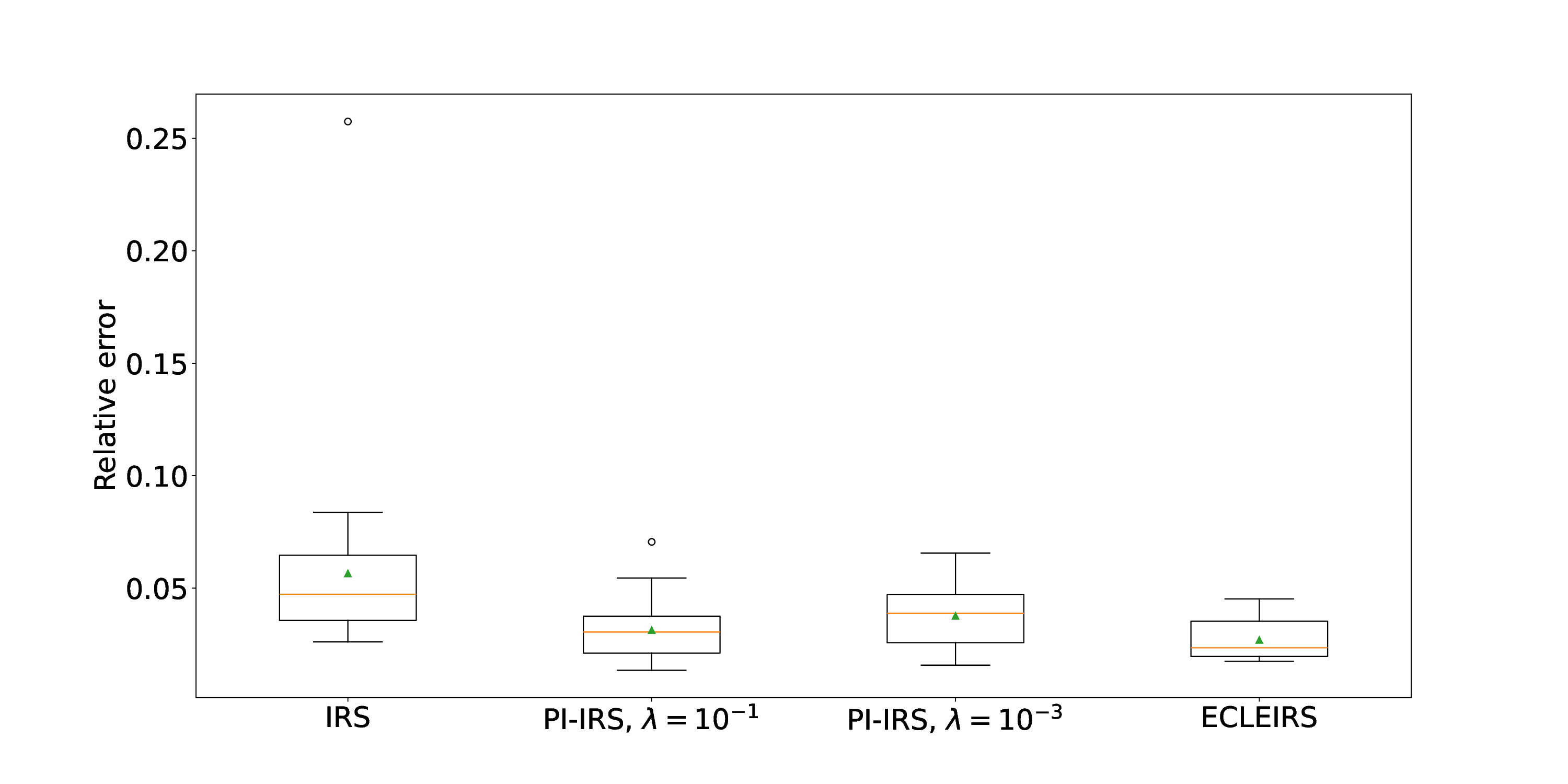}}
    \caption{2-D Euler problem: Box plots comparing relative error (defined in \eref{rel_error_def}) for the validation dataset for different reduced state dynamics approaches learned on data with (a) $40\%$ and (b) $20\%$ temporal sparsity at $1\%$ spatial sparsity and added noise with $\sigma_N = 0.1$. }
    \label{fig:Boxplot_Euler2D_testing_modelcomp}
\end{figure}

Contour plots depicting the density predicted by different reduced state dynamics approaches are shown in \figref{Euler2D_SolComp_lowNoise}. These results are for models trained using moderately dense data with $20\%$ spatial and temporal sparsity without any added noise. The results show that all models agree well with the ground truth predictions. Therefore, when large amounts of clean data are available, IRS gives similar accuracy as PI-IRS and ECLEIRS. There is some disagreement in density prediction near the shocks at the upper and bottom faces between IRS and reference results, but overall predictions qualitatively appear good. However, this behavior does not hold when much sparser and noisy data is available for learning these reduced state dynamics approaches as shown in \figref{Euler2D_SolComp}. The results indicate that IRS exhibits large oscillations and incorrect prediction of the shock front. On the contrary, both PI-IRS with $\lambda = 10^{-1}$ and ECLEIRS accurately predict density fields, with the former providing slightly more accurate results. The relative error in this prediction is quantified for different parameters in \figref{Boxplot_Euler2D_testing_modelcomp}. These results echo the previous discussion for other test cases where IRS was the least predictive model with higher errors compared to PI-IRS and ECLEIRS. We observe that while PI-IRS with $\lambda = 10^{-1}$ provided the best results for the identification of clean density fields from the sparse and noisy, PI-IRS with $\lambda = 10^{-3}$ provides more accurate results for unseen parameters. Note that these tests use the same sparse and noisy data with $1\%$ spatial sparsity, $40\%$ temporal sparsity and added noise with $\sigma_N = 0.1$. This trend reverses when $20\%$ temporal sparse data is used. These results indicate that the selection of optimal $\lambda$ for PI-IRS is also dependent on the application scenario of either clean solution identification or dynamics prediction. For $40\%$ temporal sparsity, ECLEIRS performs as well as PI-IRS with similar mean errors, but lower standard deviation and outliers in the predicted dynamics. ECLEIRS provides the best result for the scenario with $20\%$ temporal sparsity with lower mean and standard deviation of errors compared to other methods. Therefore, ECLEIRS stands out as a more robust approach that provides significantly low errors in both clean solution identification and dynamics prediction for unseen parameters, while not being dependent on a tunable parameter. 

\begin{figure}
    \centering
    \subfigure[\label{fig:Boxplot_2DEuler_ModelComp_testing}]{\includegraphics[width=0.49\linewidth, trim={0.5cm 0cm 5cm 2.5cm},clip]{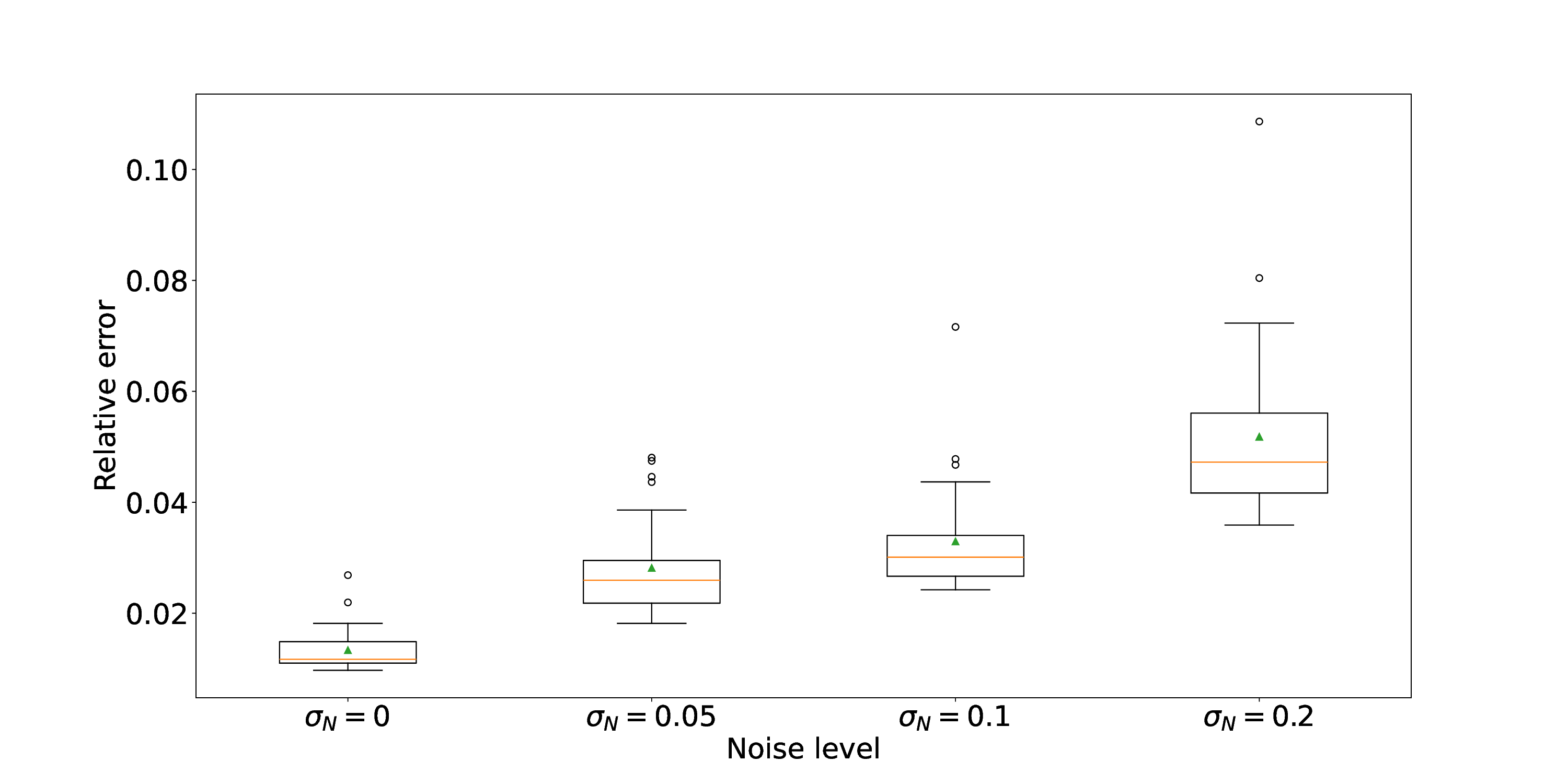}}
\subfigure[\label{fig:Boxplot_2DEuler_RhoxComp_2}]{\includegraphics[width=0.49\linewidth, trim={0.5cm 0cm 5cm 2.5cm},clip]{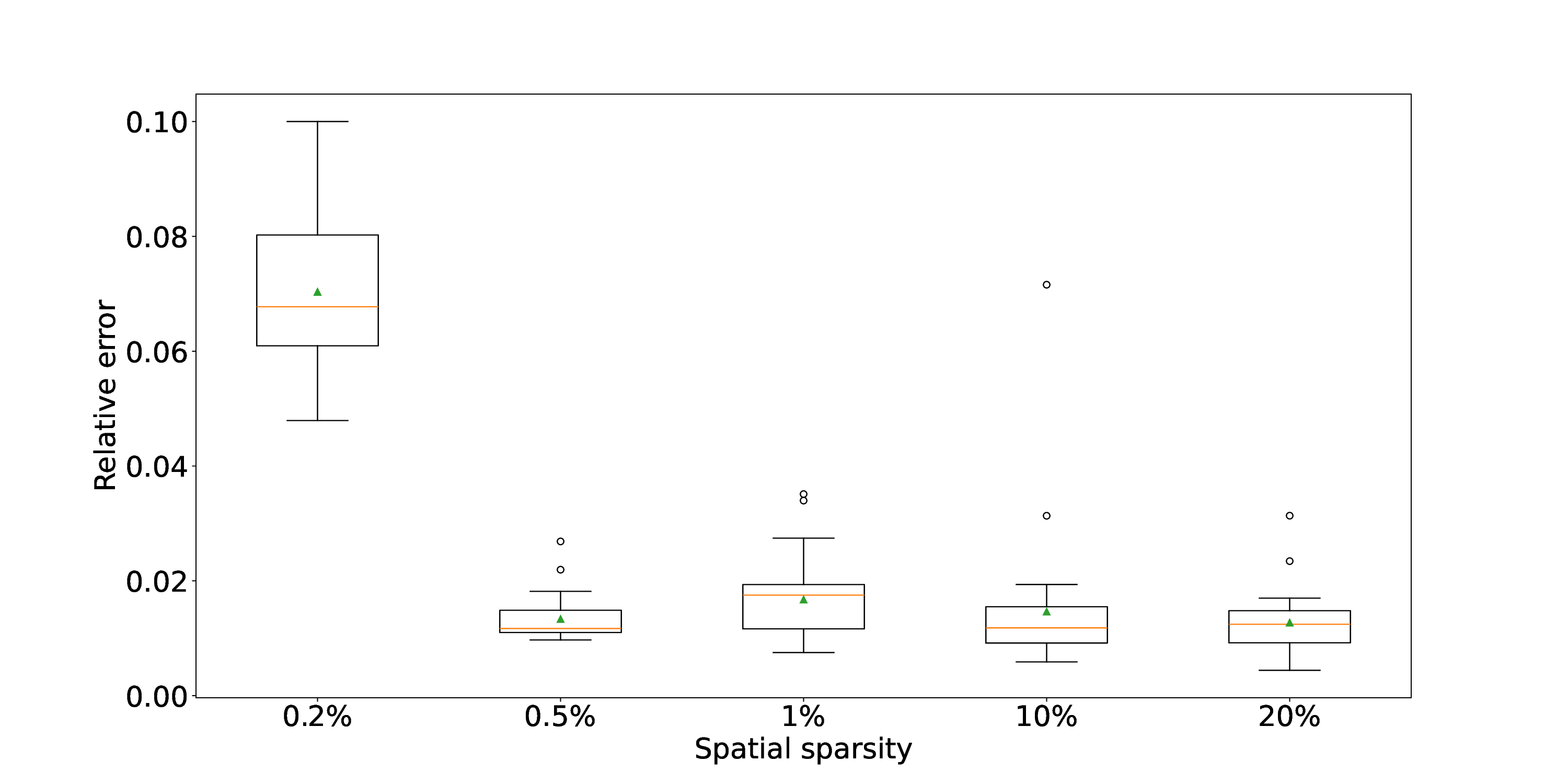}}
    \caption{2-D Euler problem: (a) Box plots of relative error (defined in \eref{rel_error_def}) in the validation dataset for ECLEIRS learned using data with $0.5\%$ spatial sparsity, $20\%$ temporal sparsity and various noise levels. (b) Box plots of relative error (defined in \eref{rel_error_def}) in the validation dataset for ECLEIRS learned using data with $20\%$ temporal sparsity, no added noise and different spatial sparsity levels.}
    \label{fig:Boxplot_Euler2D_testing_Ecleirs}
\end{figure}

We assess the robustness of ECLEIRS by comparing the performance of this method for predicting dynamics for unseen parameters when the model is trained with different levels of corrupted data. The results for the performance of ECLEIRS trained on data for different noise and sparsity levels are shown in \figref{Boxplot_Euler2D_testing_Ecleirs}. The results indicate low errors when the learning dataset is less noisy. This error increases gradually with an increase in the noise level, but the mean errors remain around $5\%$ even for a high noise level of $\sigma_N = 0.2$. Similar trends were also observed at other sparsity levels. The errors do not increase significantly until a spatial sparsity of $0.5\%$, while even sparser data leads to a more significant increase in error levels. Even this higher error at $0.2\%$ sparsity exhibits a mean of $7\%$ error, which may still be practical for certain applications. This trend remains similar at higher noise levels, although the error is shifted slightly upwards to account for the effect of higher noise in the data. 

These results indicate that ECLEIRS is an accurate and robust method for identifying clean solution representation and predicting dynamics for scenarios with sparse and noisy data. An additional benefit of ECLEIRS is the ability to satisfy the conservation laws even for parameters and time instances unseen during model training, which provides more confidence in results for these unseen scenarios. While we have excluded the comparison of conservation satisfaction between different methods for the 2-D Euler problem for brevity, similar results that demonstrate this property are shown for 1-D advection and 1-D Burgers problems. 

\section{Conclusions}
\label{sec:Conclusions}

Identifying reduced state dynamics from data for parameterized PDE problems is important for several multi-query applications, such as parameter estimation, design optimization and uncertainty quantification, where high-fidelity simulations based on common numerical methods are expensive and unaffordable. These reduced state dynamics approaches rely on vast amounts of grid-based data generated by PDE solvers. Several practical problems involve scenarios with sparse spatial placement of sensors, lower collection frequency from these sensors and unavailability of storage for high-resolution data. In such scenarios, only spatiotemporal sparse measurements may be available that could further be corrupted with some noise. Therefore, novel reduced state dynamics approaches that are robust to incomplete and corrupted data environments are important.

In this article, we present a reduced state dynamics approach, which we refer to as ECLEIRS, that is robust to sparse and noisy measurements. ECLEIRS leverages the space-time divergence-free formulation to embed the model form in an autodecoder-based INR architecture while enabling the prediction of solution and fluxes that satisfy conservation laws exactly. This feature of the model makes it suitable for two important applications: identifying a clean solution signal and estimating dynamics at unseen parameters when only sparse and noisy measurements at other parameters are available. The performance of ECLEIRS is assessed for these two applications by considering three parameterized PDE shock-propagation problems: 1) 1-D advection equation, 2) 1-D Burgers equation and 3) 2-D Euler equation. This assessment involved comparison with other reduced state dynamics approaches that either do not enforce any physical constraints or those that enforce physical constraints weakly using a penalty conservation loss term in the optimization problem. The results indicate that ECLEIRS gives the overall best performance by consistently providing the most accurate results while also not requiring any parameter tuning, which is needed for physics-informed formulations. Furthermore, the solution and fluxes predicted by ECLEIRS satisfy the underlying conservation law to machine precision even for parameters unseen during the model learning phase. The ability to ensure exact conservation and accurately predict dynamics makes ECLEIRS a reliable and robust approach for parameterized PDE problems.

While ECLEIRS is robust and reliable compared to other reduced state dynamics approaches, additional studies with more complex physics must be performed to identify the limitations of this reduced state dynamics approach. In this article, we demonstrate the robustness of the method for solving the inverse problem of identifying a clean solution field from sparse and noisy measurements. This application can be extended for parameter estimation and identifying relevant constitutive relations for PDEs when only sparse and noisy data is available. We plan to perform a detailed study of these problems while comparing the model to parameterized version of PINNs, which have recently become popular for such inverse problems. One of the main features of this approach involves estimating both solution and fluxes, which can lead to inconsistency in multi-equation coupled PDEs. While this inconsistency can be removed by suitable model construction, these strategies must be tailored for specific PDEs. Future work in this direction involves reducing this inconsistency using a loss term that penalizes this consistency error. 

\section{Acknowledgements}

We would like to thank Ryosuke Park and Marc Charest as the primary developers of the \emph{flecsim} package we used for 2-D Euler calculations.  This work was supported by Laboratory Directed Research and Development program of Los Alamos National Laboratory under project number 20230068DR. BSS was partially supported by the DOE Office of Advanced Scientific Computing Research Applied Mathematics program through Contract No. 89233218CNA000001. Los Alamos National Laboratory Report LA-UR-25-23407.

\bibliographystyle{unsrt}
\bibliography{ref_bbl}

\newpage

\end{document}